\documentclass{aa}
\usepackage{graphicx}
\usepackage{caption}
\usepackage{txfonts}
\usepackage{epstopdf}
\usepackage{color}
\usepackage{multirow}
\usepackage{lscape}
\usepackage{amsmath}
\usepackage{amssymb}
\usepackage{upgreek}
\usepackage{hyperref}

\usepackage{subcaption}
\usepackage{comment}
\usepackage{upgreek}
\usepackage[dvipsnames]{xcolor}

\begin{document}

\title{Asteroseismology of four eccentric double-lined spectroscopic eclipsing binaries}
     \author{A. Liakos
           }

   \institute{Institute for Astronomy, Astrophysics, Space Applications and Remote Sensing, National Observatory of Athens,\\
              Metaxa \& Vas. Pavlou St., GR-15236, Penteli, Athens, Greece \\
              \email{alliakos@noa.gr}
       }
           \date{Received XX April 2025; accepted XX January 2025}


\abstract
{Photometric data from the Transiting Exoplanet Survey Satellite (TESS) mission and radial velocities from the \textit{Gaia} mission and ground-based observations were used to model the light curves and calculate the physical parameters of the eccentric eclipsing systems CH~Ind, V577~Oph, CX~Phe, and TIC~35481236. The components of these systems have temperatures between 6450 and 7500~K, masses between 1.4 and 1.85~M$_{\sun}$, and radii between  1.49 and 3.05~R$_{\sun}$. The residuals of these models were further analyzed using the Fourier method to reveal the pulsational frequencies of their oscillating components. Due to the similarity of the components of each system, the eclipses were used as spatial filters in order to determine which member is the pulsating star. CH~Ind was found to pulsate in 46 frequencies; its primary component is a $\gamma$~Dor star and the secondary a $\delta$~Sct star. The primary component of V577~Oph oscillates in both the regimes of $\gamma$~Dor and $\delta$~Sct stars. Moreover, using past timings of minima, an eclipse timing variation analysis was also performed for V577~Oph, resulting in the calculation of the apsidal motion parameters and the existence of a third body around the system. Both components of CX~Phe were found to be $\delta$~Sct stars; its primary has three independent frequencies in the range of 14.5-17.4~d$^{-1}$ and its secondary has two main modes of 5.19 and 7.22~d$^{-1}$. The analysis of TIC~35481236 indicates the hybrid $\delta$~Sct-$\gamma$~Dor nature of its secondary component. The physical and pulsational properties of the $\delta$~Sct stars of these systems were compared with those of other $\delta$~Sct stars-members of binaries in evolutionary diagrams.}


\maketitle
%

\section{Introduction}
\label{Sec:Intro}

The $\delta$~Sct stars exhibit rapid, multi-periodic pulsations within a frequency range of 4-80~d$^{-1}$. Their pulsations are primarily driven by radial and low-order non-radial pressure ($p$) modes, initiated by the $\kappa$-mechanism \citep[cf.][]{ZHE63, BRE00, AER10, BAL15}. Additionally, they can oscillate in high-order non-radial modes, influenced by turbulent pressure in the hydrogen convective zone  \citep{ANT14}. These stars typically have masses between 1.5 and 2.5~M$_{\sun}$ and belong to spectral types ranging from A to early F. They are predominantly located within the classical instability strip, extending from the main sequence to the giant branch \citep{AER10}.

The $\gamma$~Dor stars are pulsating stars that oscillate within the range 0.2-4~d$^{-1}$ and lie in the lower red part of the classical instability strip. They have similar physical parameters to $\delta$~Sct stars \citep{KAY99}, but their oscillations are driven by gravity ($g$) modes triggered by the convective blocking mechanism \citep{DUP05}.

Hybrid stars are those that exhibit characteristics of different types of stars. Particularly, due to the overlap of the instability strips of $\gamma$~Dor and $\delta$~Sct stars \citep{WAR03}, many stars display pulsations of both types \citep{HAN02, UYT11}. These hybrid stars are particularly valuable for asteroseismology, as the $g$-modes observed in $\gamma$~Dor stars provide insights into the stellar core, while the $p$-modes of $\delta$~Sct stars help probe the envelope.

Eclipsing binaries (EBs) are invaluable for determining the physical parameters of their components, such as masses, radii, and luminosities, as well as their evolutionary status. Especially, when light curves (LCs) and radial velocity (RV) curves are analyzed together, these parameters can be calculated very accurately. Additionally, EBs offer another powerful tool, that is the eclipse timing variations (ETV) method, which enables the detection of mechanisms that modulate the orbital period. EBs with a pulsating component are particularly significant, as they provide valuable insights into the properties and evolution of pulsating stars. Furthermore, these systems offer a unique opportunity to study the impact of binarity, such as proximity effects and mass transfer, on pulsation evolution. Detached EBs, especially those with relatively long orbital periods, provide the opportunity to study the stellar properties in terms of physical parameters and evolution but without any significant influence by the companions.

The $\delta$~Sct stars that belong to binaries has become a challenging topic since early 00's, when \citet{MKR02} introduced the term `oEA~stars' (oscillating EBs of Algol type) to characterize the mass accreting $\delta$~Sct stars. Since then, many discoveries have been made, especially from the space missions of Kepler \citep{BOR10}, K2 \citep{KOC10}, and Transiting Exoplanet Survey Satellite \citep[TESS;][]{RIC15}. Fundamental characteristics and correlations between their properties (e.g.,~orbital-pulsation period) have been discussed in \citet{SOY06a}, \citet{LIA12}, \citet{ZHA13}, and \citet{KAH17}. The main catalogue of these systems was published by \citet{LIAN17} \citep[see also][]{LIAN15, LIAN16} and its preliminary updated version was given by \citet{LIA25}. It should to be noticed that the last few years the `Stars WIth Pulsations and Eclipses' \citep[SWIPE;][]{SOU25} project has been initiated and focuses on the pulsating stars in detached EBs. \citet{LIA20a} using the sample of $\delta$~Sct stars in detached binaries published new correlations, which were later updated by \citet{LIA25}.

The high precision (on the order of 10$^{-4}$~mag) and continuous time coverage of photometric data from the NASA/TESS mission provide an excellent foundation for detailed studies of stellar oscillations. This strength is further enhanced when combined with multi-band photometry and high-resolution spectroscopy. TESS observations offer temporal resolutions ranging from 20~seconds to 30~minutes, making them particularly effective for detecting short-period frequencies -- a defining trait of $\delta$~Sct stars. Importantly, the precision of TESS data also allows for the detection of low-amplitude pulsations down to a few $\upmu$mag \citep[cf.][]{KUR20, KIM21, LIA22}. Additionally, the mission's extended temporal coverage, often spanning several years, is crucial for accurately determining the timings of minima in EBs, thereby enabling a more comprehensive analysis of their orbital period variations.

The ESA/\textit{Gaia} mission \citep{GAIA16} was designed to create the most precise 3D map of the Milky Way by measuring the positions, distances, and motions of over a billion stars. One of its key instruments, the Radial Velocity Spectrometer \citep{CRO18} that is a near-infrared (845–872~nm), medium-resolution, integral-field spectrograph, provides crucial spectroscopic data by measuring the Doppler shifts in stellar spectra. This allows \textit{Gaia} to determine the radial velocities of stars, complementing its astrometric data and enabling a full 3D view of stellar motions. The data from this instrument \citep{GAIA22, GAIA23} are essential for studying the kinematics, dynamics, and structure of the Galaxy, as well as identifying binary systems and stellar populations.

Based on the catalogue of \citet{LIAN17}, its updated version \citep{LIA25}, and the current list\footnote{\url{https://alexiosliakos.weebly.com/catalogue.html}} hosted by the author (to be published in a forthcoming study), over 900 binaries containing $\delta$~Sct components have been identified to date. However, accurate physical and/or detailed pulsation parameters for the oscillating stars are lacking in more than 90\% of these systems. Currently, only about 77 systems have well-defined parameters, meaning they are both double-lined spectroscopic and eclipsing binaries (SB2+E). Particularly, there are only 35 detached, 41 semi-detached and one unclassified SB2+E systems hosting at least one $\delta$~Sct star. This highlights the importance of new studies delivering precise absolute parameters of pulsating components, especially when combined with high-quality photometric data from space missions, to advance Asteroseismology.

This paper is dedicated to a detailed investigation of individual detached EBs containing $\delta$~Sct component(s). We present various types of analyses of four eclipsing binaries, namely CH~Ind, V577~Oph, CX~Phe, and TIC~35481236. All these systems are known EBs exhibiting pulsations but their absolute stellar parameters as well as their detailed pulsation models have not been accurately determined to date. The goals of this work are: a)~the calculation of the physical parameters of the pulsating components of the systems, b)~the detailed frequency analysis and the estimation of the oscillation modes of the pulsators, c)~the study of the orbital period changes (if any) of the systems, and d)~the comparison of the results with other systems of similar type.

We note that in this paper the terms `primary' and `secondary' components correspond to the more and less luminous component of each system, respectively, and not to their more and less massive nor to their hotter and cooler star-members. Accordingly, the primary eclipse corresponds to the orbital phase 0.0 during which the more luminous component is eclipsed by its companion. Moreover, we note that in all tables of this work, the errors are given in parentheses alongside values and correspond to the last digit(s).

\subsection{Brief history}
\label{Sec:HIS}

All systems were observed by the TESS mission in various sectors and with different time resolutions, as well as by the \textit{Gaia} mission (particularly with the Radial Velocity Spectrometer), which determined the amplitudes of the RVs of both components. Details for these observations are given in Section~\ref{Sec:DATA}.

CH~Ind ($m_{\rm V}=$7.5~mag \citep{PAU15}, HD~204370, TIC~139699256, \textit{Gaia}~DR3~6466658483488677376) was discovered as a variable by \citet{STR66}. It was included as a target in the Radial Velocity Experiment \citep[RAVE;][]{KUN17, STE20}. Pulsations in its LC were first noticed by \citet{MKR22}, who stated that the dominant pulsational frequency has a value of 8.85~d$^{-1}$. \citet{SHI22} and \citet[][they provided approximately half of the orbital period]{KAH23} found some basic characteristics of the binary system (e.g.,~color, period). The effective temperature of the system ranges between 6780 and 7500~K in various studies \citep{OTE03, TSA22, GAIA22}, but the most recent one assigns it a value of 6900~K \citep{VER24}. \citet{GAIA22} includes the mass values 1.87~M$_{\sun}$ and 1.90~M$_{\sun}$ for the primary and secondary components, respectively.

The eclipsing system V577~Oph ($m_{\rm V}=$11.2~mag \citep{KER22}, \textit{Gaia}~DR3~4477836145791337088, TIC~416302408, HIP~89579) is a known variable since the mid 30's \citep{HOF35}. Its eccentric orbit has been studied by many researchers \citep[cf.][]{SHU85, HAG88, JEF17, KOZ19}. Its pulsational behaviour was first discovered by \citet{VOL90} and re-studied by \citet{DIE93}, \citet{ZHO01a}, and \citet{VOL10b}. The dominant frequency of the oscillating member was found as 14.39~d$^{-1}$. The most recent works on the system were published by \citet{CRE10} and \citet{JEF17}, who provided RVs of both components and calculated their mass functions. Temperatures of the system are given in various catalogues and range between 6996 and 7174~K \citep{STA19, AND22, GAIA22}. The most recent one is provided by \citet{VER24} as 7000~K. The automatic calculation of masses by \citet{GAIA22} resulted in 1.67~M$_{\sun}$ and 1.76~M$_{\sun}$ for the primary and secondary components, respectively. However, the RVs provided by \textit{Gaia} are based on wrong orbital period value (see Sect.~\ref{Sec:DATA}).

The variability of CX~Phe ($m_{\rm V}=$8.8~mag \citep{KER22}, \textit{Gaia}~DR3~4930629881808500096, TIC~158536052, HD~8093) was discovered by \citet{STR65}. Its temperature is referred in the range of 6904-7090~K \citep{DON12, STE17, TON18, STA19, PAE21, GAIA22, AND22, VER24}. The discovery of the oscillations in the system was made by \citet{KAH23}, who resulted in a dominant frequency of 5.186~d$^{-1}$. The automatic calculation of the masses of the components by \citet{GAIA22} resulted in 1.54~M$_{\sun}$ and 1.38~M$_{\sun}$ for the primary and secondary components, respectively.

The binarity of TIC~35481236 ($m_{\rm V}=$10.4~mag \citep{NAS16}, \textit{Gaia}~DR3~2946311659129414144, ASAS~J064753-1642.9) was discovered by The All Sky Automated Survey \citep[ASAS;][]{POJ97}. According to the catalogues of \citet{STA19}, \citet{AND22}, \citet{GAIA22}, and \citet{DOY24}, its temperature is between 7249 and 7800~K. \citet{SHI22} discovered the pulsations in the system and proposed a dominant frequency of $\sim25$~d$^{-1}$. Later on, \citet{KAH23} re-examined the system and concluded that the main oscillation mode has a frequency of $\sim6$~d$^{-1}$. Mass values of 1.52~M$_{\sun}$ and 1.59~M$_{\sun}$ for the primary and secondary components, respectively, are referred in \citet{GAIA22}.


\section{Observational data}
\label{Sec:DATA}

The TESS data of the selected EBs were obtained from the Mikulski Archive for Space Telescopes (MAST\footnote{\url{https://mast.stsci.edu}}). The Pre-search Data Conditioning Simple Aperture Photometry (PDCSAP) flux—typically corrected for long-term instrumental trends—was used when available. In cases where the PDC pipeline introduced distortions in the LCs, the Simple Aperture Photometry (SAP) flux was used instead. The systems' magnitudes ($T$), assumed to represent their maximum brightness, were taken from \citet{FET23} and \citet{SCH19} and used for converting flux to magnitude. A summary of the TESS observations for each system is provided in Table~\ref{Tab:TESSlog}.

This study aims to determine the pulsational frequencies of the oscillating members of the systems. The frequency analysis is very sensitive to the: a)~time resolution and continuity (i.e.~time-gaps) of the data sample, b)~total duration of the observations, and c)~photometric accuracy of the data. Thus, for the following analyses, we selected the TESS data sets with the highest time resolution, the longest duration, the least number of observation gaps, and the best quality. For CH~Ind, we selected the 2-min cadence data set of sector~68 (17664 data points within 25.3~d) instead of the respective one of sector~1, because the data of the latter appear distorted. For V577~Oph, the only available data set is that of sector~80 that contains a total number of 10463 data points with a cadence of 200~s within a 26.4~day interval. The best data set for CX~Phe was found to be that of sector~2 (2-min cadence, 18279 data points within 26.8~days). Although there are two more data sets with the same time resolution (sectors~29 and 69), the selected one has the least number of time gaps and due to the long orbital period of the system (i.e.~$\sim20$~d), it provides the best LC coverage. Finally, TIC~35481236 was observed in two successive sectors (6 and 7) with a time resolution of 2~min. We combined these data sets, thus we used a total time coverage of 47.8~d and 30878 data points. The TESS LCs of all systems are plotted in Fig.~\ref{fig:LCs}.

All systems were observed with the Radial Velocity Spectrometer \citep{CRO18} of the \textit{Gaia} mission; they were identified as SB2 systems, with the \textit{Gaia} pipeline providing RV measurements for both components. More specific, \textit{Gaia} provides the semi-amplitudes $K$ of the RVs of both components in the \textit{Gaia}~DR3 part~3 \citep{GAIA22, GAIA23} based on the Non-Single-Star (NSS) model adopted from the NSS solution. These $K$ values are listed in Table~\ref{Tab:LCmodels} and used for the mass ratio calculation (Sect.~\ref{Sec:Models}). It should to be noted that these $K$ values were used for all systems except for V577~Oph. For the latter system, the \textit{Gaia} NSS model reports wrong orbital period value, therefore, the $K$ values might not be trustworthy. However, instead of \textit{Gaia} measurements, we used the $K$ values published by \citet{JEF17}.

\begin{table}
\begin{center}
\caption{TESS observations log for the selected EBs.}
\label{Tab:TESSlog}
\begin{tabular}{ccc| ccc}
\hline\hline											
\multicolumn{3}{c}{\textbf{CH~Ind}}			&	\multicolumn{3}{c}{\textbf{TIC~35481236}}	\\
\hline											
Sector	&	Cadence	&	$T$ 	&	Sector	&	Cadence	&	$T$ 	\\
	&	(s)	&	(mag)	&		&	(s)	&	(mag)	\\
\hline										
1	&	120, 1800	&	\multirow{ 4}{*}{7.15}	&	6	&	120, 1800	&	\multirow{ 4}{*}{9.97}	\\
27	&	600	&		&	7	&	120, 1800	&		\\
28	&	600	&		&	33	&	600	&		\\
68	&	120, 200	&		&	87	&	120, 200	&		\\
\hline											
\multicolumn{3}{c}{\textbf{V577~Oph}	}				&	\multicolumn{3}{c}{\textbf{CX~Phe}}				\\
\hline											
80	&	200	&	10.35	&	2	&	120, 1800	&	\multirow{3}{*}{8.46}	\\
	&		&		&	29	&	120, 600	&		\\
	&		&		&	69	&	120, 200	&		\\
\hline											
						
\end{tabular}
\end{center}																										
\end{table}


\section{Light curve modelling}
\label{Sec:Models}

To mitigate the influence of pulsations on the binary modeling, phased LCs were employed in the subsequent iterative process, rather than modeling each cycle individually, as the pulsation amplitudes are non-negligible relative to the eclipse depths. For all cases, and given that their dominant pulsation frequency (see Sec.~\ref{Sec:Puls}) is not an integer multiple of their orbital period (or frequency), the phased data can be used as mean magnitude values for all phases.

The phases of the data points of all systems, except for CX~Phe, were calculated based on linear ephemerides (i.e.~time of primary minimum $T_0$ and orbital period $P$), which were derived from a linear fit on the timings of primary minima of the data sets used in the analyses. For CX~Phe, due to its long period value ($\sim$20~d), only one orbital cycle is covered by a TESS sector. Thus, we used all the available TESS data (Table~\ref{Tab:TESSlog}) for calculating its ephemeris. The times of minima of the systems were calculated using the \citet{KWE56} method and they are listed in Table~\ref{Tab:MIN}.

\begin{table*}
\begin{center}
\caption{Light curve modelling parameters of the studied EBs.}
\label{Tab:LCmodels}
\begin{tabular}{l cc cc cc cc}
\hline\hline																	
System:	&	\multicolumn{2}{c}{CH~Ind}			&	\multicolumn{2}{c}{V577~Oph}			&	\multicolumn{2}{c}{CX~Phe}			&	\multicolumn{2}{c}{TIC~35481236}			\\
\hline																	
	&	\multicolumn{8}{c}{System parameters}															\\
\hline																	
$T_0$~(BJD)	&	\multicolumn{2}{c}{2460155.84289(2)}			&	\multicolumn{2}{c}{2460482.3375(4)}			&	\multicolumn{2}{c}{2458364.5609(2)}			&	\multicolumn{2}{c}{2458469.22630(6)}			\\
$P$~(d)	&	\multicolumn{2}{c}{5.95257(1)}			&	\multicolumn{2}{c}{6.0793(2)}			&	\multicolumn{2}{c}{19.97589(1)}			&	\multicolumn{2}{c}{1.843433(4)}			\\
$i$~(deg)	&	\multicolumn{2}{c}{89.41(1)}			&	\multicolumn{2}{c}{89.24(4)}			&	\multicolumn{2}{c}{88.67(1)}			&	\multicolumn{2}{c}{84.20(8)}			\\
$e$	&	\multicolumn{2}{c}{0.058(5)}			&	\multicolumn{2}{c}{0.18(6)}			&	\multicolumn{2}{c}{0.36(1)}			&	\multicolumn{2}{c}{0.13(1)}			\\
$\omega$~(deg)	&	\multicolumn{2}{c}{222(1)}			&	\multicolumn{2}{c}{50(1)}			&	\multicolumn{2}{c}{87(1)}			&	\multicolumn{2}{c}{42(1)}			\\
$q$	&	\multicolumn{2}{c}{1.020(1)}			&	\multicolumn{2}{c}{0.871(4)}			&	\multicolumn{2}{c}{0.890(1)}			&	\multicolumn{2}{c}{1.037(1)}			\\
\hline																	
	&	\multicolumn{8}{c}{Components parameters}															\\
\hline																	
	&	$Primary$	&	$Secondary$	&	$Primary$	&	$Secondary$	&	$Primary$	&	$Secondary$	&	$Primary$	&	$Secondary$	\\
\hline																	
$T_{\rm eff}$~(K)	&	6900(200)$^a$	&	6908(75)	&	7000(200)$^a$	&	6901(99)	&	7000(200)$^a$	&	6478(81)	&	7400(200)$^a$	&	7463(84)	\\
$\Omega$	&	8.122(1)	&	8.887(1)	&	11.39(1)	&	11.33(1)	&	24.02(1)	&	29.91(1)	&	6.898(1)	&	7.551(1)	\\
$F$	&	1.0(1)	&	1.0(1)	&	1.0(1)	&	1.0(1)	&	1.0(1)	&	1.0(1)	&	1.0(1)	&	1.0(1)	\\
$A^a$	&	1	&	1	&	1	&	1	&	1	&	0.5	&	1	&	1	\\
$g^a$	&	1	&	1	&	1	&	1	&	1	&	0.32	&	1	&	1	\\
$r_{\rm mean}$	&	0.143(1)	&	0.131(1)	&	0.097(1)	&	0.087(1)	&	0.0442(5)	&	0.0315(5)	&	0.177(1)	&	0.162(1)	\\
$x$	&	0.301	&	0.301	&	0.298	&	0.301	&	0.298	&	0.320	&	0.257	&	0.257	\\
$L/(L_1+L_2)$	&	0.543(1)	&	0.457(1)	&	0.565(1)	&	0.435(1)	&	0.712(1)	&	0.288(1)	&	0.535(1)	&	0.465(1)	\\
$K$~(km~s$^{-1}$)	&	91.6(2)$^b$	&	90.0(2)$^b$	&	79(2)$^c$	&	91.7(2)$^c$	&	59(1)$^b$	&	66(2)$^b$	&	130(1)$^b$	&	124(1)$^b$	\\
\hline	
\end{tabular}
\end{center}
\textbf{Notes.} $^a$assumed, $^b$\citet{GAIA22, GAIA23}, $^c$\citet{JEF17}																													
\end{table*}

\begin{figure*}
\centering
\begin{tabular}{cc}
\includegraphics[width=8.9cm]{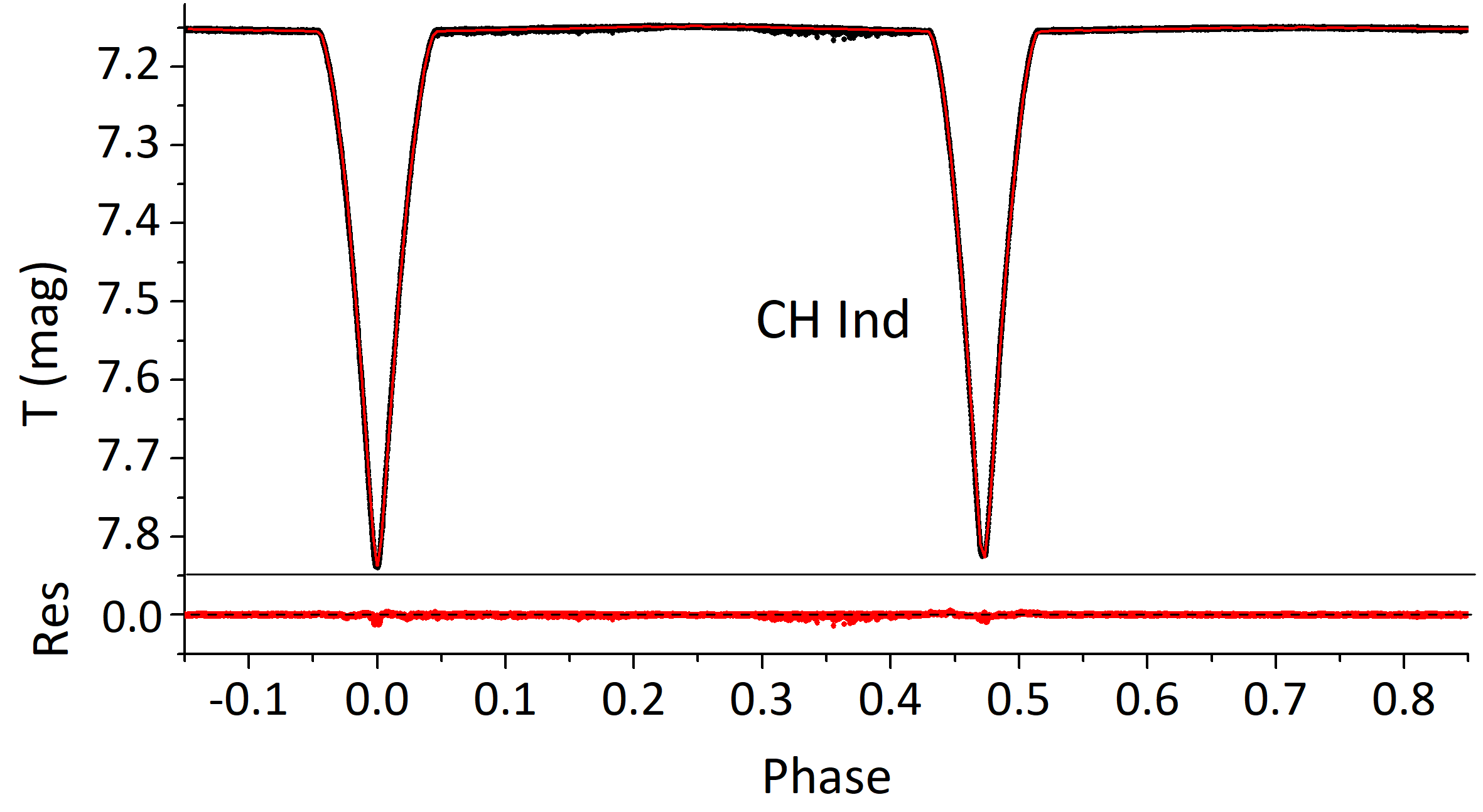}&\includegraphics[width=8.9cm]{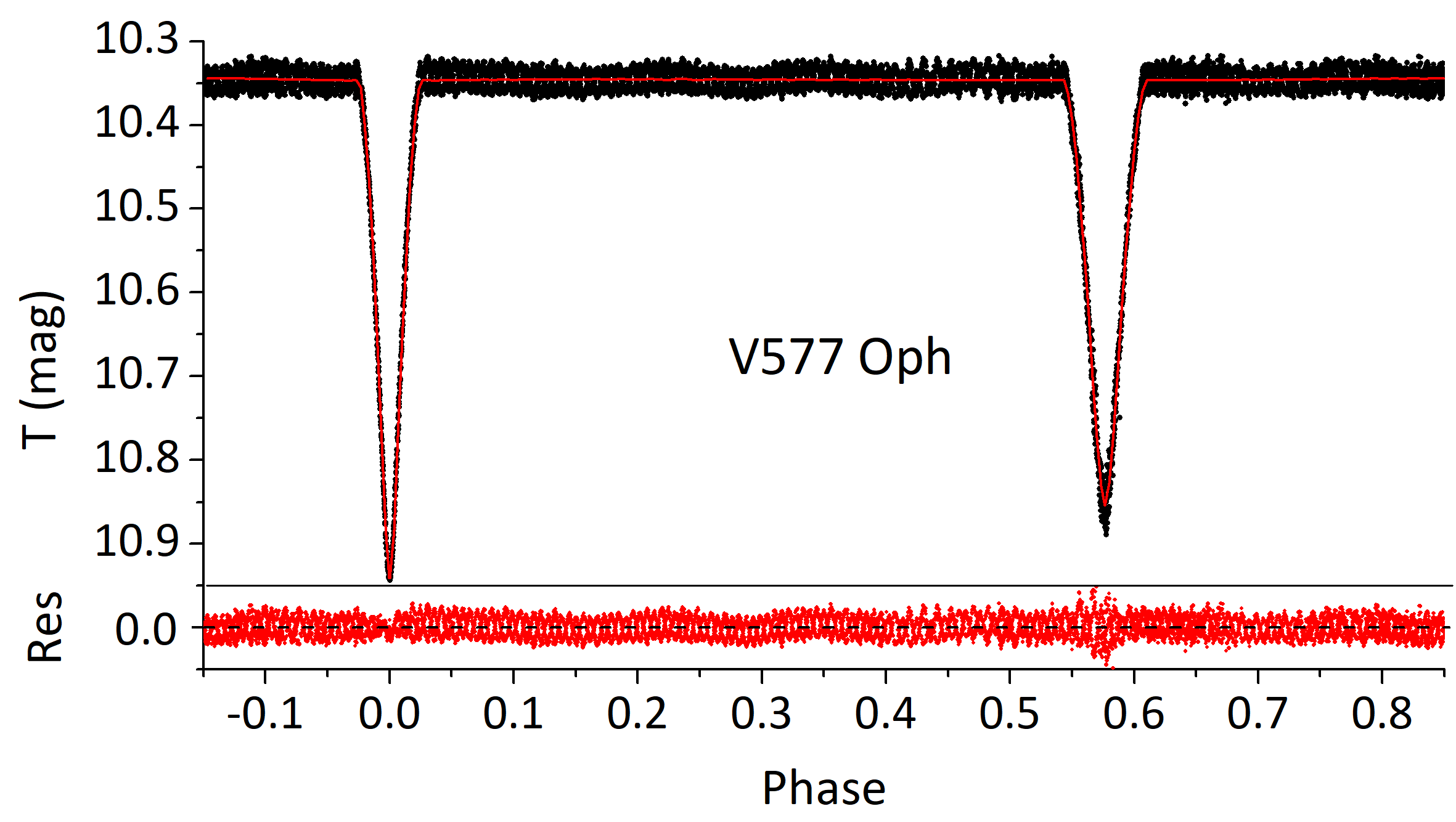}\\
\includegraphics[width=8.9cm]{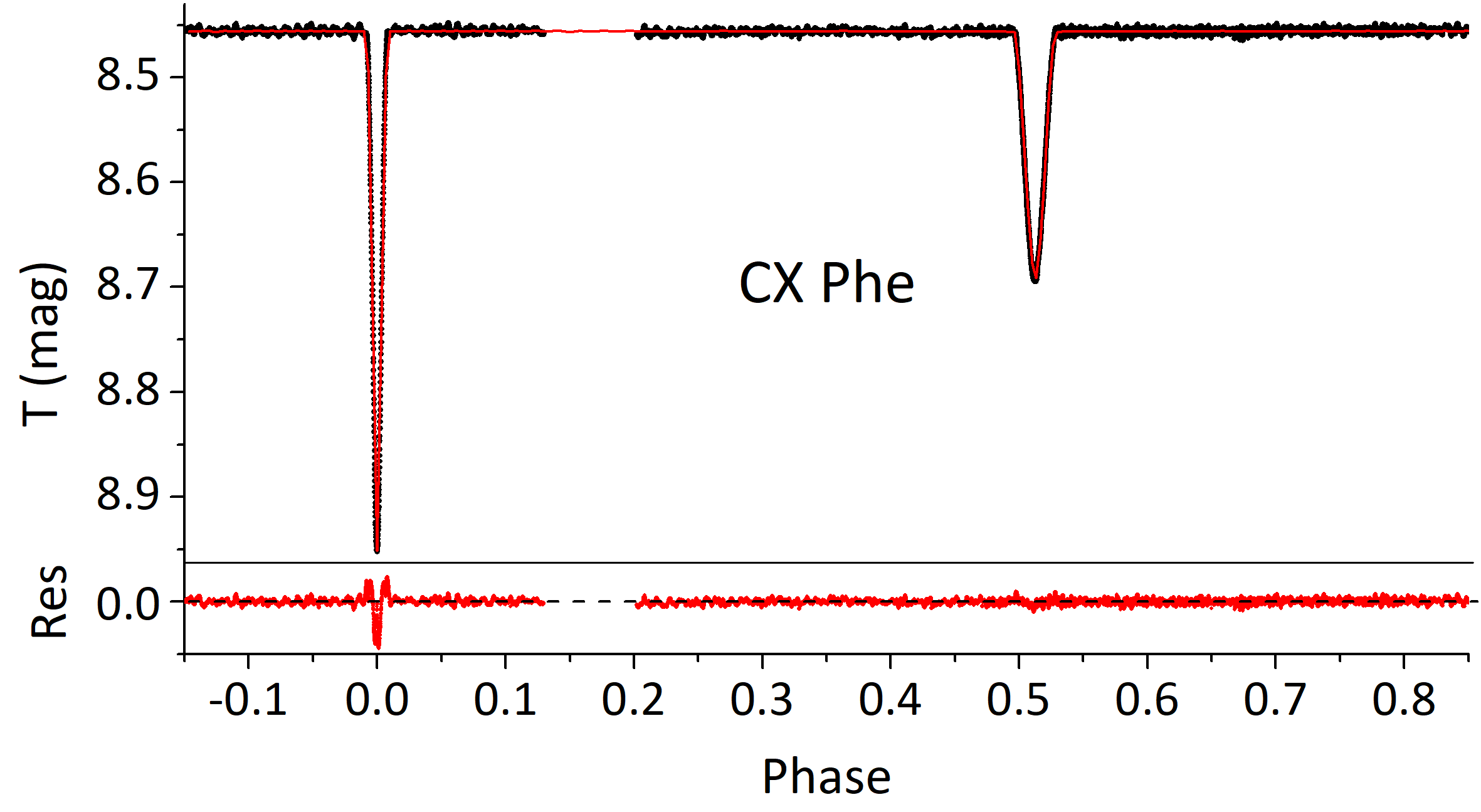}&\includegraphics[width=8.9cm]{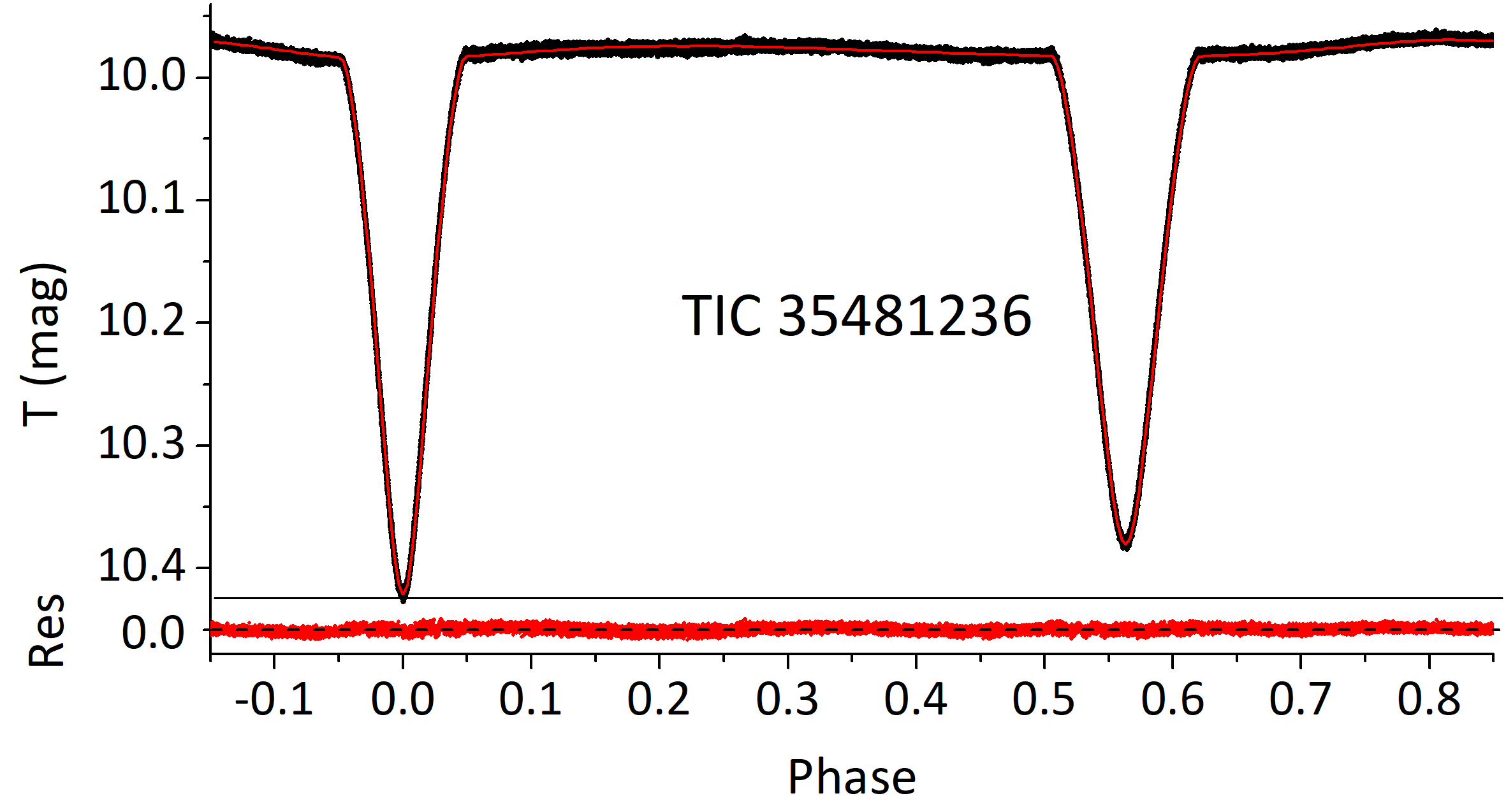}\\
\end{tabular}
\caption{$TESS$ phased (black points) and synthetic (red lines) light curves and residuals of the binary modelling (lower panels--red points) of all studied cases.}
\label{fig:LCMs}
\end{figure*}

The TESS LCs of the systems were analyzed using the PHOEBE v.0.31 software \citep{PRS05}, which incorporates the 2003 version of the Wilson-Devinney (WD) code \citep{WIL71}, and applies the Markov Chain Monte Carlo (MCMC) method for the errors calculation. Temperatures ($T_{\rm eff, 1}$) of the primary components were given values derived from the literature (see Sect.~\ref{Sec:HIS}). Particularly, the $T_{\rm eff, 1}$ of CH~Ind and V577~Oph were assigned values catalogued in the work of \citet{VER24}. In that study, the authors employed a fitting procedure where \textit{Gaia} BP/RP spectra were matched against a grid of synthetic spectra. These synthetic spectra are generated based on stellar atmosphere models that span a range of stellar parameters, notably effective temperature, surface gravity, and metallicity [Fe/H]. By identifying the best-fitting synthetic spectrum to each observed BP/RP spectrum, the corresponding stellar parameters, including $T_{\rm eff}$, are inferred. The similarity in $T_{\rm eff}$ between the components of each system (i.e.~similar eclipse depths; see Fig.~\ref{fig:LCMs}) makes the $T_{\rm eff}$  assumptions of \citet{VER24} for the primaries of these two systems highly plausible. For the other two systems, namely CX~Phe and TIC~35481236, the values are the average of those reported in recent catalogues (Sec.~\ref{Sec:HIS}). Similarly to the previous two cases, the components of TIC~35481236 have very similar temperatures. The average of the $T_{\rm eff}$ values of the various catalogues is $\sim7400$~K, which is, again, plausible for assigning it to the primary component. The average $T_{\rm eff}$ of CX~Phe is $7000\pm100$~K according to eight catalogues; its components have a relatively high temperature difference, but the primary dominates the spectrum. Therefore, maybe its spectrum, hence its $T_{\rm eff, 1}$  is slightly affected by the secondary. For all cases a reasonable error in $T_{\rm eff, 1}$ of 200~K was assumed. The primary temperatures were held fixed throughout the analysis, while the secondary temperatures ($T_{\rm eff, 2}$) were treated as adjustable parameters. The mass ratio values ($q$) of the systems were calculated based on the semi-amplitude values ($K$; see Sect.~\ref{Sec:DATA}) of the RVs of their components (i.e.~$q=M_2/M_1=K_1/K_2$) and were allowed to vary within their respective uncertainty range. For all systems, the albedos ($A$) and gravity darkening coefficients ($g$) of the components were assigned values based on their spectral types \citep{RUC69, ZEI24, LUC67}. As the secondary minima of the LCs are displaced from the orbital phase 0.5, the eccentricity ($e$) and the argument of periastron ($\omega$) were also enabled and adjusted. The limb darkening coefficients ($\chi$) were taken from the tables of \citet{CLA18}. The dimensionless potentials ($\Omega$), the fractional luminosity of the primary component ($L_1$), the inclination of the system ($i$), the synchronicity parameters ($F$), and the relative radii ($r$) were also adjusted during the modelling. Moreover, in the case of V577~Oph, the third light parameter ($l_3$) was tested initially due to possible existence of a third body around the system (Sect.~\ref{Sec:ETV}). Nevertheless, it converged to a zero value and, therefore, it was excluded from further modelling.

For all systems, we detected no asymmetries in the LC maxima between successive cycles. Therefore, we solved the respective phased LCs including all the data points. However, we note that TIC~35481236 exhibits an asymmetry between its maxima in every cycle, which was modelled using the $e$ and $\omega$ parameters. Given, that all systems have eccentric orbits, the mode~2 of the WD code (i.e.~detached binary) was selected for the fit. The modelling results are listed in Table~\ref{Tab:LCmodels}, the fittings on the observational points are illustrated in Fig.~\ref{fig:LCMs}, while the LC residuals are plotted against phase and time in the lower panels of Figs.~\ref{fig:LCMs} and ~\ref{fig:LCs}, respectively. It should to be noted that the $q$ errors in this table are those derived directly from the LC fittings and not the propagating errors from the $K$ values.


\section{Physical parameters and evolutionary status}
\label{Sec:AbsPar}

Using the LC modelling results, which are based on almost fixed mass ratios due to the use of the $K$ values, we are able to calculate the absolute parameters of both components of each system. For this, we employed the AbsParEB software under mode~1 \citep[i.e.~information from both spectroscopy and photometry;][]{LIA15}. The results are given in Table~\ref{Tab:AbsPar}, with the listed parameters to denote the standard astrophysical quantities (i.e.~mass, radius, luminosity, gravity acceleration, semi-major axis, and bolometric magnitude). The errors in this table were calculated using the error propagation method from the values and their respective errors of the $i$, $P$, $e$, $r_{\rm mean}$, and $K$ parameters listed in Table~\ref{Tab:LCmodels}. Additionally, by utilizing the absolute luminosities of the components, along with the extinction \citep[$A_{\rm V}$; taken from][]{KHA24}, the TESS magnitude (Table~\ref{Tab:TESSlog}), and the TESS bolometric correction \citep[$BC_{\rm TESS}$; taken from the grid models of the Modules for Experiments in Stellar Astrophysics (MESA) Isochrones \& Stellar Tracks (MIST);][]{PAX18} for the components' temperatures, $\log~g$, and solar metallicity, we can determine the TESS extinction \citep[$A_{\rm TESS}\approx0.77~A_{\rm V}$;][]{STA19} and the absolute magnitude of the system. Subsequently, the distance ($D$) of the system is derived using the distance modulus, with the results provided in the same table.

\begin{figure}
\centering
\includegraphics[width=\columnwidth]{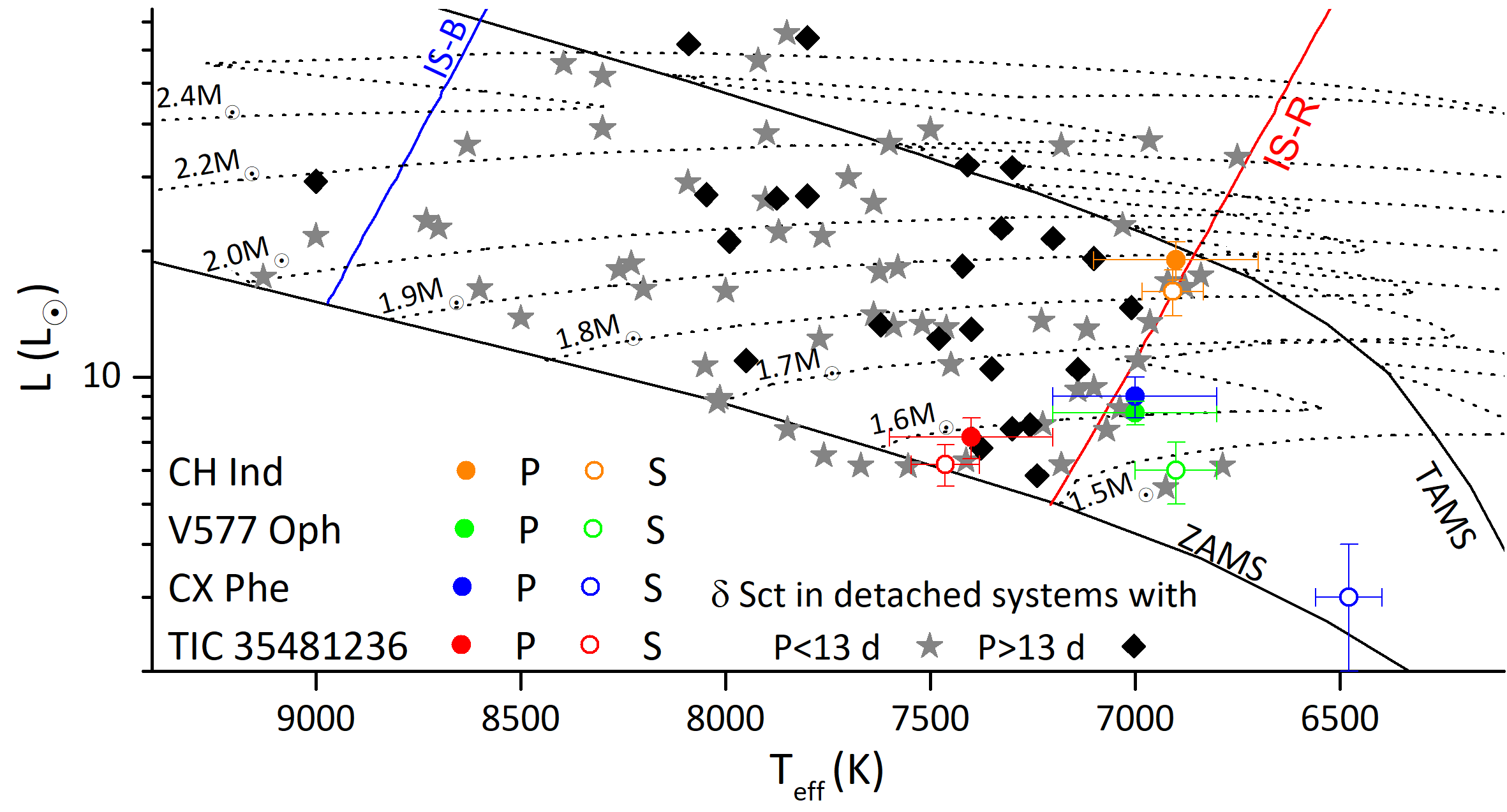}\\
\includegraphics[width=\columnwidth]{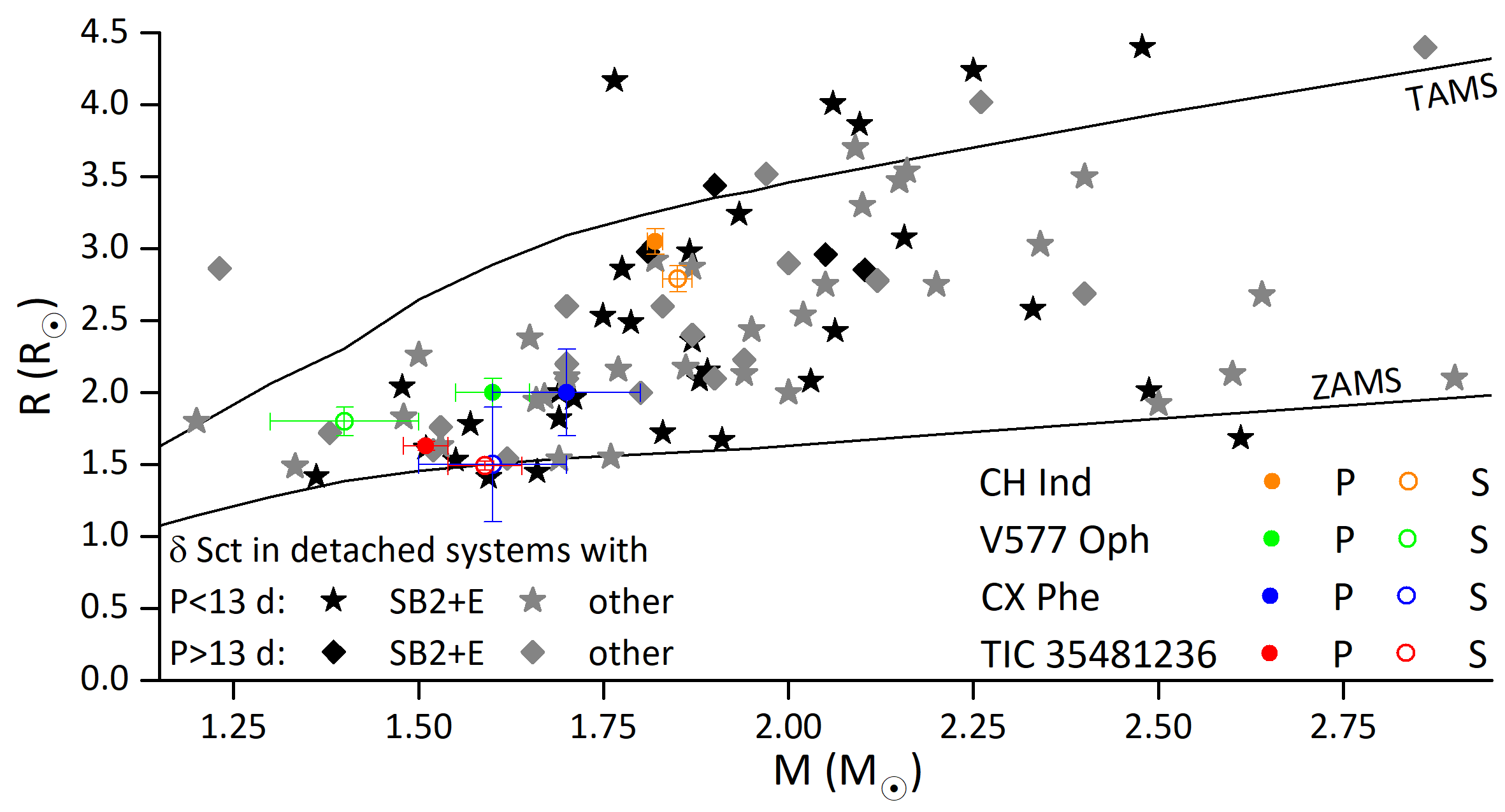}\\
\caption{Hertzsprung–Russell (top panel) and mass--radius (bottom panel) diagrams including the components of all studied systems (colored filled and empty circles) along with other $\delta$~Sct components of detached binaries with $P_{\rm orb}<13$~d (star symbols) and $P_{\rm orb}>13$~d (diamond symbols). Moreover, in the mass-radius diagram the $\delta$~Sct stars of SB2+E systems (black star and diamond symbols) are distinguished from those that are members of other type of binaries (gray star and diamond symbols). The data of the binary $\delta$~Sct stars were taken from \citet{LIAN17}, \citet{LIA20a}, and \citet{LIA25}. The boundaries of the instability strip (IS) were taken from \citet{SOY06b} and the theoretical stellar evolutionary tracks from \citet{GIR00}.}
\label{fig:EVOL}
\end{figure}

\begin{table}									
\begin{center}									
\caption{Physical parameters of the components of all EBs.}									
\label{Tab:AbsPar}									
\begin{tabular}{l cc cc}									
\hline\hline									
	&	\multicolumn{2}{c}{CH~Ind}			&	\multicolumn{2}{c}{V577~Oph}			\\
\hline									
$M$~(M$_{\sun}$)	&	1.82(1)	&	1.85(2)	&	1.60(5)	&	1.4(1)	\\
$R$~(R$_{\sun}$)	&	3.05(9)	&	2.79(9)	&	2.0(1)	&	1.8(1)	\\
$L$~(L$_{\sun}$)	&	19(2)	&	16(2)	&	8.2(5)	&	6(1)	\\
$\log g$~(cm~s$^{-2}$)	&	3.73(3)	&	3.82(3)	&	4.06(6)	&	4.09(8)	\\
$a$~(R$_{\sun}$)	&	10.76(2)	&	10.57(2)	&	9.3(3)	&	10.8(1)	\\
$M_{\rm bol}$ (mag)	&	1.6(7)	&	1.7(5)	&	2.5(9)	&	2.8(9)	\\
$A_{\rm V}$~(mag)	&	\multicolumn{2}{c}{0.137}			&	\multicolumn{2}{c}{0.526}			\\
$BC_{\rm TESS}$ (mag)	&	\multicolumn{2}{c}{0.316}			&	\multicolumn{2}{c}{0.052}			\\
$D$ (pc)	&	\multicolumn{2}{c}{195(8)}			&	\multicolumn{2}{c}{401(17)}			\\
\hline									
	&	\multicolumn{2}{c}{CX~Phe}			&	\multicolumn{2}{c}{TIC~35481236}			\\
\hline									
$M$~(M$_{\sun}$)	&	1.7(1)	&	1.6(1)	&	1.51(3)	&	1.59(5)	\\
$R$~(R$_{\sun}$)	&	2.0(3)	&	1.5(4)	&	1.63(3)	&	1.49(3)	\\
$L$~(L$_{\sun}$)	&	9(1)	&	3(1)	&	7.2(8)	&	6.2(7)	\\
$\log g$~(cm~s$^{-2}$)	&	4.1(3)	&	4.3(5)	&	4.19(2)	&	4.29(2)	\\
$a$~(R$_{\sun}$)	&	21.7(4)	&	24.3(7)	&	4.72(4)	&	4.50(4)	\\
$M_{\rm bol}$ (mag)	&	2.4(8)	&	3.4(9)	&	2.6(7)	&	2.8(4)	\\
$A_{\rm V}$~(mag)	&	\multicolumn{2}{c}{0.016}			&	\multicolumn{2}{c}{0.294}			\\
$BC_{\rm TESS}$ (mag)	&	\multicolumn{2}{c}{0.397}			&	\multicolumn{2}{c}{0.116}		\\
$D$ (pc)	&	\multicolumn{2}{c}{229(14)}		&	\multicolumn{2}{c}{375(13)}			\\
\hline									
\end{tabular}									
\end{center}									
\end{table}

The positions of all systems' star-members are plotted in the Hertzsprung–Russell (HR) and mass-radius diagrams in Fig.~\ref{fig:EVOL}. The components of all systems, except those of CX~Phe and the secondary of CH~Ind, follow well the single-star evolutionary tracks in the HR diagram. On the contrary, the primary of CX~Phe appears less luminous than expected for its mass, while its secondary deviates a lot (much less luminous for its mass) from the single-star evolutionary track of $M$=1.5~M$_\sun$. These discrepancies potentially indicate mass exchange between the components or mass loss from the system in the past. The secondary of CH~Ind ($M$=1.85(2)~M$_{\sun}$) is very close to the single-star evolutionary track of 1.8~M$_{\sun}$, thus, it appears slightly less luminous than expected. All stars, except the secondary of CX~Phe, are located very close to the red edge of the classical instability strip. TIC~35481236 has the less evolved components on the main sequence compared to the other systems. Its star-members, along with the secondaries of CX~Phe and V577~Oph, are located very close to the zero age main sequence (ZAMS). The primaries of CX~Phe and V577~Oph are almost in the middle of the main sequence plane. On the contrary, the components of CH~Ind appear as the more evolved stars of the systems studied and they are positioned very close to the terminal age main sequence (TAMS). All the aforementioned conclusions stand also for the positions of all stars in the mass-radius diagram. In both diagrams, we also include other $\delta$~Sct members of detached binary systems for comparison \citep[taken from][and the personal list of the author]{LIAN17, LIA20a, LIA25}. In the HR diagram, 62 pulsators in detached binaries with $P_{\rm orb}<13$~d are observed, along with 24 additional pulsators in similar systems with $P_{\rm orb}>13$~d. The mass-radius diagram contains, 31+4 $\delta$~Sct stars in detached SB2+E systems with shorter and longer orbital period than 13~d, respectively, and 31 (with $P_{\rm orb}<13$~d) + 19 (with $P_{\rm orb}>13$~d) in detached binaries, whose physical properties were calculated with less accurate methods (e.g.,~SB1+E).

Masses and radii of CH~Ind and TIC~35481236 have an accuracy of more than 3.2\%, the masses of V577~Oph and CX~Phe more than 7\%, and the radii of V577~Oph better than 6\%. Finally, the accuracy for the radii of the components of CX~Phe range between 15 and 26.7\%. Our results for the masses of these stars agree totally with the \textit{Gaia} automatic calculations (see Sect.~\ref{Sec:HIS}) only for the case of TIC~35481236. We have close agreement for both components of CH~Ind (difference 2.1-2.7\%) and the primary of V577~Oph (difference 4.2\%). On the contrary, there is a discrepancy of 10.9-11.7\% for the components of CX~Phe (we result in higher mass values) and a difference of 21.6\% for the secondary of V577~Oph (we resulted in lower mass value). Regarding the distances, the \textit{Gaia} DR3 Lite Distances \citep{BAI21} refer 202, 615, 226, and 405~pc for CH~Ind, V577~Oph, CX~Phe, and TIC~35481236, respectively. The percentage differences between these values and those in Table~\ref{Tab:AbsPar} are 4\%, 42\%, 1\%, and 8\%, respectively for each system. Our results agree perfectly with \textit{Gaia} for CH~Ind and CX~Phe, marginally for TIC~35481236, and show a large difference for V577~Oph. For the latter system, this significant discrepancy may arise from possible error in the adopted $BC_{\rm TESS}$ value. However, further investigation on this issue is beyond the scope of this paper.

\section{Eclipse Timing Variation analysis}
\label{Sec:ETV}	

The only system, among those examined in this study, with available historical times of minima is V577~Oph. Particularly, there are 20 old minima timings in the time interval 1928-1999. These timings were downloaded from the VarAstro\footnote{\url{https://var.astro.cz/en}} web database and the `Bob Nelson's $O-C$ files\footnote{\url{https://binaries.boulder.swri.edu/binaries/omc/}}' and were combined with those calculated in this work (seven in total; Table~\ref{Tab:MIN}) to analyze the orbital period variations of the system. The minima between 1928 and 1964 were based on photographic plates measurements and they present too much scatter in the ETV diagram. Thus, these timings were neglected and we used only those that are based on photoeletcric and CCD observations (inc. TESS data). In total, we used 14 primary and six secondary times of minima in the following ETV analysis.

For this analysis, we employed the dedicated code APSIDAL MOTION + LITE\footnote{\url{https://sirrah.troja.mff.cuni.cz/~zasche/Programs.html}} \citep{ZAS09}. The code uses statistical weights on the times of minima based on the method followed for the observations of the eclipses. However, since all the minima used are trustworthy, we assigned them the same statistical weight. The code is able to perform a fit to the ETV points with Light-Time Effect \citep[LITE;][]{IRW59}, parabolic, and apsidal motion curves. Given that the primary and secondary minima present symmetric behaviour, the apsidal motion parameters were enabled. The results of the LC modelling (Sect.~\ref{Sec:Models}) yielded that the system is detached, thus, the parabola, which is connected to mass transfer between the components or mass loss from the system, was neglected. The LITE curve was also selected for fitting since there is periodical behaviour of the ETV points. The apsidal motion curve has five free parameters; the ephemeris of the EB ($T_0$ and $P$; taken from the VarAstro database), the orbital eccentricity~($e$), the argument of periastron~($\omega$), and its first derivative~($\dot{\omega}$). The LITE curve includes seven free parameters: the ephemeris of the binary, the variation amplitude~($A$), the orbital period~($P_3$), the time of periastron passage~($JD_0$), the argument of periastron~($\omega_3$), and the orbital eccentricity~($e_3$) of the tertiary component.

\begin{figure}
\centering
\includegraphics[width=\columnwidth]{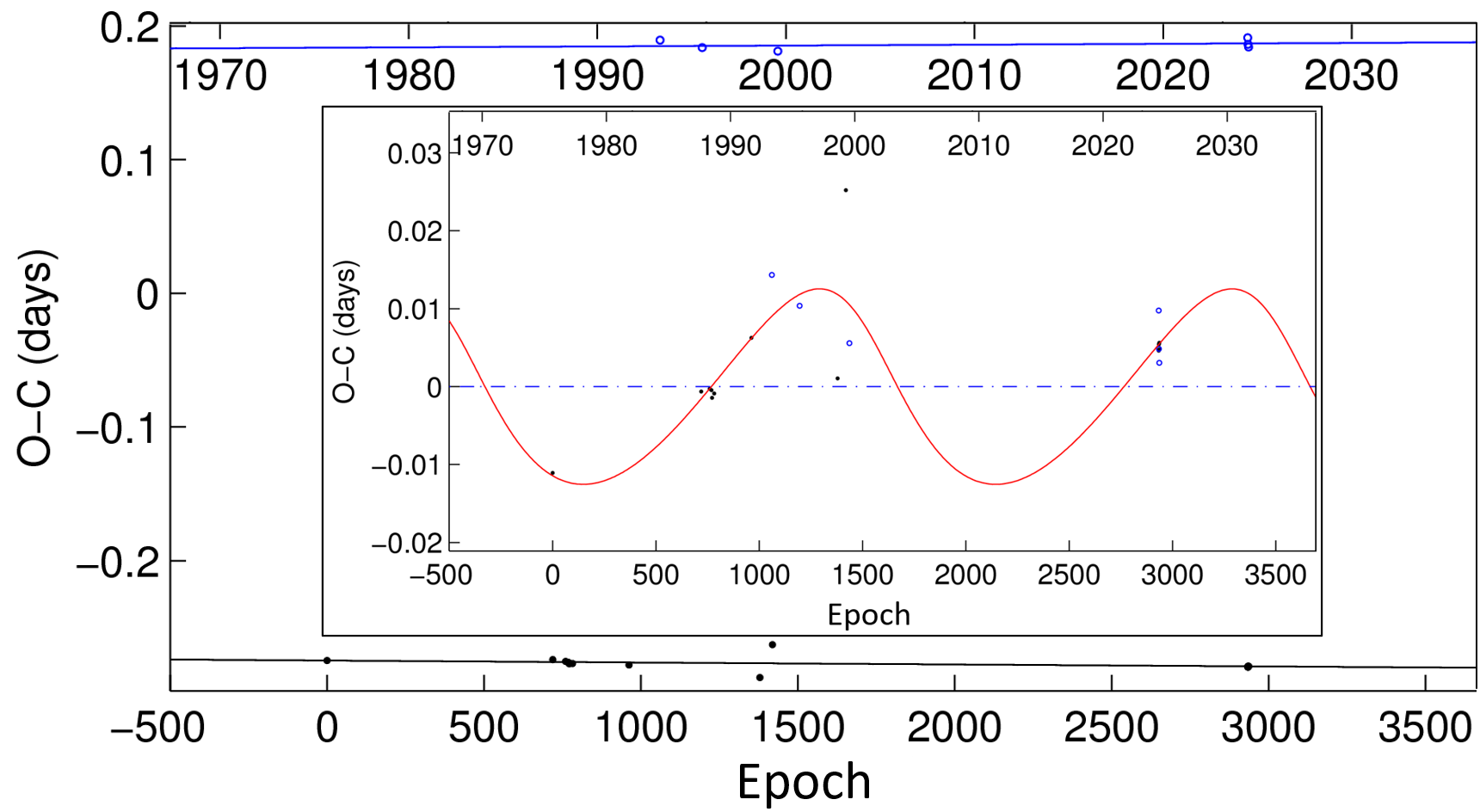}
\caption{Fitting of apsidal motion curves (black and blue solid lines) on the primary (black points) and secondary (blue points) minima timings of V577~Oph. The inner plot shows the fit on the residuals of this model by a LITE curve (red solid line).}
\label{fig:ETV}
\end{figure}

\begin{table}					
\begin{center}					
\caption{Parameters of the ETV analysis of V577~Oph.}					
\label{Tab:ETV}
\scalebox{0.95}{
\begin{tabular}{lc|lc}
\hline\hline							
	\multicolumn{2}{c}{Apsidal motion}		&	\multicolumn{2}{c}{LITE}			\\
\hline							
$T_0$~(HJD)	&	2442652.57(1)	&	$P_3$~(yr)	&	33(2)	\\
$P$~(d)	&	6.07910(6)	&	$JD_0$~(HJD)	&	2451953(278)	\\
$e$	&	0.12(9)	&	$A$~(d)	&	0.013(3)	\\
$\omega$~($\degr$)	&	347(74)	&	$\omega_3$~($\degr$)	&	147(22)	\\
$\dot{\omega}$~($\degr$~cycle$^{-1}$)	&	0.0012(5)	&	$e_3$	&	0.28(8)	\\
$U$~(yr)	&	4956(115)	&	$f(m_3)$~(M$_{\sun}$)	&	0.010(2)	\\
	&		&	$M_{3, \rm min}$~(M$_{\sun}$)	&	0.50(6)	\\
\hline																						
\end{tabular}}
\end{center}					
\end{table}					
	
The results are listed in Table~\ref{Tab:ETV}, while the ETV diagrams for each detected orbital period modulating mechanism are illustrated in Fig.~\ref{fig:ETV}. Regarding the EB, we found an apsidal motion with a period ($U$) of $\sim5000$~yr. The $e$ and $\omega$ derived from this analysis come in agreement within the error ranges with the respective results from the LC modelling (Table~\ref{Tab:LCmodels}). A potential third body with a minimal mass of 0.5~M$_{\sun}$ was found to orbit the eclipsing pair. Using the InPeVeb software under the mode `Light-Time effect' \citep{LIA15}, we found that the tertiary component, assuming to be a main sequence star, should contribute with $\sim0.6\%$ to the total luminosity. Therefore, the non-detection of a third light in the LC modelling (Sect.~\ref{Sec:Models}) seems reasonable.

\section{Pulsation Analyses}
\label{Sec:Puls}

According to the physical parameters of the components of the studied cases and their positions in the evolutionary diagrams (Sect.~\ref{Sec:AbsPar}), all of them are candidates to be pulsating stars either of $\delta$~Sct or $\gamma$~Dor type or even hybrids of these types. Unfortunately, no total eclipses occur in any of these systems, which could provide direct proof regarding which component pulsates. Nevertheless, the eclipses still play the role of a spatial filter and can be used to determine, at least roughly, if one or both components oscillate. Thus, for all systems, we performed frequency analyses in both the in-eclipse and out-of-eclipse data to detect all the pulsation frequencies and to attribute them to the correct component. Since the duration of the eclipses varies for each system and the secondary minimum is shifted from the orbital phase ($\Phi_{\rm orb}$) 0.5 due to the eccentricity, Table~\ref{Tab:datalog} presents the different phase intervals of each system used for frequency analyses. We note that the subsequent pulsation models are based solely on the analysis of out-of-eclipse data to prevent amplitude variations in the frequencies, which result from the total light fluctuations caused by eclipses.

The classical Fourier analysis technique was applied in the different phase parts of each binary using the software PERIOD04 \citep{LEN05} to extract the pulsation frequencies. We note that, due to the significantly lower number of data points of the eclipses of the systems, only the strong frequencies were derived for these phase parts and used only for the determination of the oscillating component(s). The main frequency search, aimed at constructing a complete pulsation model, was conducted using the out-of-eclipse data. Pulsations were analyzed within the frequency range of 0–80~d$^{-1}$, encompassing the frequency domains of both $\delta$~Sct \citep{BRE00, BOW18} and $\gamma$~Dor stars \citep{HEN07, QIA19}. For the detection of reliable frequencies and since many frequencies are very close to each other (i.e.~the local background appears increased), we used the method proposed by \citet{LIA17} \citep[cf.][]{LIAN20} regarding the signal-to-noise ratio $S/N$ calculation. Briefly, we calculated the background noise of the data sets in a region where no frequencies exist within a 2~d$^{-1}$ range using a box size of 2. Thus, the calculated $S/N$ of each detected frequency is the ratio of its amplitude over the background noise. After determining each frequency, the residuals were pre-whitened before identifying the next one. The search continued until the $S/N$ of the detected frequency reached approximately 5 \citep{BAR15, BOW21}. The number of points ($N$), the background noise ($N_{\rm bgd}$), the Nyquist frequency ($f_{\rm Nyq}$), and the frequency resolution  \citep[$\delta f=1.5/\delta t$, where $\delta t$ is the observations time range in days; cf.][]{LOU78} of each used data set are listed in Table~\ref{Tab:datalog}.

\begin{figure*}
\centering
\begin{tabular}{cc}		
\includegraphics[width=8.8cm]{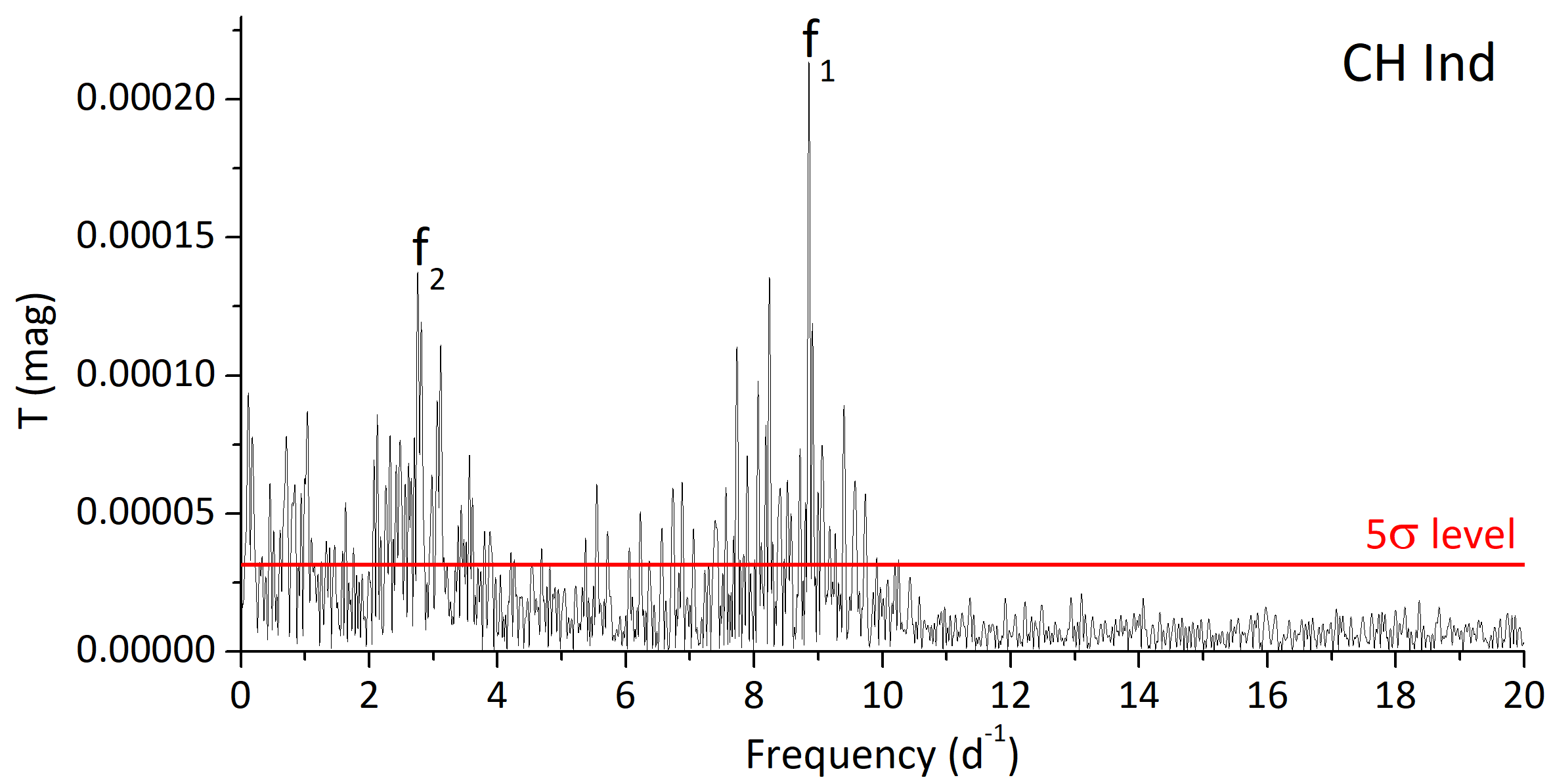}&\includegraphics[width=8.8cm]{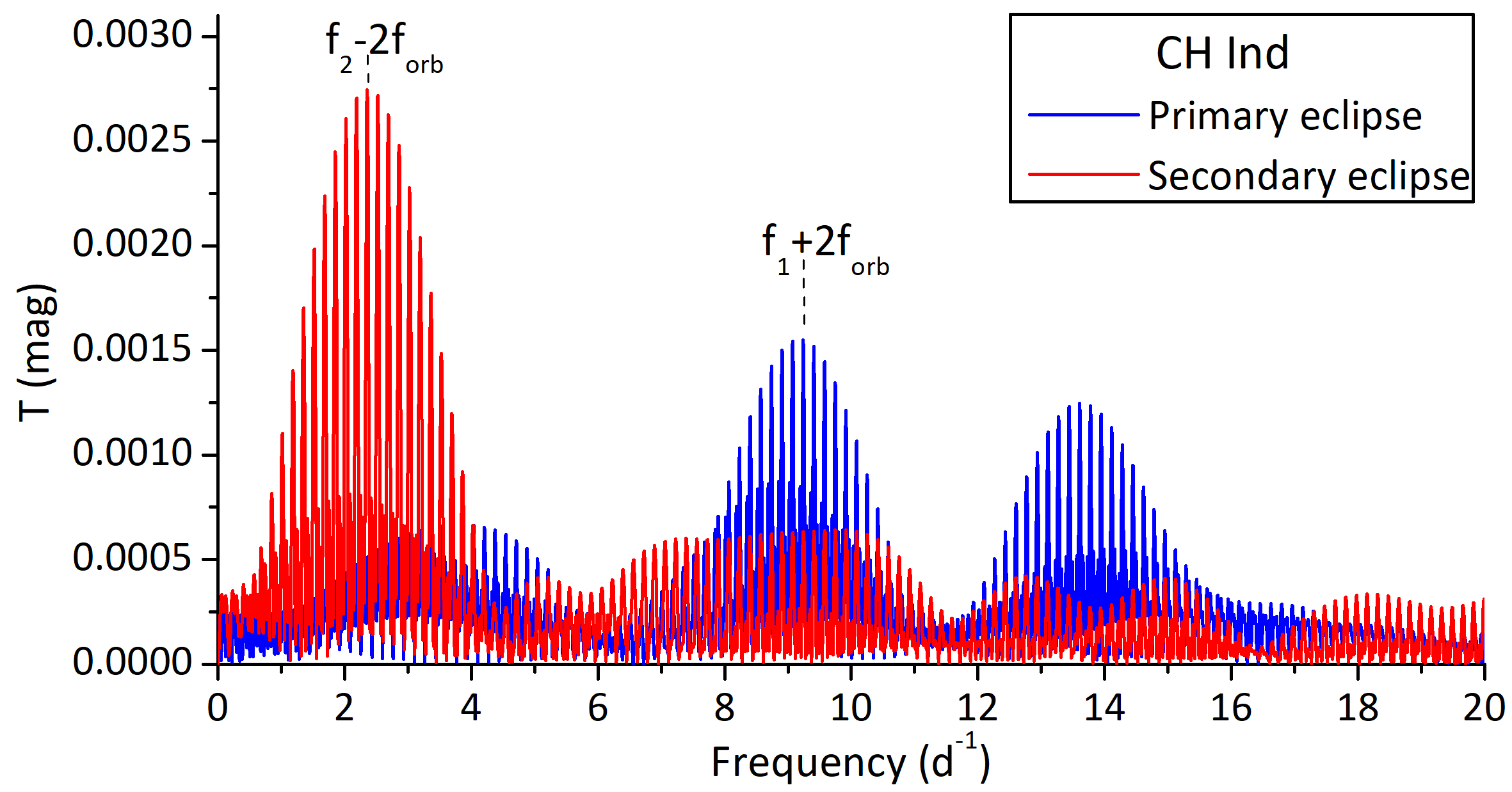}\\
\includegraphics[width=8.8cm]{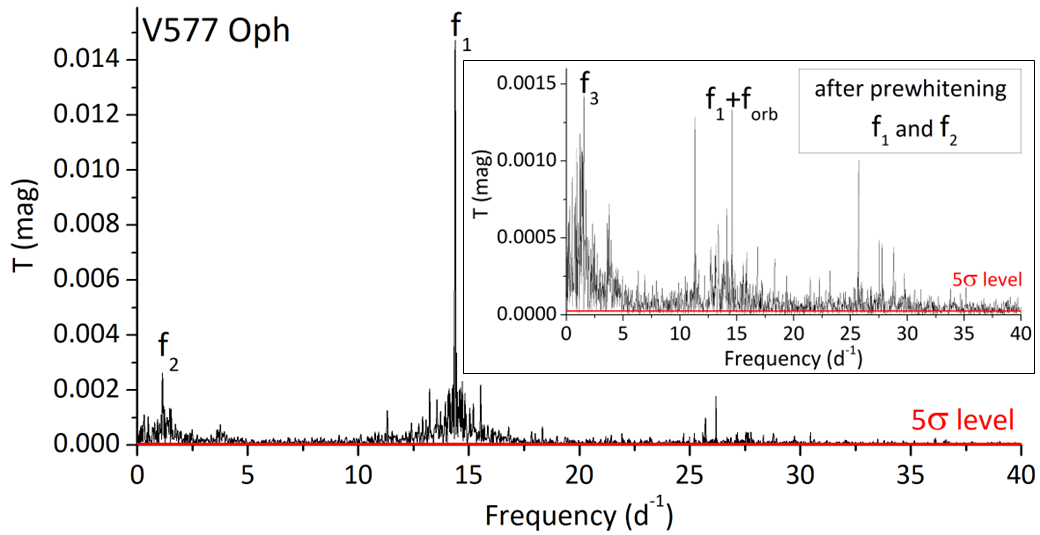}&\includegraphics[width=8.8cm]{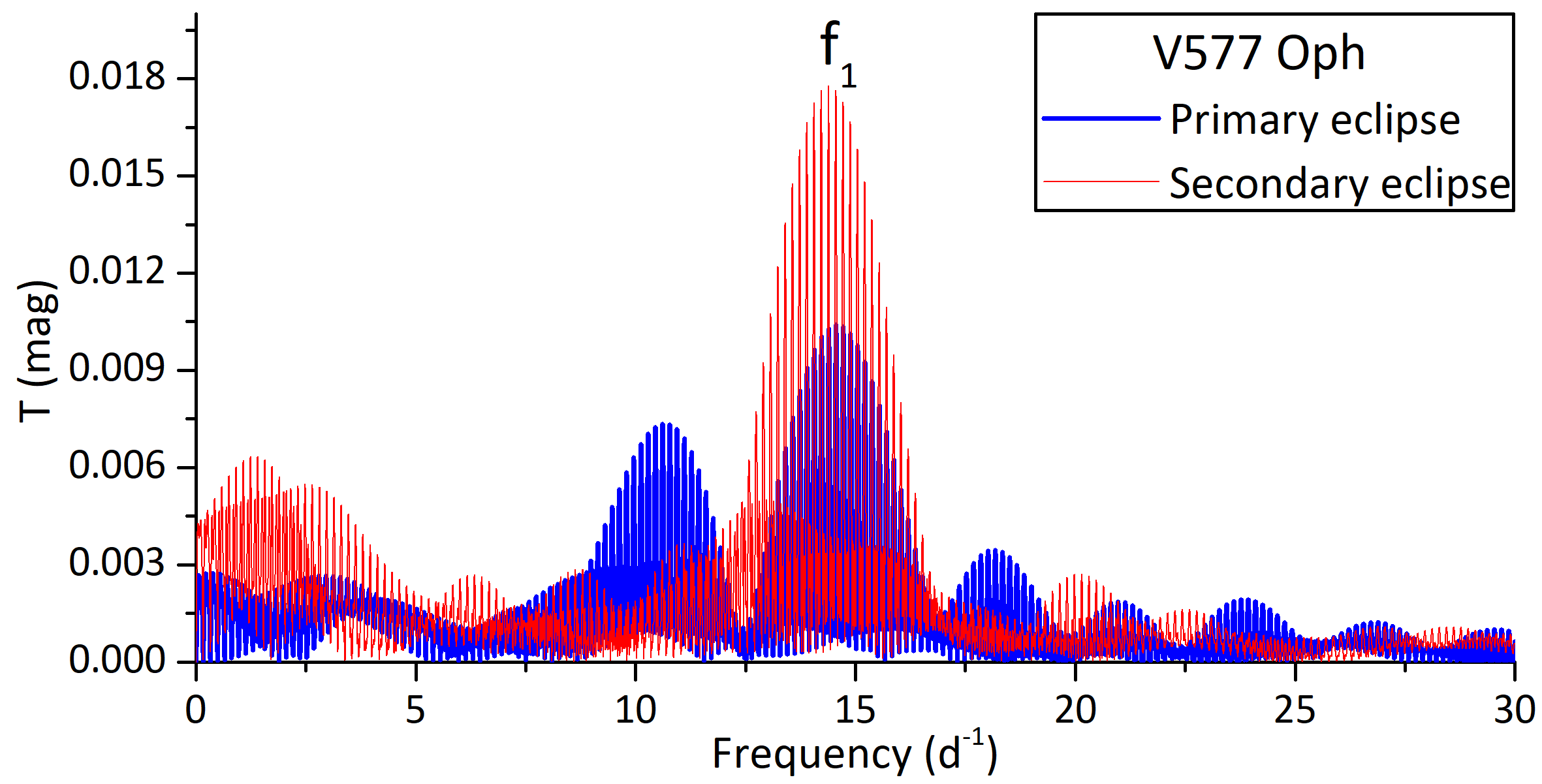}\\
\includegraphics[width=8.8cm]{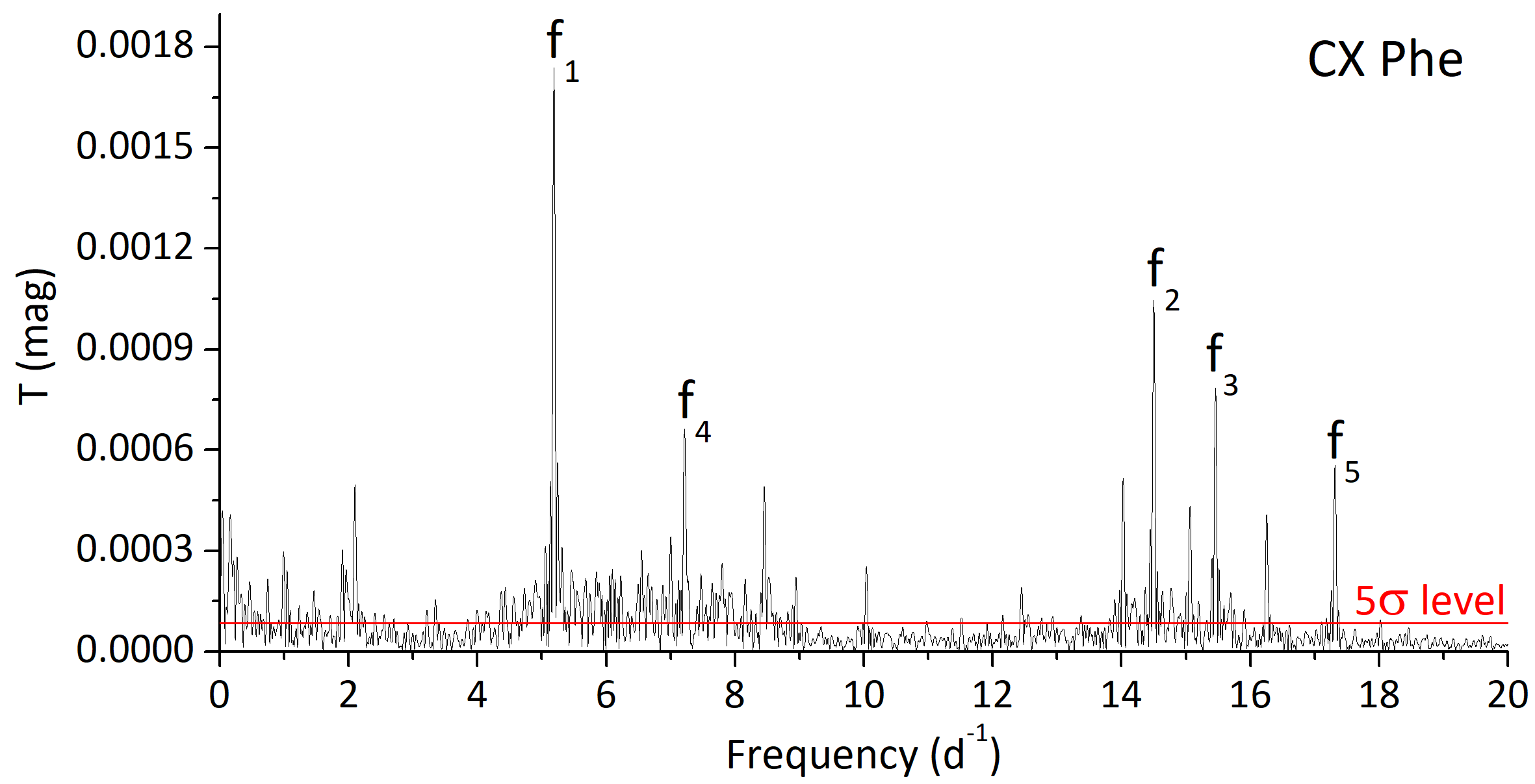}&\includegraphics[width=8.8cm]{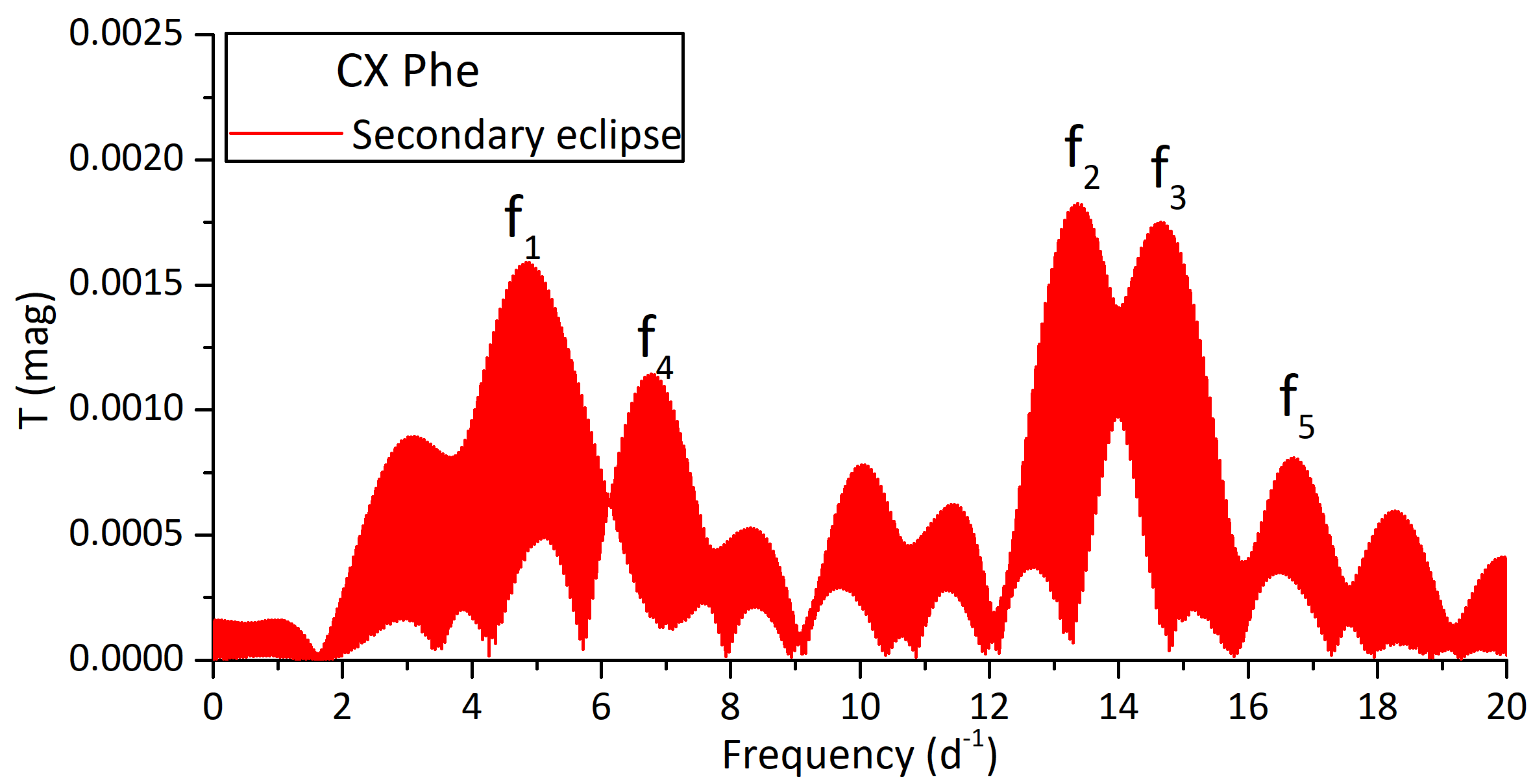}\\
\includegraphics[width=8.8cm]{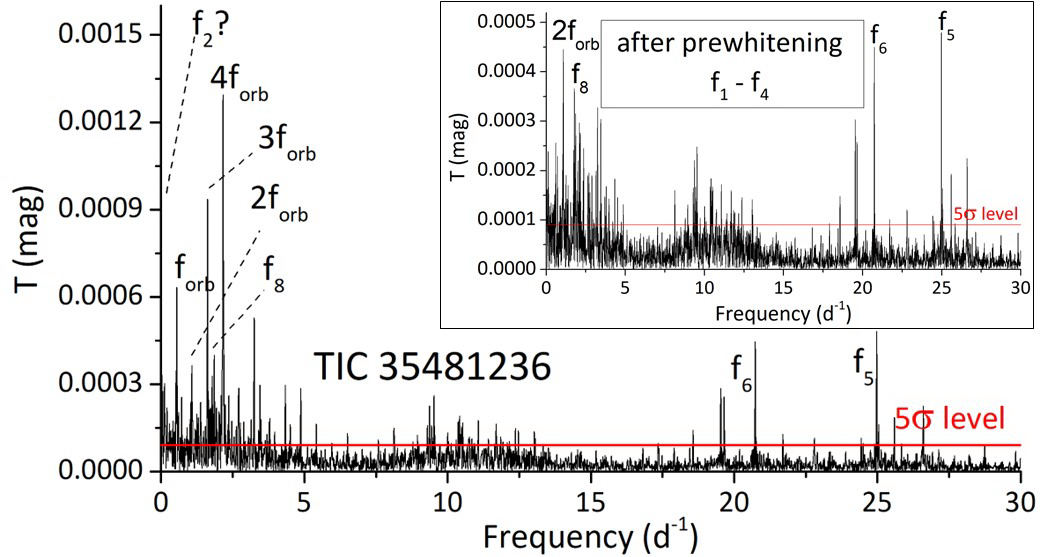}&\includegraphics[width=8.8cm]{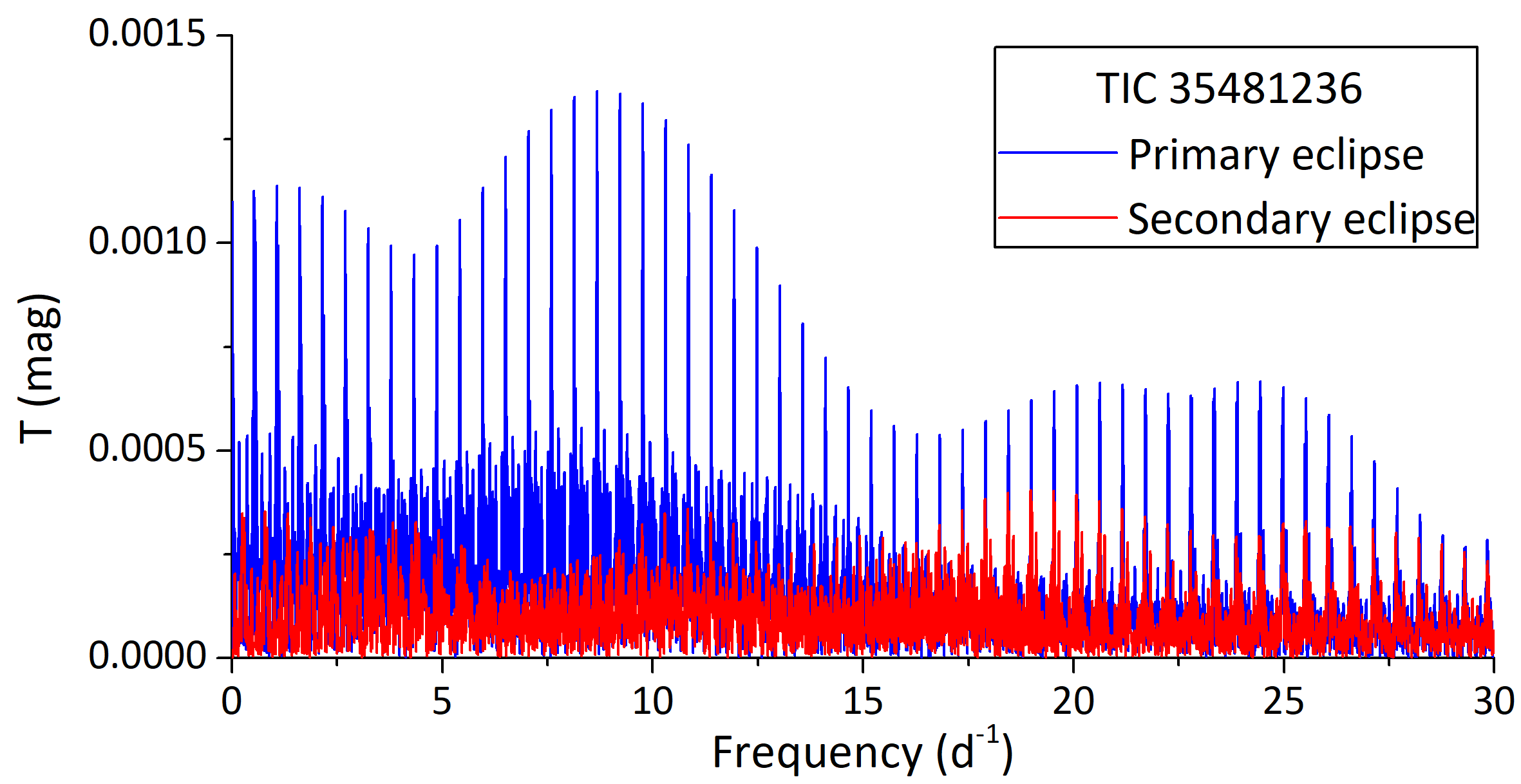}\\
\end{tabular}
\caption{Periodograms of the out-of-eclipse data points (left panels) and the in-eclipse data points (right panels) of all systems studied. The internal panels for V577~Oph and TIC~35481236 are the periodograms after prewhitening the very strong frequencies. }
\label{fig:FS}
\end{figure*}

\begin{figure*}
\centering
\includegraphics[width=18cm]{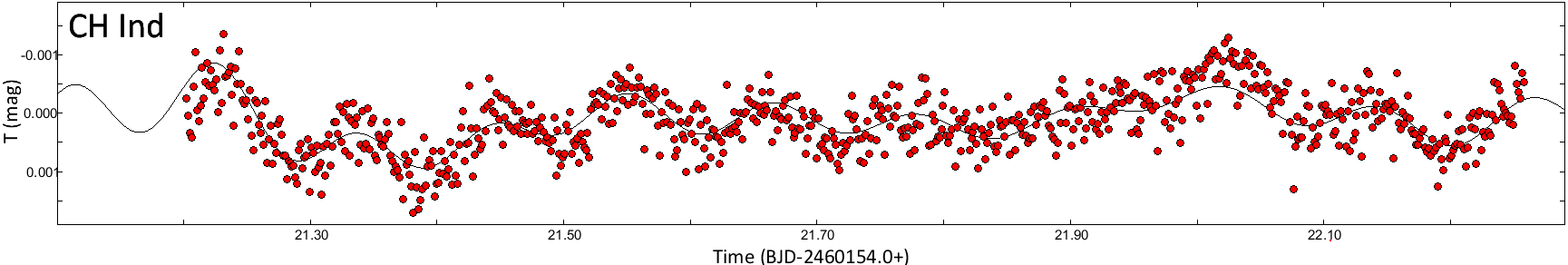}\\
\includegraphics[width=18cm]{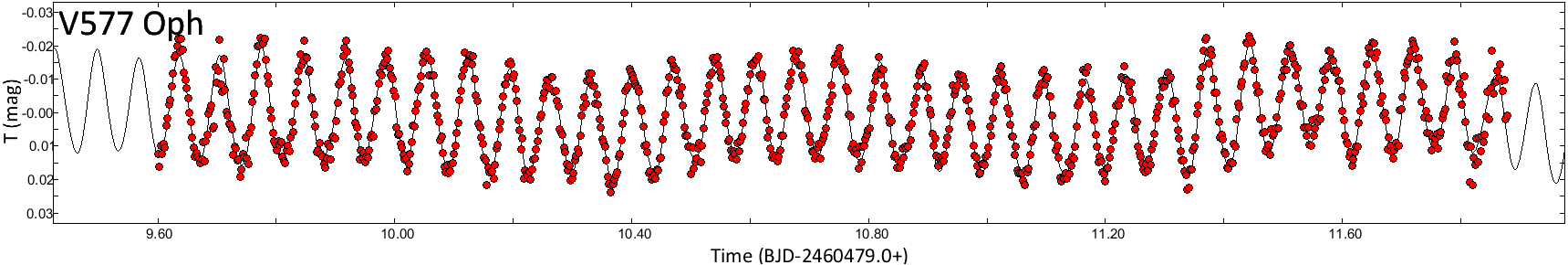}\\
\includegraphics[width=18cm]{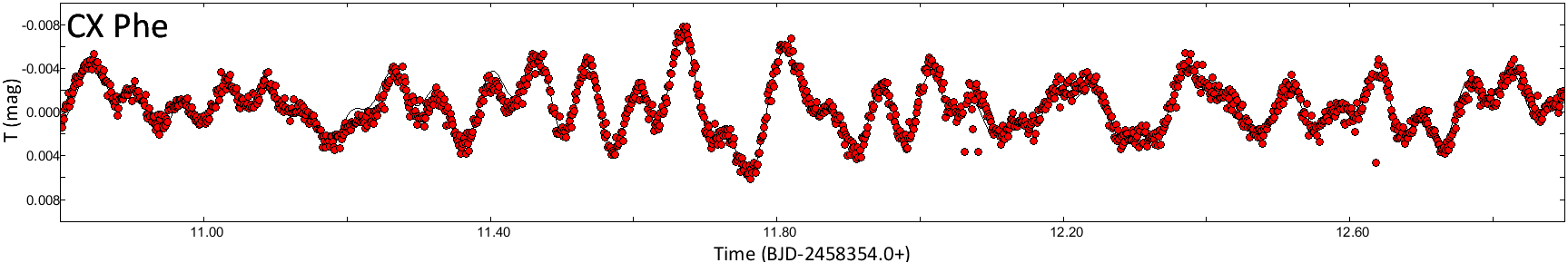}\\
\includegraphics[width=18cm]{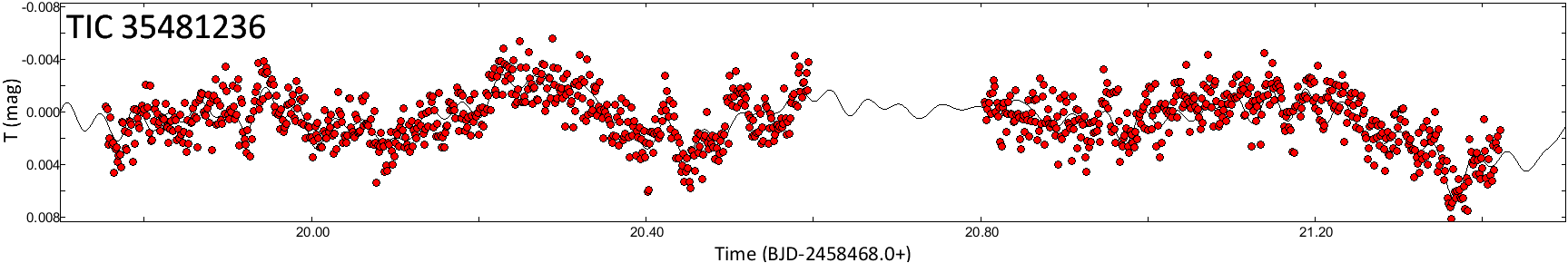}\\
\caption{Fourier fit on sample of data points of all studied cases.}
\label{fig:FF}
\end{figure*}

The analysis results are categorized into two main groups: independent frequencies and combination frequencies. In addition to Fourier modeling of the pulsations, we estimated the most likely $l$-degrees of the independent frequencies. This was done using the method of \citet{BRE00} to calculate the pulsation constant $Q$ for each frequency \citep[see also][]{LIA22, LIA24a}, leveraging the known absolute properties of the pulsating star (Sect.~\ref{Sec:AbsPar}). The computed $Q$ values were then compared with the models of \citet{FIT81} to determine the most probable oscillation modes. The determination of the oscillating member(s) of each system based on the comparison of the strong frequencies detected in the in- and out-of-eclipse data as well as further comments on their pulsational behaviour are given in the following subsections. Table~\ref{Tab:FreqsInd} presents the results for the independent frequencies detected in each system's component, listing: the component of the system, frequency index ($i$), frequency value ($f_{\rm i}$), amplitude ($A$), phase ($\Phi$), $S/N$, $Q$, and $l$-degrees. Table~\ref{tab:fcombo} provides the same information as Table~\ref{Tab:FreqsInd}, except for the $Q$ values and $l$-degrees, but focuses on the combination frequencies. It should to be noted that in the latter table, the combinations of frequencies are based mostly on the independent frequencies of Table~\ref{Tab:FreqsInd} and the orbital frequency ($f_{\rm orb}$) of each system. Moreover, we also note that since the LC models (Table~\ref{Tab:LCmodels}) resulted in almost complete tidal locked configurations, we assume for the components of each system the same rotational frequency with their $f_{\rm orb}$. The periodograms of the in- and out-of-eclipse frequency searches for all cases are illustrated in Fig.~\ref{fig:FS}. The latter plots, for scaling reasons, show only the regimes where frequencies were detected and not all the region covered by the frequency search (i.e.~0-80~d$^{-1}$). The Fourier fittings on sample of out-of-eclipse data points of each EB are plotted in Fig.~\ref{fig:FF}, and the respective distribution of frequencies in Fig.~\ref{fig:FD}.

\begin{table}	
\begin{center}															
\caption{Specifications of the data sets used for frequency analysis of all studied cases.}										
\label{Tab:datalog}	
\scalebox{0.95}{
\begin{tabular}{l ccccc}		
\hline\hline											
Phase part	&	$\Phi_{\rm orb}$	&	$N$	&	$f_{\rm Nyq}$	&	$\delta f$	&	$N_{\rm bgd}$	\\
	&		&		&	     (d$^{-1}$)	&	     (d$^{-1}$)	&	($\upmu$mag)	\\
\hline											
	&	\multicolumn{5}{c}{CH Ind}									\\
\hline											
1ry eclipse	&	0.955-0.045	&	1686	&		&		&		\\
2ry eclipse	&	0.430-0.515	&	1460	&		&		&		\\
\multirow{2}{*}{Quadratures}	&	0.045-0.430	&	\multirow{2}{*}{14518}	&	\multirow{2}{*}{360}	&	\multirow{2}{*}{0.057}	&	\multirow{2}{*}{6.2}	\\
	&	0.515-0.955	&		&		&		&		\\
\hline											
	&	\multicolumn{5}{c}{CX Phe}									\\
\hline											
1ry eclipse	&	0.991-0.009	&	257	&		&		&		\\
2ry eclipse	&	0.495-0.530	&	1007	&		&		&		\\
\multirow{2}{*}{Quadratures}	&	0.009-0.495	&	\multirow{2}{*}{17015}	&	\multirow{2}{*}{360}	&	\multirow{2}{*}{0.057}	&	\multirow{2}{*}{16}	\\
	&	0.530-0.991	&		&		&		&		\\
\hline											
		\multicolumn{5}{c}{V577 Oph}									\\
\hline											
1ry eclipse	&	0.970-0.030	&	629	&		&		&		\\
2ry eclipse	&	0.540-0.610	&	743	&		&		&		\\
\multirow{2}{*}{Quadratures}	&	0.030-0.540	&	\multirow{2}{*}{9091}	&	\multirow{2}{*}{210}	&	\multirow{2}{*}{0.057}	&	\multirow{2}{*}{51}	\\
	&	0.610-0.970	&		&		&		&		\\
\hline											
	&	\multicolumn{5}{c}{TIC 35481236}									\\
\hline											
1ry eclipse	&	0.960-0.040	&	2516	&		&		&		\\
2ry eclipse	&	0.505-0.620	&	3422	&		&		&		\\
\multirow{2}{*}{Quadratures}	&	0.040-0.505	&	\multirow{2}{*}{24940}	&	\multirow{2}{*}{360}	&	\multirow{2}{*}{0.030}	&	\multirow{2}{*}{18}	\\
	&	0.620-0.960	&		&		&		&		\\
\hline																																																																																															\end{tabular}}															
\end{center}															
\end{table}

\begin{table*}	
\begin{center}															
\caption{Independent oscillation frequencies of the components of the studied EBs.}										
\label{Tab:FreqsInd}	
\scalebox{0.95}{
\begin{tabular}{ll ccc cccc}															
\hline\hline																	
Component	&	$i$	&	  $f_{\rm i}$	&	$A$	&	  $\Phi$	&	S/N	  & 	$Q$	&	$l$-degree	&	Mode	\\
	&		&	     (d$^{-1}$)	&	(mmag)	&	(2$\pi$~rad)	&		  & 	(d)	&		&		\\
\hline																	
		\multicolumn{9}{c}{CH Ind}															\\
\hline																	
Secondary	&	1	&	8.8527(5)	&	0.202(5)	&	0.885(4)	&	32.2	&	0.033(1)	&	3	&	NR $F$	\\
Primary	&	2	&	2.7486(8)	&	0.135(5)	&	0.298(5)	&	21.5	&		&		&		\\
\hline																	
		\multicolumn{9}{c}{V577 Oph}															\\
\hline																	
\multirow{2}{*}{Primary}	&	1	&	14.3903(1)	&	14.68(4)	&	0.090(1)	&	287.2	&	0.032(3)	&	0	&	R $F$	\\
	&	3	&	1.5426(6)	&	1.43(4)	&	0.458(4)	&	27.9	&		&		&		\\
\hline																	
		\multicolumn{9}{c}{CX Phe}															\\
\hline																	
Secondary	&	1	&	5.1855(1)	&	1.745(9)	&	0.353(1)	&	103.2	&	0.14(3)	&		&		\\
Primary	&	2	&	14.5002(2)	&	1.063(9)	&	0.516(1)	&	62.8	&	0.031(4)	&	0	&	R $F$	\\
Primary	&	3	&	15.4579(2)	&	0.795(9)	&	0.418(2)	&	47.0	&	0.029(4)	&	1	&	NR $p_1$	\\
Secondary	&	4	&	7.2150(3)	&	0.666(9)	&	0.477(2)	&	39.4	&	0.010(2)	&		&		\\
Primary	&	5	&	17.3145(3)	&	0.544(9)	&	0.789(3)	&	32.1	&	0.026(3)	&	2	&	NR $p_1$	\\
\hline																	
		\multicolumn{9}{c}{TIC 35481236}															\\
\hline																	
\multirow{3}{*}{Secondary}	&	5	&	24.9829(3)	&	0.480(12)	&	0.434(4)	&	26.5	&	0.028(1)	&	1	&	NR $p_1$	\\
	&	6	&	20.7429(3)	&	0.450(12)	&	0.146(4)	&	24.9	&	0.033(1)	&	3	&	NR $F$	\\
	&	8	&	1.7911(4)	&	0.367(12)	&	0.874(5)	&	20.3	&		&		&		\\
\hline																																																	
\end{tabular}}									
\end{center}
\textbf{Notes.} R=radial, NR=non radial, $F$=Fundamental, $p_{\rm n}$=pressure where $n$=1, 2, 3 ...N the overtones of the fundamental mode 																
\end{table*}

\subsection{CH~Ind}
\label{Sec:Puls_Ind}

The pulsation analysis for this system resulted in a total of 46 frequencies, two of which are independent modes, namely $f_1$ and $f_2$ in Table~\ref{Tab:FreqsInd}. The distribution of frequencies (Fig.~\ref{fig:FD}) as well as the periodogram of the analysis (Fig.~\ref{fig:FS}) show that there are two main groups of frequencies; one up to 4~d$^{-1}$ and a second between 6 and 10~d$^{-1}$. The frequency analyses of the in-eclipse data revealed that during the primary eclipse the strongest frequency found has a value of $f_1+2f_{\rm orb}$, while the $f_2$ is almost vanished. On the other hand, during the secondary eclipse, we clearly detected as the strongest frequency the $f_2-2f_{\rm orb}$, while the amplitude of $f_1$ appeared decreased. Therefore, we conclude that the lower frequency ($f_2$) is originated from the primary component, which is a $\gamma$~Dor type star, and the higher frequency ($f_1$) from the system's secondary component, which is a $\delta$~Sct star. The models of \citet{FIT81} suggest that $f_1$ is a non-radial fundamental oscillation mode.

\subsection{V577~Oph}
\label{Sec:Puls_Oph}

The Fourier spectrum of V577~Oph (Fig.~\ref{fig:FS}) is dominated by two parent frequencies, one at 14.39~d$^{-1}$ and the other at 1.54~d$^{-1}$. It should to be noted that $f_2$ (=7$f_{\rm orb}$) is the second strongest detected frequency. Another 51 frequencies were detected with their vast majority to exist in the range 0-4~d$^{-1}$ (Fig.~\ref{fig:FD}). The rest frequencies are almost homogeneously spread in the range 4 and 32~d$^{-1}$, with a peak between 12 and 16~d$^{-1}$, where $f_1$ and its rotational splitting sidelobes lie within. During the primary eclipse, all prominent frequencies appear weaker compared to their counterparts in both the out-of-eclipse data and the secondary eclipse data. Thus, it is plausible to conclude that the oscillating star of the system is the primary component and exhibits hybrid $\delta$~Sct-$\gamma$~Dor behaviour. It should be noted that $f_1$ is revealed as a radial fundamental pulsation mode.

\subsection{CX~Phe}
\label{Sec:Puls_Phe}

Due to the relatively long orbital period of this EB ($\sim 20$~d), only one primary eclipse is included in the TESS data of sector~2, hence the frequency analysis for this phase part was not reliable. Its Fourier spectrum is rather complex (Fig.~\ref{fig:FS}) but we can recognize at least two forests of frequencies, one that covers the range 4-10~d$^{-1}$ and another between 13 and 18~d$^{-1}$. Particularly, two frequencies, namely $f_1$ and $f_4$, and another three, namely $f_2$, $f_3$, and $f_5$, were revealed as independent oscillation modes in the aforementioned regimes. We detected a total of 84 frequencies with their majority to have values up to 8~d$^{-1}$ (Fig.~\ref{fig:FD}). The comparison of the periodograms of the out-of-eclipse and the secondary eclipse data yields that the amplitudes of $f_2$, $f_3$, $f_4$, and $f_5$ are amplified during the eclipse, while that of $f_1$ remains almost intact. Thus, $f_2$, $f_3$, and $f_5$ definitely originate from the primary component, since they also have relatively close values. The frequency $f_1$ is likely associated with a pulsation mode of the secondary component. Meanwhile, although $f_4$ exhibits an increase during the secondary eclipse, it also has a ratio of approximately 0.72 with $f_1$ (i.e.~$f_1/f_4$), which is a typical ratio for the radial fundamental and first overtone modes \citep{OAS06}. Therefore, while the origin star cannot be determined with absolute certainty, it is more likely that $f_4$ also comes from the secondary component. It should to be noted that the models of \citet{FIT81} do not predict the respective $Q$ values for these two frequencies. On the contrary, they suggest that $f_2$ is a radial fundamental mode, while $f_3$ and $f_5$ are the first overtones of non-radial pressure modes with $l$-degrees of 1 and 2, respectively.

\subsection{TIC~35481236}
\label{Sec:Puls_TIC}

A total of 73 frequencies were identified for this binary (Tables~\ref{Tab:FreqsInd} and \ref{tab:fcombo}). However, in contrast with the previous cases, the first four strongest frequencies have values less than 2.17~d$^{-1}$, with the three of them ($f_1$, $f_3$, and $f_4$) to be connected to the $f_{\rm orb}$. The frequency $f_2\sim 0.014$~d$^{-1}$ corresponds to a time interval of approximately 71.4~d and is not typical for these types of pulsating stars. Therefore, we consider it as an artifact. The periodogram of the system after removing $f_1$-$f_4$ has two main regimes of frequencies, that are 0-4~d$^{-1}$ and 8-12~d$^{-1}$, as well as many peaks between 17 and 27~d$^{-1}$ (Figs.~\ref{fig:FS} and \ref{fig:FD}). Three frequencies were revealed as parent frequencies; $f_5$ and $f_6$ lie well inside the $\delta$~Sct region of frequencies, and $f_8$ that has a typical value for a $\gamma$~Dor star. In this case the Fourier analyses of the in-eclipse data were rather enlightening. The amplitudes of all frequencies are significantly increased during the primary eclipse and present the exact opposite behaviour during the secondary eclipse. Hence, we conclude that all pulsation modes originate from the secondary component. Therefore, the latter component is a hybrid $\delta$~Sct-$\gamma$~Dor star. Both $f_5$ and $f_6$ are identified as non-radial oscillation pressure modes with $f_5$ corresponding to the first overtone and $f_6$ the fundamental modes.

\section{Summary, discussion, and conclusions}
\label{Sec:Disc}

This work presented for the first time accurate determination of the physical parameters and pulsation properties of four detached EBs in eccentric orbits. TESS photometric data along with spectroscopic observations from \textit{Gaia} and ground-based facilities were used to calculate the binary models and derive the absolute parameters of the components. The median values of the uncertainties in masses and radii were found to be 3\% and 4\%, respectively. The LC residuals of all systems were analyzed using Fourier techniques in order to detect the oscillation modes that the systems' component(s) exhibit. Moreover, using the eclipses as spatial filters and by performing frequency search in their data, we managed to identify which component pulsates in each system and to classify them accordingly. The current work identified in total three pure $\delta$~Sct and one $\gamma$~Dor stars, as well as two hybrid $\delta$~Sct-$\gamma$~Dor pulsatots within these four EBs. Therefore, as all of them are confirmed SB2+E systems, our study expands the sample of detached systems hosting $\delta$~Sct members and with well-constrained absolute properties (i.e.~35 in total SB2+E systems; see Sect.~\ref{Sec:AbsPar}) by $\sim$14\%.

CH~Ind is an EB that hosts two pulsating stars. These stars are very similar in mass, radius, and temperature. Its primary (less massive) component is a $\gamma$~Dor star with a dominant frequency of 2.75~d$^{-1}$. It is the most evolved star in our sample and is located very close to the intersection of the red edge of the classical instability strip and the TAMS line. The secondary component is a $\delta$~Sct star that oscillates in a non-radial pressure mode with a dominant frequency of 8.85~d$^{-1}$ and it is located very close to its companion in the evolutionary diagrams. In total, we detected another 44 combination frequencies. Our results regarding the pulsation frequency of the $\delta$~Sct star come in perfect agrement with those of \citet{MKR22}, while the derived masses are very close to the values given by \citet{GAIA22}.

The primary (more massive) component of V577~Oph was identified as a hybrid $\delta$~Sct-$\gamma$~Dor pulsator and it is located slightly outside the red edge of the classical instability strip, within the region where $\delta$~Sct and $\gamma$~Dor stars overlap \citep{GRI10, UYT11}. Its main frequencies are 14.39~d$^{-1}$ and 1.54~d$^{-1}$, but another 51 were also detected. The former frequency was revealed as a radial fundamental mode, with our results to agree totally with those of previous studies (Sect.~\ref{Sec:HIS}). The latter frequency is a typical $g$~mode. In presence of past decades minima timings of the system, we determined the period of its apsidal motion to be $\sim5000$~yr. Moreover, a periodic modulation of $\sim33$~yr of the orbital period was detected and attributed to a low-mass star orbiting the eclipsing pair that is too faint to be detected. Our derived masses (1.60~M$_\sun$ and 1.4~M$_\sun$) have discrepancies with those of \citet{GAIA22}, i.e.~1.67~M$_\sun$ and 1.76~M$_\sun$, respectively. We contend that these discrepancies stem from an incorrect automatic solution of the NSS model by \citet{GAIA22}, as it relies on an erroneous orbital period value. In contrast, we used the correct orbital period and RV semi-amplitude values, derived from TESS and ground-based observations, respectively.

The frequency analysis of CX~Phe was the most complex in comparison with the rest studied systems. Although there were not enough data to perform a reliable frequency analysis for the primary eclipse, the results showed that there are at least two regimes of frequencies (i.e.~two main frequency concentration ranges). According to the comparison of the results of the frequency searches in the secondary eclipse and the out-of-eclipse data, we propose that both components are $\delta$~Sct stars. The primary (more massive and hotter) component has three independent frequencies ranging between 14.5 and 17.32~d$^{-1}$. The dominant frequency ($f_2=14.5$~d$^{-1}$) of this star was identified as a radial fundamental mode, while the other two ($f_3$ and $f_5$) as non-radial pressure modes. The secondary component pulsates in two parent modes with frequencies $\sim$5.19~d$^{-1}$ and $\sim$7.22~d$^{-1}$. The primary component is positioned just beyond the red edge of the classical instability strip, while the secondary is far away from it. Both components do not follow well the theoretical evolutionary tracks, indicating that past mass transfer between them or mass loss from the system might have been occurred. \citet{KAH23} determined only the dominant frequency of the system and resulted in the same value as we did in the present study ($f_1=5.186$~d$^{-1}$). Significant discrepancies of approximately 11.5\% exist in the mass values between our results and those of \citet{GAIA22}. Nevertheless, we argue that our solution that is based on detailed LC modelling provides more accurate results.

The frequency search results for TIC~35481236 were more distinct. Their comparison for both eclipses data indicated that all frequencies were enhanced during the primary eclipse, implying they originate from the secondary (more massive and hotter) component. Particularly, three main independent frequencies were detected, two between 20 and 25~d$^{-1}$ and one at $\sim$1.79~d$^{-1}$. Therefore, we fairly conclude its hybrid $\delta$~Sct-$\gamma$~Dor nature. The $\delta$~Sct frequencies were revealed as non-radial pressure modes (first overtone and fundamental modes). The pulsator is located on the ZAMS boundary, inside the classical instability strip, but closer to its red edge and well within the overlap region of $\delta$~Sct-$\gamma$~Dor stars. The primary component is similar to its companion regarding its physical parameters and evolutionary status, but it does not exhibit any pulsations to date. In total, 73 frequencies were detected, with one (i.e.~$f_2$) to be considered as an artifact. Our findings on the dominant pulsation align with those of \citet{SHI22} but differ from those of \citet{KAH23}. Our derived mass values agree totally with the automatic solution of \citet{GAIA22}.

\begin{figure}
\centering
\includegraphics[width=\columnwidth]{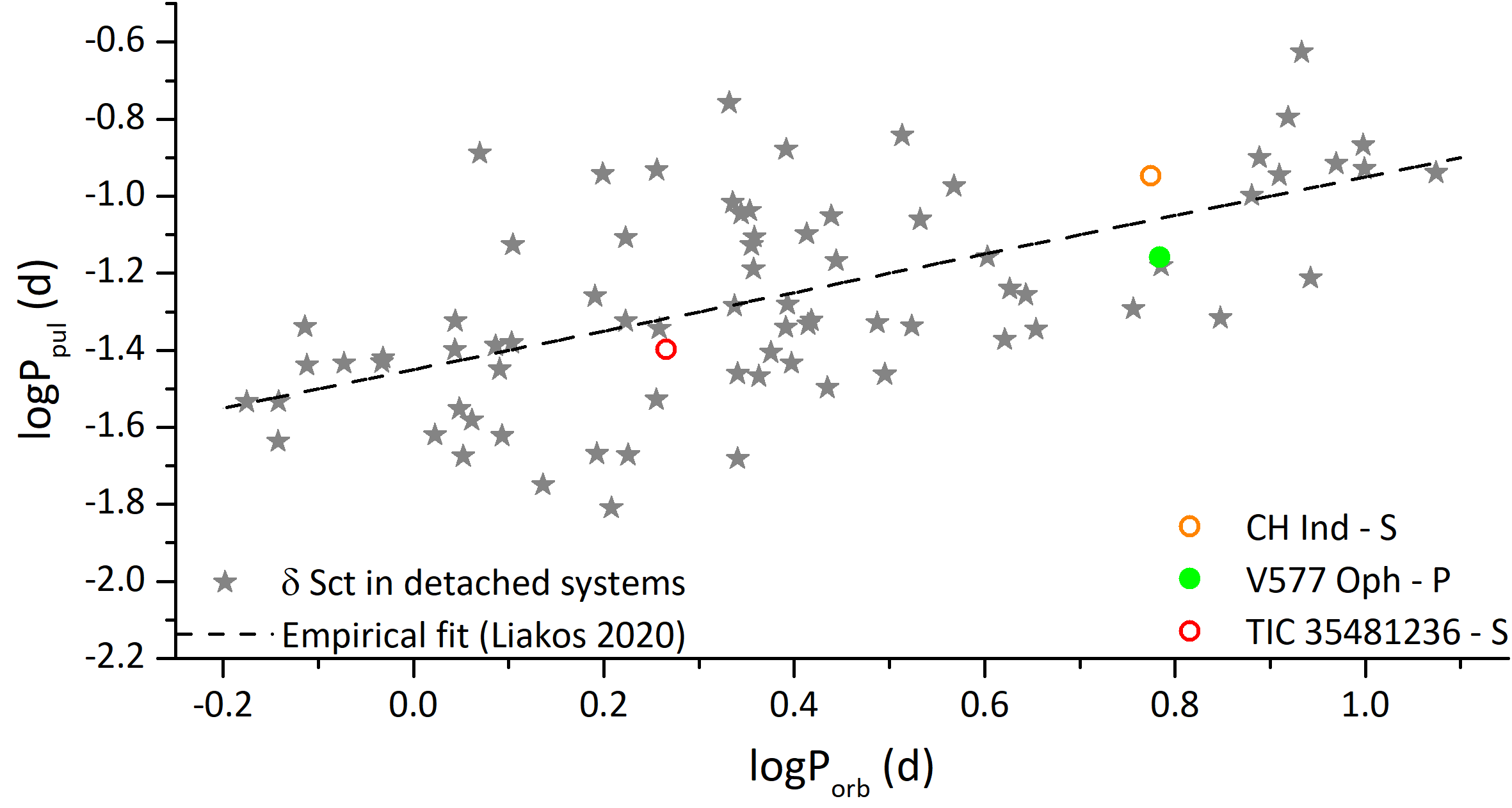}
\caption{$P_{\rm orb}-P_{\rm puls}$ correlation for $\delta$~Sct stars of detached systems (star symbols) and the locations of the $\delta$~Scuti components of the studied systems. Black solid line represents the linear fitting of \citet{LIA25}. We note that this correlation extends up to $P_{\rm orb}=13$~d.}
\label{fig:PP}
\vspace{0.3cm}
\centering
\includegraphics[width=\columnwidth]{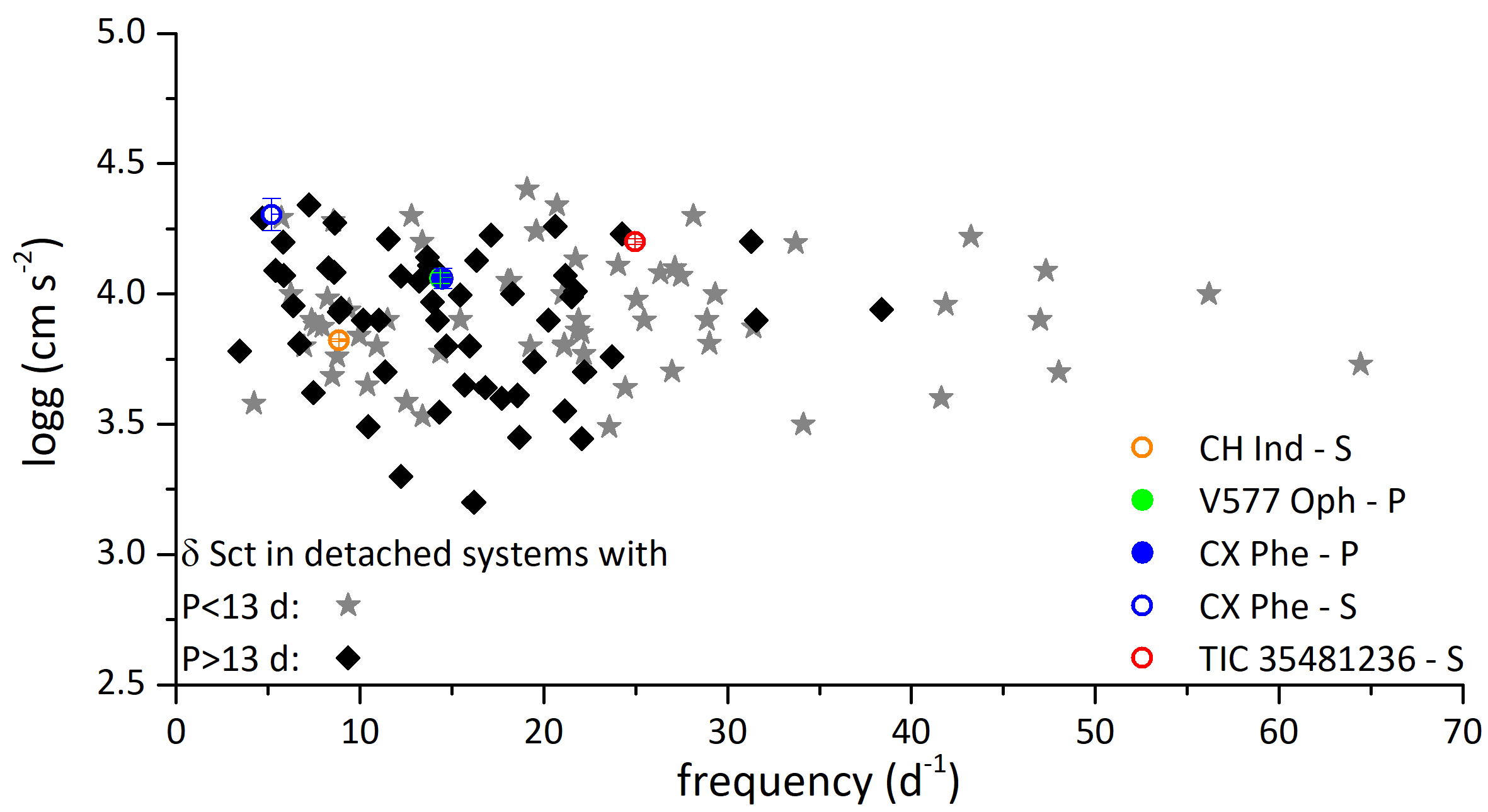}
\caption{$f-\log g$ plot for $\delta$~Sct stars-members of binary systems with $P_{\rm orb}<13$~d (grey star symbols) and $P_{\rm orb}>13$~d (black diamond symbols) and the locations of the $\delta$~Sct components of the studied cases (filled and empty colored circles). We note that the primaries of CX~Phe and V577~Oph have very similar $\log g$ and frequency values, thus there is an overlap between the symbols.}
\label{fig:GP}
\end{figure}

To compare the properties of the $\delta$~Sct components in our studied systems with similar cases, we positioned them among other $\delta$~Sct stars of detached systems in two diagrams. Fig.~\ref{fig:PP} represents the relationship between orbital and pulsation periods ($P_{\rm orb}-P_{\rm puls}$), while Fig.~\ref{fig:GP} illustrates the correlation between the dominant frequency and evolutionary stage \citep[$f-\log g$; cf.][]{SOY06a, LIA12, LIAN15, LIAN17, LIA20a, LIA25}. The first plot includes a sample of 77 $\delta$~Sct stars in binaries, while the second plot comprises 60 stars in binary systems with $P_{\rm orb}<13$~d and 54 stars in systems with $P_{\rm orb}>13$~d. The $\delta$~Sct stars of CH~Ind, V577~Oph, and TIC~35481236 follow very well the empirical trend of \citet{LIA20a} \citep[cf.][]{LIA25}. Although \citet{LIA20a, LIA25} mentioned about a potential correlation between $\log g$ and dominant pulsation period for this kind of stars, the new sample (taken from these studies and the personal updated catalogue of the author) does not clear support for it. \citet{LIA25} reported low correlation coefficients between these quantities, and, therefore, we avoided to present their empirical fit in Fig.~\ref{fig:GP}, which includes only the sample stars. A detailed discussion of this issue is beyond the scope of the present study. The secondary component of CX~Phe is one of the youngest and slowest pulsators of the current sample. The primary component of CH~Ind as well as the $\delta$~Sct stars of the other two studied systems are located well within the majority of the sample stars.

Further studies of similar systems are highly recommended. Combining \textit{Gaia} data from the present Data Release-3 (DR3), and DR4 and DR5 catalogues in the following years, with data from the TESS, Kepler and, in the near future, the PLAnetary Transits and Oscillations of stars \citep[PLATO;][]{RAU14} missions, provides the means to increase drastically our knowledge about the physical and pulsational properties of the current sample of $\delta$~Sct stars in binaries, thus, our knowledge regarding the evolution of these pulsators.

\begin{acknowledgements}
This research has made use of NASA’s Astrophysics Data System Bibliographic Services and the SIMBAD database, operated at CDS, Strasbourg, France. We made use of data of the TESS mission via the Mikulski Archive for Space Telescopes (MAST). Funding for the TESS mission is provided by the NASA Explorer Program. Moreover, we used data from the European Space Agency (ESA) mission \textit{Gaia} (\url{https://www.cosmos.esa.int/Gaia}), processed by the {\it \textit{Gaia}} Data Processing and Analysis Consortium (DPAC,\url{https://www.cosmos.esa.int/web/Gaia/dpac/consortium}). Funding for the DPAC has been provided by national institutions, in particular the institutions participating in the \textit{Gaia} Multilateral Agreement. Minima timings were taken from the VarAstro web database. The author acknowledges financial support from the NOA's internal fellowship `SPECIES' (No.~5094) and thanks the anonymous reviewer for the valuable comments that improved the quality of this work.
\end{acknowledgements}

%
%

\bibliographystyle{aa} 
\bibliography{references.bib} 

\begin{thebibliography}{94}
\expandafter\ifx\csname natexlab\endcsname\relax\def\natexlab#1{#1}\fi

\bibitem[{{Aerts} {et~al.}(2010){Aerts}, {Christensen-Dalsgaard}, \&
  {Kurtz}}]{AER10}
{Aerts}, C., {Christensen-Dalsgaard}, J., \& {Kurtz}, D.~W. 2010,
  {Asteroseismology} (Springer Netherlands)

\bibitem[{{Anders} {et~al.}(2022){Anders}, {Khalatyan}, {Queiroz}, {Chiappini},
  {Ard{\`e}vol}, {Casamiquela}, {Figueras}, {Jim{\'e}nez-Arranz}, {Jordi},
  {Mongui{\'o}}, {Romero-G{\'o}mez}, {Altamirano}, {Antoja}, {Assaad},
  {Cantat-Gaudin}, {Castro-Ginard}, {Enke}, {Girardi}, {Guiglion}, {Khan},
  {Luri}, {Miglio}, {Minchev}, {Ramos}, {Santiago}, \& {Steinmetz}}]{AND22}
{Anders}, F., {Khalatyan}, A., {Queiroz}, A.~B.~A., {et~al.} 2022, \aap, 658,
  A91

\bibitem[{{Antoci} {et~al.}(2014){Antoci}, {Cunha}, {Houdek}, {Kjeldsen},
  {Trampedach}, {Handler}, {L{\"u}ftinger}, {Arentoft}, \& {Murphy}}]{ANT14}
{Antoci}, V., {Cunha}, M., {Houdek}, G., {et~al.} 2014, \apj, 796, 118

\bibitem[{{Bailer-Jones} {et~al.}(2021){Bailer-Jones}, {Rybizki}, {Fouesneau},
  {Demleitner}, \& {Andrae}}]{BAI21}
{Bailer-Jones}, C.~A.~L., {Rybizki}, J., {Fouesneau}, M., {Demleitner}, M., \&
  {Andrae}, R. 2021, \aj, 161, 147

\bibitem[{{Balona} {et~al.}(2015){Balona}, {Daszy{\'n}ska-Daszkiewicz}, \&
  {Pamyatnykh}}]{BAL15}
{Balona}, L.~A., {Daszy{\'n}ska-Daszkiewicz}, J., \& {Pamyatnykh}, A.~A. 2015,
  \mnras, 452, 3073

\bibitem[{{Baran} {et~al.}(2015){Baran}, {Koen}, \& {Pokrzywka}}]{BAR15}
{Baran}, A.~S., {Koen}, C., \& {Pokrzywka}, B. 2015, \mnras, 448, L16

\bibitem[{{Borucki} {et~al.}(2010){Borucki}, {Koch}, {Basri}, {Batalha},
  {Brown}, {Caldwell}, {Caldwell}, {Christensen-Dalsgaard}, {Cochran},
  {DeVore}, {Dunham}, {Dupree}, {Gautier}, {Geary}, {Gilliland}, {Gould},
  {Howell}, {Jenkins}, {Kondo}, {Latham}, {Marcy}, {Meibom}, {Kjeldsen},
  {Lissauer}, {Monet}, {Morrison}, {Sasselov}, {Tarter}, {Boss}, {Brownlee},
  {Owen}, {Buzasi}, {Charbonneau}, {Doyle}, {Fortney}, {Ford}, {Holman},
  {Seager}, {Steffen}, {Welsh}, {Rowe}, {Anderson}, {Buchhave}, {Ciardi},
  {Walkowicz}, {Sherry}, {Horch}, {Isaacson}, {Everett}, {Fischer}, {Torres},
  {Johnson}, {Endl}, {MacQueen}, {Bryson}, {Dotson}, {Haas}, {Kolodziejczak},
  {Van Cleve}, {Chandrasekaran}, {Twicken}, {Quintana}, {Clarke}, {Allen},
  {Li}, {Wu}, {Tenenbaum}, {Verner}, {Bruhweiler}, {Barnes}, \& {Prsa}}]{BOR10}
{Borucki}, W.~J., {Koch}, D., {Basri}, G., {et~al.} 2010, Science, 327, 977

\bibitem[{{Bowman} \& {Kurtz}(2018)}]{BOW18}
{Bowman}, D.~M. \& {Kurtz}, D.~W. 2018, \mnras, 476, 3169

\bibitem[{{Bowman} \& {Michielsen}(2021)}]{BOW21}
{Bowman}, D.~M. \& {Michielsen}, M. 2021, \aap, 656, A158

\bibitem[{{Breger}(2000)}]{BRE00}
{Breger}, M. 2000, in Astronomical Society of the Pacific Conference Series,
  Vol. 210, Delta Scuti and Related Stars, ed. M.~{Breger} \& M.~{Montgomery},
  3

\bibitem[{{Claret}(2018)}]{CLA18}
{Claret}, A. 2018, \aap, 618, A20

\bibitem[{{Creevey} {et~al.}(2010){Creevey}, {Telting}, {Belmonte}, {Brown},
  {Handler}, {Metcalfe}, {Pinheiro}, {Sousa}, {Terrell}, \& {Zhou}}]{CRE10}
{Creevey}, O.~L., {Telting}, J., {Belmonte}, J.~A., {et~al.} 2010,
  Astronomische Nachrichten, 331, 952

\bibitem[{{Cropper} {et~al.}(2018){Cropper}, {Katz}, {Sartoretti}, {Prusti},
  {de Bruijne}, {Chassat}, {Charvet}, {Boyadjian}, {Perryman}, {Sarri}, {Gare},
  {Erdmann}, {Munari}, {Zwitter}, {Wilkinson}, {Arenou}, {Vallenari},
  {G{\'o}mez}, {Panuzzo}, {Seabroke}, {Allende Prieto}, {Benson}, {Marchal},
  {Huckle}, {Smith}, {Dolding}, {Jan{\ss}en}, {Viala}, {Blomme}, {Baker},
  {Boudreault}, {Crifo}, {Soubiran}, {Fr{\'e}mat}, {Jasniewicz}, {Guerrier},
  {Guy}, {Turon}, {Jean-Antoine-Piccolo}, {Th{\'e}venin}, {David}, {Gosset}, \&
  {Damerdji}}]{CRO18}
{Cropper}, M., {Katz}, D., {Sartoretti}, P., {et~al.} 2018, \aap, 616, A5

\bibitem[{{Diethelm}(1993)}]{DIE93}
{Diethelm}, R. 1993, Information Bulletin on Variable Stars, 3894

\bibitem[{{Doyle} {et~al.}(2024){Doyle}, {Armstrong}, {Bayliss}, {Rodel}, \&
  {Kunovac}}]{DOY24}
{Doyle}, L., {Armstrong}, D.~J., {Bayliss}, D., {Rodel}, T., \& {Kunovac}, V.
  2024, \mnras, 529, 1802

\bibitem[{{Dupret} {et~al.}(2005){Dupret}, {Grigahc{\`e}ne}, {Garrido},
  {Gabriel}, \& {Scuflaire}}]{DUP05}
{Dupret}, M.~A., {Grigahc{\`e}ne}, A., {Garrido}, R., {Gabriel}, M., \&
  {Scuflaire}, R. 2005, \aap, 435, 927

\bibitem[{{Fetherolf} {et~al.}(2023){Fetherolf}, {Pepper}, {Simpson}, {Kane},
  {Mo{\v{c}}nik}, {English}, {Antoci}, {Huber}, {Jenkins}, {Stassun},
  {Twicken}, {Vanderspek}, \& {Winn}}]{FET23}
{Fetherolf}, T., {Pepper}, J., {Simpson}, E., {et~al.} 2023, \apjs, 268, 4

\bibitem[{{Fitch}(1981)}]{FIT81}
{Fitch}, W.~S. 1981, \apj, 249, 218

\bibitem[{{Gaia Collaboration} {et~al.}(2022){Gaia Collaboration}, {Klioner},
  {Lindegren}, {Mignard}, {Hern{\'a}ndez}, {Ramos-Lerate}, {Bastian},
  {Biermann}, {Bombrun}, {de Torres}, {Gerlach}, {Geyer}, {Hilger}, {Hobbs},
  {Lammers}, {McMillan}, {Steidelm{\"u}ller}, {Teyssier}, {Raiteri},
  {Bartolom{\'e}}, {Bernet}, {Casta{\~n}eda}, {Clotet}, {Davidson},
  {Fabricius}, {Garralda Torres}, {Gonz{\'a}lez-Vidal}, {Portell}, {Rowell},
  {Torra}, {Torra}, {Brown}, {Vallenari}, {Prusti}, {de Bruijne}, {Arenou},
  {Babusiaux}, {Creevey}, {Ducourant}, {Evans}, {Eyer}, {Guerra}, {Hutton},
  {Jordi}, {Luri}, {Panem}, {Pourbaix}, {Randich}, {Sartoretti}, {Soubiran},
  {Tanga}, {Walton}, {Bailer-Jones}, {Drimmel}, {Jansen}, {Katz}, {Lattanzi},
  {van Leeuwen}, {Bakker}, {Cacciari}, {De Angeli}, {Fouesneau}, {Fr{\'e}mat},
  {Galluccio}, {Guerrier}, {Heiter}, {Masana}, {Messineo}, {Mowlavi},
  {Nicolas}, {Nienartowicz}, {Pailler}, {Panuzzo}, {Riclet}, {Roux},
  {Seabroke}, {Sordo}, {Th{\'e}venin}, {Gracia-Abril}, {Altmann}, {Andrae},
  {Audard}, {Bellas-Velidis}, {Benson}, {Berthier}, {Blomme}, {Burgess},
  {Busonero}, {Busso}, {C{\'a}novas}, {Carry}, {Cellino}, {Cheek},
  {Clementini}, {Damerdji}, {de Teodoro}, {Nu{\~n}ez Campos}, {Delchambre},
  {Dell'Oro}, {Esquej}, {Fern{\'a}ndez-Hern{\'a}ndez}, {Fraile}, {Garabato},
  {Garc{\'\i}a-Lario}, {Gosset}, {Haigron}, {Halbwachs}, {Hambly}, {Harrison},
  {Hestroffer}, {Hodgkin}, {Holl}, {Jan{\ss}en}, {Jevardat de Fombelle},
  {Jordan}, {Krone-Martins}, {Lanzafame}, {L{\"o}ffler}, {Marchal}, {Marrese},
  {Moitinho}, {Muinonen}, {Osborne}, {Pancino}, {Pauwels}, {Recio-Blanco},
  {Reyl{\'e}}, {Riello}, {Rimoldini}, {Roegiers}, {Rybizki}, {Sarro}, {Siopis},
  {Smith}, {Sozzetti}, {Utrilla}, {van Leeuwen}, {Abbas}, {{\'A}brah{\'a}m},
  {Abreu Aramburu}, {Aerts}, {Aguado}, {Ajaj}, {Aldea-Montero}, {Altavilla},
  {{\'A}lvarez}, {Alves}, {Anderson}, {Anglada Varela}, {Antoja}, {Baines},
  {Baker}, {Balaguer-N{\'u}{\~n}ez}, {Balbinot}, {Balog}, {Barache}, {Barbato},
  {Barros}, {Barstow}, {Bassilana}, {Bauchet}, {Becciani}, {Bellazzini},
  {Berihuete}, {Bertone}, {Bianchi}, {Binnenfeld}, {Blanco-Cuaresma}, {Boch},
  {Bossini}, {Bouquillon}, {Bragaglia}, {Bramante}, {Breedt}, {Bressan},
  {Brouillet}, {Brugaletta}, {Bucciarelli}, {Burlacu}, {Butkevich}, {Buzzi},
  {Caffau}, {Cancelliere}, {Cantat-Gaudin}, {Carballo}, {Carlucci},
  {Carnerero}, {Carrasco}, {Casamiquela}, {Castellani}, {Castro-Ginard},
  {Chaoul}, {Charlot}, {Chemin}, {Chiaramida}, {Chiavassa}, {Chornay},
  {Comoretto}, {Contursi}, {Cooper}, {Cornez}, {Cowell}, {Crifo}, {Cropper},
  {Crosta}, {Crowley}, {Dafonte}, {Dapergolas}, {David}, {de Laverny}, {De
  Luise}, {De March}, {De Ridder}, {de Souza}, {del Peloso}, {del Pozo},
  {Delbo}, {Delgado}, {Delisle}, {Demouchy}, {Dharmawardena}, {Diakite},
  {Diener}, {Distefano}, {Dolding}, {Enke}, {Fabre}, {Fabrizio}, {Faigler},
  {Fedorets}, {Fernique}, {Fienga}, {Figueras}, {Fournier}, {Fouron},
  {Fragkoudi}, {Gai}, {Garcia-Gutierrez}, {Garcia-Reinaldos},
  {Garc{\'\i}a-Torres}, {Garofalo}, {Gavel}, {Gavras}, {Giacobbe}, {Gilmore},
  {Girona}, {Giuffrida}, {Gomel}, {Gomez}, {Gonz{\'a}lez-N{\'u}{\~n}ez},
  {Gonz{\'a}lez-Santamar{\'\i}a}, {Granvik}, {Guillout}, {Guiraud},
  {Guti{\'e}rrez-S{\'a}nchez}, {Guy}, {Hatzidimitriou}, {Hauser}, {Haywood},
  {Helmer}, {Helmi}, {Sarmiento}, {Hidalgo}, {H{\l}adczuk}, {Holland},
  {Huckle}, {Jardine}, {Jasniewicz}, {Jean-Antoine Piccolo},
  {Jim{\'e}nez-Arranz}, {Juaristi Campillo}, {Julbe}, {Karbevska}, {Kervella},
  {Khanna}, {Kordopatis}, {Korn}, {K{\'o}sp{\'a}l}, {Kostrzewa-Rutkowska},
  {Kruszy{\'n}ska}, {Kun}, {Laizeau}, {Lambert}, {Lanza}, {Lasne}, {Le
  Campion}, {Lebreton}, {Lebzelter}, {Leccia}, {Leclerc}, {Lecoeur-Taibi},
  {Liao}, {Licata}, {Lindstr{\o}m}, {Lister}, {Livanou}, {Lobel}, {Lorca},
  {Loup}, {Madrero Pardo}, {Magdaleno Romeo}, {Managau}, {Mann}, {Manteiga},
  {Marchant}, {Marconi}, {Marcos}, {Santos}, {Mar{\'\i}n Pina}, {Marinoni},
  {Marocco}, {Marshall}, {Polo}, {Mart{\'\i}n-Fleitas}, {Marton}, {Mary},
  {Masip}, {Massari}, {Mastrobuono-Battisti}, {Mazeh}, {Messina}, {Michalik},
  {Millar}, {Mints}, {Molina}, {Molinaro}, {Moln{\'a}r}, {Monari},
  {Mongui{\'o}}, {Montegriffo}, {Montero}, {Mor}, {Mora}, {Morbidelli},
  {Morel}, {Morris}, {Muraveva}, {Murphy}, {Musella}, {Nagy}, {Noval},
  {Oca{\~n}a}, {Ogden}, {Ordenovic}, {Osinde}, {Pagani}, {Pagano}, {Palaversa},
  {Palicio}, {Pallas-Quintela}, {Panahi}, {Payne-Wardenaar}, {Pe{\~n}alosa
  Esteller}, {Penttil{\"a}}, {Pichon}, {Piersimoni}, {Pineau}, {Plachy},
  {Plum}, {Poggio}, {Pr{\v{s}}a}, {Pulone}, {Racero}, {Ragaini}, {Rainer},
  {Rambaux}, {Ramos}, {Re Fiorentin}, {Regibo}, {Richards}, {Diaz}, {Ripepi},
  {Riva}, {Rix}, {Rixon}, {Robichon}, {Robin}, {Robin}, {Roelens}, {Rogues},
  {Rohrbasser}, {Romero-G{\'o}mez}, {Royer}, {Ruz Mieres}, {Rybicki},
  {Sadowski}, {S{\'a}ez N{\'u}{\~n}ez}, {Sagrist{\`a} Sell{\'e}s}, {Sahlmann},
  {Salguero}, {Samaras}, {Sanchez Gimenez}, {Sanna}, {Santove{\~n}a},
  {Sarasso}, {Schultheis}, {Sciacca}, {Segol}, {Segovia}, {S{\'e}gransan},
  {Semeux}, {Shahaf}, {Siddiqui}, {Siebert}, {Siltala}, {Silvelo}, {Slezak},
  {Slezak}, {Smart}, {Snaith}, {Solano}, {Solitro}, {Souami}, {Souchay},
  {Spagna}, {Spina}, {Spoto}, {Steele}, {Stephenson}, {S{\"u}veges}, {Surdej},
  {Szabados}, {Szegedi-Elek}, {Taris}, {Taylor}, {Teixeira}, {Tolomei},
  {Tonello}, {Torralba Elipe}, {Trabucchi}, {Tsounis}, {Turon}, {Ulla},
  {Unger}, {Vaillant}, {van Dillen}, {van Reeven}, {Vanel}, {Vecchiato},
  {Viala}, {Vicente}, {Voutsinas}, {Weiler}, {Wevers}, {Wyrzykowski}, {Yoldas},
  {Yvard}, {Zhao}, {Zorec}, {Zucker}, \& {Zwitter}}]{GAIA22}
{Gaia Collaboration}, {Klioner}, S.~A., {Lindegren}, L., {et~al.} 2022, \aap,
  667, A148

\bibitem[{{Gaia Collaboration} {et~al.}(2016){Gaia Collaboration}, {Prusti},
  {de Bruijne}, {Brown}, {Vallenari}, {Babusiaux}, {Bailer-Jones}, {Bastian},
  {Biermann}, {Evans}, {Eyer}, {Jansen}, {Jordi}, {Klioner}, {Lammers},
  {Lindegren}, {Luri}, {Mignard}, {Milligan}, {Panem}, {Poinsignon},
  {Pourbaix}, {Randich}, {Sarri}, {Sartoretti}, {Siddiqui}, {Soubiran},
  {Valette}, {van Leeuwen}, {Walton}, {Aerts}, {Arenou}, {Cropper}, {Drimmel},
  {H{\o}g}, {Katz}, {Lattanzi}, {O'Mullane}, {Grebel}, {Holland}, {Huc},
  {Passot}, {Bramante}, {Cacciari}, {Casta{\~n}eda}, {Chaoul}, {Cheek}, {De
  Angeli}, {Fabricius}, {Guerra}, {Hern{\'a}ndez}, {Jean-Antoine-Piccolo},
  {Masana}, {Messineo}, {Mowlavi}, {Nienartowicz}, {Ord{\'o}{\~n}ez-Blanco},
  {Panuzzo}, {Portell}, {Richards}, {Riello}, {Seabroke}, {Tanga},
  {Th{\'e}venin}, {Torra}, {Els}, {Gracia-Abril}, {Comoretto},
  {Garcia-Reinaldos}, {Lock}, {Mercier}, {Altmann}, {Andrae}, {Astraatmadja},
  {Bellas-Velidis}, {Benson}, {Berthier}, {Blomme}, {Busso}, {Carry},
  {Cellino}, {Clementini}, {Cowell}, {Creevey}, {Cuypers}, {Davidson}, {De
  Ridder}, {de Torres}, {Delchambre}, {Dell'Oro}, {Ducourant}, {Fr{\'e}mat},
  {Garc{\'\i}a-Torres}, {Gosset}, {Halbwachs}, {Hambly}, {Harrison}, {Hauser},
  {Hestroffer}, {Hodgkin}, {Huckle}, {Hutton}, {Jasniewicz}, {Jordan},
  {Kontizas}, {Korn}, {Lanzafame}, {Manteiga}, {Moitinho}, {Muinonen},
  {Osinde}, {Pancino}, {Pauwels}, {Petit}, {Recio-Blanco}, {Robin}, {Sarro},
  {Siopis}, {Smith}, {Smith}, {Sozzetti}, {Thuillot}, {van Reeven}, {Viala},
  {Abbas}, {Abreu Aramburu}, {Accart}, {Aguado}, {Allan}, {Allasia},
  {Altavilla}, {{\'A}lvarez}, {Alves}, {Anderson}, {Andrei}, {Anglada Varela},
  {Antiche}, {Antoja}, {Ant{\'o}n}, {Arcay}, {Atzei}, {Ayache}, {Bach},
  {Baker}, {Balaguer-N{\'u}{\~n}ez}, {Barache}, {Barata}, {Barbier}, {Barblan},
  {Baroni}, {Barrado y Navascu{\'e}s}, {Barros}, {Barstow}, {Becciani},
  {Bellazzini}, {Bellei}, {Bello Garc{\'\i}a}, {Belokurov}, {Bendjoya},
  {Berihuete}, {Bianchi}, {Bienaym{\'e}}, {Billebaud}, {Blagorodnova},
  {Blanco-Cuaresma}, {Boch}, {Bombrun}, {Borrachero}, {Bouquillon}, {Bourda},
  {Bouy}, {Bragaglia}, {Breddels}, {Brouillet}, {Br{\"u}semeister},
  {Bucciarelli}, {Budnik}, {Burgess}, {Burgon}, {Burlacu}, {Busonero}, {Buzzi},
  {Caffau}, {Cambras}, {Campbell}, {Cancelliere}, {Cantat-Gaudin}, {Carlucci},
  {Carrasco}, {Castellani}, {Charlot}, {Charnas}, {Charvet}, {Chassat},
  {Chiavassa}, {Clotet}, {Cocozza}, {Collins}, {Collins}, \&
  {Costigan}}]{GAIA16}
{Gaia Collaboration}, {Prusti}, T., {de Bruijne}, J.~H.~J., {et~al.} 2016,
  \aap, 595, A1

\bibitem[{{Gaia Collaboration} {et~al.}(2023){Gaia Collaboration}, {Vallenari},
  {Brown}, {Prusti}, {de Bruijne}, {Arenou}, {Babusiaux}, {Biermann},
  {Creevey}, {Ducourant}, {Evans}, {Eyer}, {Guerra}, {Hutton}, {Jordi},
  {Klioner}, {Lammers}, {Lindegren}, {Luri}, {Mignard}, {Panem}, {Pourbaix},
  {Randich}, {Sartoretti}, {Soubiran}, {Tanga}, {Walton}, {Bailer-Jones},
  {Bastian}, {Drimmel}, {Jansen}, {Katz}, {Lattanzi}, {van Leeuwen}, {Bakker},
  {Cacciari}, {Casta{\~n}eda}, {De Angeli}, {Fabricius}, {Fouesneau},
  {Fr{\'e}mat}, {Galluccio}, {Guerrier}, {Heiter}, {Masana}, {Messineo},
  {Mowlavi}, {Nicolas}, {Nienartowicz}, {Pailler}, {Panuzzo}, {Riclet}, {Roux},
  {Seabroke}, {Sordo}, {Th{\'e}venin}, {Gracia-Abril}, {Portell}, {Teyssier},
  {Altmann}, {Andrae}, {Audard}, {Bellas-Velidis}, {Benson}, {Berthier},
  {Blomme}, {Burgess}, {Busonero}, {Busso}, {C{\'a}novas}, {Carry}, {Cellino},
  {Cheek}, {Clementini}, {Damerdji}, {Davidson}, {de Teodoro}, {Nu{\~n}ez
  Campos}, {Delchambre}, {Dell'Oro}, {Esquej}, {Fern{\'a}ndez-Hern{\'a}ndez},
  {Fraile}, {Garabato}, {Garc{\'\i}a-Lario}, {Gosset}, {Haigron}, {Halbwachs},
  {Hambly}, {Harrison}, {Hern{\'a}ndez}, {Hestroffer}, {Hodgkin}, {Holl},
  {Jan{\ss}en}, {Jevardat de Fombelle}, {Jordan}, {Krone-Martins}, {Lanzafame},
  {L{\"o}ffler}, {Marchal}, {Marrese}, {Moitinho}, {Muinonen}, {Osborne},
  {Pancino}, {Pauwels}, {Recio-Blanco}, {Reyl{\'e}}, {Riello}, {Rimoldini},
  {Roegiers}, {Rybizki}, {Sarro}, {Siopis}, {Smith}, {Sozzetti}, {Utrilla},
  {van Leeuwen}, {Abbas}, {{\'A}brah{\'a}m}, {Abreu Aramburu}, {Aerts},
  {Aguado}, {Ajaj}, {Aldea-Montero}, {Altavilla}, {{\'A}lvarez}, {Alves},
  {Anders}, {Anderson}, {Anglada Varela}, {Antoja}, {Baines}, {Baker},
  {Balaguer-N{\'u}{\~n}ez}, {Balbinot}, {Balog}, {Barache}, {Barbato},
  {Barros}, {Barstow}, {Bartolom{\'e}}, {Bassilana}, {Bauchet}, {Becciani},
  {Bellazzini}, {Berihuete}, {Bernet}, {Bertone}, {Bianchi}, {Binnenfeld},
  {Blanco-Cuaresma}, {Blazere}, {Boch}, {Bombrun}, {Bossini}, {Bouquillon},
  {Bragaglia}, {Bramante}, {Breedt}, {Bressan}, {Brouillet}, {Brugaletta},
  {Bucciarelli}, {Burlacu}, {Butkevich}, {Buzzi}, {Caffau}, {Cancelliere},
  {Cantat-Gaudin}, {Carballo}, {Carlucci}, {Carnerero}, {Carrasco},
  {Casamiquela}, {Castellani}, {Castro-Ginard}, {Chaoul}, {Charlot}, {Chemin},
  {Chiaramida}, {Chiavassa}, {Chornay}, {Comoretto}, {Contursi}, {Cooper},
  {Cornez}, {Cowell}, {Crifo}, {Cropper}, {Crosta}, {Crowley}, {Dafonte},
  {Dapergolas}, {David}, {David}, {de Laverny}, {De Luise}, \& {De
  March}}]{GAIA23}
{Gaia Collaboration}, {Vallenari}, A., {Brown}, A.~G.~A., {et~al.} 2023, \aap,
  674, A1

\bibitem[{{Girardi} {et~al.}(2000){Girardi}, {Bressan}, {Bertelli}, \&
  {Chiosi}}]{GIR00}
{Girardi}, L., {Bressan}, A., {Bertelli}, G., \& {Chiosi}, C. 2000, \aaps, 141,
  371

\bibitem[{{Grigahc{\`e}ne} {et~al.}(2010){Grigahc{\`e}ne}, {Antoci}, {Balona},
  {Catanzaro}, {Daszy{\'n}ska-Daszkiewicz}, {Guzik}, {Handler}, {Houdek},
  {Kurtz}, {Marconi}, {Monteiro}, {Moya}, {Ripepi}, {Su{\'a}rez},
  {Uytterhoeven}, {Borucki}, {Brown}, {Christensen-Dalsgaard}, {Gilliland},
  {Jenkins}, {Kjeldsen}, {Koch}, {Bernabei}, {Bradley}, {Breger}, {Di
  Criscienzo}, {Dupret}, {Garc{\'\i}a}, {Garc{\'\i}a Hern{\'a}ndez},
  {Jackiewicz}, {Kaiser}, {Lehmann}, {Mart{\'\i}n-Ruiz}, {Mathias},
  {Molenda-{\.Z}akowicz}, {Nemec}, {Nuspl}, {Papar{\'o}}, {Roth}, {Szab{\'o}},
  {Suran}, \& {Ventura}}]{GRI10}
{Grigahc{\`e}ne}, A., {Antoci}, V., {Balona}, L., {et~al.} 2010, \apjl, 713,
  L192

\bibitem[{{Hagedus}(1988)}]{HAG88}
{Hagedus}, T. 1988, Bulletin d'Information du Centre de Donnees Stellaires, 35,
  15

\bibitem[{{Handler} \& {Shobbrook}(2002)}]{HAN02}
{Handler}, G. \& {Shobbrook}, R.~R. 2002, \mnras, 333, 251

\bibitem[{{Henry} {et~al.}(2007){Henry}, {Fekel}, \& {Henry}}]{HEN07}
{Henry}, G.~W., {Fekel}, F.~C., \& {Henry}, S.~M. 2007, \aj, 133, 1421

\bibitem[{{Hoffmeister}(1935)}]{HOF35}
{Hoffmeister}, C. 1935, Astronomische Nachrichten, 255, 401

\bibitem[{{Irwin}(1959)}]{IRW59}
{Irwin}, J.~B. 1959, \aj, 64, 149

\bibitem[{{Jeffery} {et~al.}(2017){Jeffery}, {Barnes}, {Skillen}, \&
  {Montemayor}}]{JEF17}
{Jeffery}, E.~J., {Barnes}, III, T.~G., {Skillen}, I., \& {Montemayor}, T.~J.
  2017, \aj, 154, 127

\bibitem[{{Kahraman Ali{\c c}avu{\c s}} {et~al.}(2017){Kahraman Ali{\c c}avu{\c
  s}}, {Soydugan}, {Smalley}, \& {Kub{\'a}t}}]{KAH17}
{Kahraman Ali{\c c}avu{\c s}}, F., {Soydugan}, E., {Smalley}, B., \&
  {Kub{\'a}t}, J. 2017, \mnras, 470, 915

\bibitem[{{Kahraman Ali{\c{c}}avu{\c{s}}} {et~al.}(2023){Kahraman
  Ali{\c{c}}avu{\c{s}}}, {{\c{C}}oban}, {{\c{C}}elik}, {Dogan}, {Ekinci}, \&
  {Ali{\c{c}}avu{\c{s}}}}]{KAH23}
{Kahraman Ali{\c{c}}avu{\c{s}}}, F., {{\c{C}}oban}, {\c{C}}.~G., {{\c{C}}elik},
  E., {et~al.} 2023, \mnras, 524, 619

\bibitem[{{Kaye} {et~al.}(1999){Kaye}, {Handler}, {Krisciunas}, {Poretti}, \&
  {Zerbi}}]{KAY99}
{Kaye}, A.~B., {Handler}, G., {Krisciunas}, K., {Poretti}, E., \& {Zerbi},
  F.~M. 1999, \pasp, 111, 840

\bibitem[{{Kervella} {et~al.}(2022){Kervella}, {Arenou}, \&
  {Th{\'e}venin}}]{KER22}
{Kervella}, P., {Arenou}, F., \& {Th{\'e}venin}, F. 2022, \aap, 657, A7

\bibitem[{{Khalatyan} {et~al.}(2024){Khalatyan}, {Anders}, {Chiappini},
  {Queiroz}, {Nepal}, {dal Ponte}, {Jordi}, {Guiglion}, {Valentini}, {Torralba
  Elipe}, {Steinmetz}, {Pantaleoni-Gonz{\'a}lez}, {Malhotra},
  {Jim{\'e}nez-Arranz}, {Enke}, {Casamiquela}, \& {Ard{\`e}vol}}]{KHA24}
{Khalatyan}, A., {Anders}, F., {Chiappini}, C., {et~al.} 2024, \aap, 691, A98

\bibitem[{{Kim} {et~al.}(2021){Kim}, {Lee}, {Lee}, {Lee}, {Lee}, {Hong}, {Cha},
  {Kim}, \& {Park}}]{KIM21}
{Kim}, S.-L., {Lee}, J.~W., {Lee}, C.-U., {et~al.} 2021, \aj, 162, 212

\bibitem[{{Koch} {et~al.}(2010){Koch}, {Borucki}, {Basri}, {Batalha}, {Brown},
  {Caldwell}, {Christensen-Dalsgaard}, {Cochran}, {DeVore}, {Dunham},
  {Gautier}, {Geary}, {Gilliland}, {Gould}, {Jenkins}, {Kondo}, {Latham},
  {Lissauer}, {Marcy}, {Monet}, {Sasselov}, {Boss}, {Brownlee}, {Caldwell},
  {Dupree}, {Howell}, {Kjeldsen}, {Meibom}, {Morrison}, {Owen}, {Reitsema},
  {Tarter}, {Bryson}, {Dotson}, {Gazis}, {Haas}, {Kolodziejczak}, {Rowe}, {Van
  Cleve}, {Allen}, {Chandrasekaran}, {Clarke}, {Li}, {Quintana}, {Tenenbaum},
  {Twicken}, \& {Wu}}]{KOC10}
{Koch}, D.~G., {Borucki}, W.~J., {Basri}, G., {et~al.} 2010, \apjl, 713, L79

\bibitem[{{Kozyreva} {et~al.}(2019){Kozyreva}, {Kusakin}, {Krajci}, \&
  {Bogomazov}}]{KOZ19}
{Kozyreva}, V.~S., {Kusakin}, A.~V., {Krajci}, T., \& {Bogomazov}, A.~I. 2019,
  Astrophysical Bulletin, 74, 424

\bibitem[{{Kunder} {et~al.}(2017){Kunder}, {Kordopatis}, {Steinmetz},
  {Zwitter}, {McMillan}, {Casagrande}, {Enke}, {Wojno}, {Valentini},
  {Chiappini}, {Matijevi{\v{c}}}, {Siviero}, {de Laverny}, {Recio-Blanco},
  {Bijaoui}, {Wyse}, {Binney}, {Grebel}, {Helmi}, {Jofre}, {Antoja}, {Gilmore},
  {Siebert}, {Famaey}, {Bienaym{\'e}}, {Gibson}, {Freeman}, {Navarro},
  {Munari}, {Seabroke}, {Anguiano}, {{\v{Z}}erjal}, {Minchev}, {Reid},
  {Bland-Hawthorn}, {Kos}, {Sharma}, {Watson}, {Parker}, {Scholz}, {Burton},
  {Cass}, {Hartley}, {Fiegert}, {Stupar}, {Ritter}, {Hawkins}, {Gerhard},
  {Chaplin}, {Davies}, {Elsworth}, {Lund}, {Miglio}, \& {Mosser}}]{KUN17}
{Kunder}, A., {Kordopatis}, G., {Steinmetz}, M., {et~al.} 2017, \aj, 153, 75

\bibitem[{{Kurtz} {et~al.}(2020){Kurtz}, {Handler}, {Rappaport}, {Saio},
  {Fuller}, {Jacobs}, {Schmitt}, {Jones}, {Vanderburg}, {LaCourse}, {Nelson},
  {Kahraman Ali{\c{c}}avu{\c{s}}}, \& {Giarrusso}}]{KUR20}
{Kurtz}, D.~W., {Handler}, G., {Rappaport}, S.~A., {et~al.} 2020, \mnras, 494,
  5118

\bibitem[{{Kwee} \& {van Woerden}(1956)}]{KWE56}
{Kwee}, K.~K. \& {van Woerden}, H. 1956, \bain, 12, 327

\bibitem[{{Lenz} \& {Breger}(2005)}]{LEN05}
{Lenz}, P. \& {Breger}, M. 2005, Communications in Asteroseismology, 146, 53

\bibitem[{{Liakos}(2015)}]{LIA15}
{Liakos}, A. 2015, in Astronomical Society of the Pacific Conference Series,
  Vol. 496, Living Together: Planets, Host Stars and Binaries, ed. S.~M.
  {Rucinski}, G.~{Torres}, \& M.~{Zejda}, 286

\bibitem[{{Liakos}(2017)}]{LIA17}
{Liakos}, A. 2017, \aap, 607, A85

\bibitem[{{Liakos}(2020)}]{LIA20a}
{Liakos}, A. 2020, \aap, 642, A91

\bibitem[{{Liakos}(2025)}]{LIA25}
{Liakos}, A. 2025, Contributions of the Astronomical Observatory Skalnate
  Pleso, 55, 172

\bibitem[{{Liakos} {et~al.}(2022){Liakos}, {Moriarty}, {Blackford}, {West},
  {Evans}, {Moriarty}, \& {Sweet}}]{LIA22}
{Liakos}, A., {Moriarty}, D.~J.~W., {Blackford}, M.~G., {et~al.} 2022, \aap,
  663, A137

\bibitem[{{Liakos} {et~al.}(2024){Liakos}, {Moriarty}, {Erdem}, {West}, \&
  {Evans}}]{LIA24a}
{Liakos}, A., {Moriarty}, D.~J.~W., {Erdem}, A., {West}, J.~F., \& {Evans}, P.
  2024, \aap, 691, A260

\bibitem[{{Liakos} \& {Niarchos}(2015)}]{LIAN15}
{Liakos}, A. \& {Niarchos}, P. 2015, in Astronomical Society of the Pacific
  Conference Series, Vol. 496, Living Together: Planets, Host Stars and
  Binaries, ed. S.~M. {Rucinski}, G.~{Torres}, \& M.~{Zejda}, 195

\bibitem[{{Liakos} \& {Niarchos}(2016)}]{LIAN16}
{Liakos}, A. \& {Niarchos}, P. 2016, in 12th Hellenic Astronomical Conference

\bibitem[{{Liakos} \& {Niarchos}(2017)}]{LIAN17}
{Liakos}, A. \& {Niarchos}, P. 2017, \mnras, 465, 1181

\bibitem[{{Liakos} \& {Niarchos}(2020)}]{LIAN20}
{Liakos}, A. \& {Niarchos}, P. 2020, Galaxies, 8, 75

\bibitem[{{Liakos} {et~al.}(2012){Liakos}, {Niarchos}, {Soydugan}, \&
  {Zasche}}]{LIA12}
{Liakos}, A., {Niarchos}, P., {Soydugan}, E., \& {Zasche}, P. 2012, \mnras,
  422, 1250

\bibitem[{{Loumos} \& {Deeming}(1978)}]{LOU78}
{Loumos}, G.~L. \& {Deeming}, T.~J. 1978, \apss, 56, 285

\bibitem[{{Lucy}(1967)}]{LUC67}
{Lucy}, L.~B. 1967, \zap, 65, 89

\bibitem[{{McDonald} {et~al.}(2012){McDonald}, {Zijlstra}, \& {Boyer}}]{DON12}
{McDonald}, I., {Zijlstra}, A.~A., \& {Boyer}, M.~L. 2012, \mnras, 427, 343

\bibitem[{{Mkrtichian} {et~al.}(2022){Mkrtichian}, {Gunsriviwat}, {Lehmann},
  {Engelbrecht}, {Tkachenko}, \& {Nazarenko}}]{MKR22}
{Mkrtichian}, D., {Gunsriviwat}, K., {Lehmann}, H., {et~al.} 2022, Galaxies,
  10, 97

\bibitem[{{Mkrtichian} {et~al.}(2002){Mkrtichian}, {Kusakin}, {Gamarova}, \&
  {Nazarenko}}]{MKR02}
{Mkrtichian}, D.~E., {Kusakin}, A.~V., {Gamarova}, A.~Y., \& {Nazarenko}, V.
  2002, in Astronomical Society of the Pacific Conference Series, Vol. 259, IAU
  Colloq. 185: Radial and Nonradial Pulsationsn as Probes of Stellar Physics,
  ed. C.~{Aerts}, T.~R. {Bedding}, \& J.~{Christensen-Dalsgaard}, 96

\bibitem[{{Nascimbeni} {et~al.}(2016){Nascimbeni}, {Piotto}, {Ortolani},
  {Giuffrida}, {Marrese}, {Magrin}, {Ragazzoni}, {Pagano}, {Rauer}, {Cabrera},
  {Pollacco}, {Heras}, {Deleuil}, {Gizon}, \& {Granata}}]{NAS16}
{Nascimbeni}, V., {Piotto}, G., {Ortolani}, S., {et~al.} 2016, \mnras, 463,
  4210

\bibitem[{Nelson(2005)}]{NEL05}
Nelson, R. 2005, Batch Minima

\bibitem[{{Oaster} {et~al.}(2006){Oaster}, {Smith}, \& {Kinemuchi}}]{OAS06}
{Oaster}, L., {Smith}, H.~A., \& {Kinemuchi}, K. 2006, \pasp, 118, 405

\bibitem[{{Otero}(2003)}]{OTE03}
{Otero}, S.~A. 2003, Information Bulletin on Variable Stars, 5480, 1

\bibitem[{{Paegert} {et~al.}(2021){Paegert}, {Stassun}, {Collins}, {Pepper},
  {Torres}, {Jenkins}, {Twicken}, \& {Latham}}]{PAE21}
{Paegert}, M., {Stassun}, K.~G., {Collins}, K.~A., {et~al.} 2021, arXiv
  e-prints, arXiv:2108.04778

\bibitem[{{Paunzen}(2015)}]{PAU15}
{Paunzen}, E. 2015, \aap, 580, A23

\bibitem[{{Paxton} {et~al.}(2018){Paxton}, {Schwab}, {Bauer}, {Bildsten},
  {Blinnikov}, {Duffell}, {Farmer}, {Goldberg}, {Marchant}, {Sorokina},
  {Thoul}, {Townsend}, \& {Timmes}}]{PAX18}
{Paxton}, B., {Schwab}, J., {Bauer}, E.~B., {et~al.} 2018, \apjs, 234, 34

\bibitem[{{Pojmanski}(1997)}]{POJ97}
{Pojmanski}, G. 1997, \actaa, 47, 467

\bibitem[{{Pr{\v{s}}a} \& {Zwitter}(2005)}]{PRS05}
{Pr{\v{s}}a}, A. \& {Zwitter}, T. 2005, \apj, 628, 426

\bibitem[{{Qian} {et~al.}(2019){Qian}, {Li}, {He}, {Zhang}, {Zhu}, \&
  {Han}}]{QIA19}
{Qian}, S.-B., {Li}, L.-J., {He}, J.-J., {et~al.} 2019, Research in Astronomy
  and Astrophysics, 19, 001

\bibitem[{{Rauer} {et~al.}(2014){Rauer}, {Catala}, {Aerts}, {Appourchaux},
  {Benz}, {Brandeker}, {Christensen-Dalsgaard}, {Deleuil}, {Gizon}, {Goupil},
  {G{\"u}del}, {Janot-Pacheco}, {Mas-Hesse}, {Pagano}, {Piotto}, {Pollacco},
  {Santos}, {Smith}, {Su{\'a}rez}, {Szab{\'o}}, {Udry}, {Adibekyan}, {Alibert},
  {Almenara}, {Amaro-Seoane}, {Eiff}, {Asplund}, {Antonello}, {Barnes},
  {Baudin}, {Belkacem}, {Bergemann}, {Bihain}, {Birch}, {Bonfils}, {Boisse},
  {Bonomo}, {Borsa}, {Brand{\~a}o}, {Brocato}, {Brun}, {Burleigh}, {Burston},
  {Cabrera}, {Cassisi}, {Chaplin}, {Charpinet}, {Chiappini}, {Church},
  {Csizmadia}, {Cunha}, {Damasso}, {Davies}, {Deeg}, {D{\'\i}az}, {Dreizler},
  {Dreyer}, {Eggenberger}, {Ehrenreich}, {Eigm{\"u}ller}, {Erikson}, {Farmer},
  {Feltzing}, {de Oliveira Fialho}, {Figueira}, {Forveille}, {Fridlund},
  {Garc{\'\i}a}, {Giommi}, {Giuffrida}, {Godolt}, {Gomes da Silva}, {Granzer},
  {Grenfell}, {Grotsch-Noels}, {G{\"u}nther}, {Haswell}, {Hatzes},
  {H{\'e}brard}, {Hekker}, {Helled}, {Heng}, {Jenkins}, {Johansen},
  {Khodachenko}, {Kislyakova}, {Kley}, {Kolb}, {Krivova}, {Kupka}, {Lammer},
  {Lanza}, {Lebreton}, {Magrin}, {Marcos-Arenal}, {Marrese}, {Marques},
  {Martins}, {Mathis}, {Mathur}, {Messina}, {Miglio}, {Montalban}, {Montalto},
  {Monteiro}, {Moradi}, {Moravveji}, {Mordasini}, {Morel}, {Mortier},
  {Nascimbeni}, {Nelson}, {Nielsen}, {Noack}, {Norton}, {Ofir}, {Oshagh},
  {Ouazzani}, {P{\'a}pics}, {Parro}, {Petit}, {Plez}, {Poretti}, {Quirrenbach},
  {Ragazzoni}, {Raimondo}, {Rainer}, {Reese}, {Redmer}, {Reffert},
  {Rojas-Ayala}, {Roxburgh}, {Salmon}, {Santerne}, {Schneider}, {Schou},
  {Schuh}, {Schunker}, {Silva-Valio}, {Silvotti}, {Skillen}, {Snellen}, {Sohl},
  {Sousa}, {Sozzetti}, {Stello}, {Strassmeier}, {{\v{S}}vanda}, {Szab{\'o}},
  {Tkachenko}, {Valencia}, {Van Grootel}, {Vauclair}, {Ventura}, {Wagner},
  {Walton}, {Weingrill}, {Werner}, {Wheatley}, \& {Zwintz}}]{RAU14}
{Rauer}, H., {Catala}, C., {Aerts}, C., {et~al.} 2014, Experimental Astronomy,
  38, 249

\bibitem[{{Ricker} {et~al.}(2015){Ricker}, {Winn}, {Vanderspek}, {Latham},
  {Bakos}, {Bean}, {Berta-Thompson}, {Brown}, {Buchhave}, {Butler}, {Butler},
  {Chaplin}, {Charbonneau}, {Christensen-Dalsgaard}, {Clampin}, {Deming},
  {Doty}, {De Lee}, {Dressing}, {Dunham}, {Endl}, {Fressin}, {Ge}, {Henning},
  {Holman}, {Howard}, {Ida}, {Jenkins}, {Jernigan}, {Johnson}, {Kaltenegger},
  {Kawai}, {Kjeldsen}, {Laughlin}, {Levine}, {Lin}, {Lissauer}, {MacQueen},
  {Marcy}, {McCullough}, {Morton}, {Narita}, {Paegert}, {Palle}, {Pepe},
  {Pepper}, {Quirrenbach}, {Rinehart}, {Sasselov}, {Sato}, {Seager},
  {Sozzetti}, {Stassun}, {Sullivan}, {Szentgyorgyi}, {Torres}, {Udry}, \&
  {Villasenor}}]{RIC15}
{Ricker}, G.~R., {Winn}, J.~N., {Vanderspek}, R., {et~al.} 2015, Journal of
  Astronomical Telescopes, Instruments, and Systems, 1, 014003

\bibitem[{{Ruci{\'n}ski}(1969)}]{RUC69}
{Ruci{\'n}ski}, S.~M. 1969, \actaa, 19, 245

\bibitem[{{Schofield} {et~al.}(2019){Schofield}, {Chaplin}, {Huber},
  {Campante}, {Davies}, {Miglio}, {Ball}, {Appourchaux}, {Basu}, {Bedding},
  {Christensen-Dalsgaard}, {Creevey}, {Garc{\'\i}a}, {Handberg}, {Kawaler},
  {Kjeldsen}, {Latham}, {Lund}, {Metcalfe}, {Ricker}, {Serenelli}, {Silva
  Aguirre}, {Stello}, \& {Vanderspek}}]{SCH19}
{Schofield}, M., {Chaplin}, W.~J., {Huber}, D., {et~al.} 2019, \apjs, 241, 12

\bibitem[{{Shi} {et~al.}(2022){Shi}, {Qian}, \& {Li}}]{SHI22}
{Shi}, X.-d., {Qian}, S.-b., \& {Li}, L.-J. 2022, \apjs, 259, 50

\bibitem[{{Shugarov}(1985)}]{SHU85}
{Shugarov}, S.~Y. 1985, Astronomicheskij Tsirkulyar, 1359, 4

\bibitem[{{Southworth}(2025)}]{SOU25}
{Southworth}, J. 2025, arXiv e-prints, arXiv:2501.17068

\bibitem[{{Soydugan} {et~al.}(2006{\natexlab{a}}){Soydugan}, {{\.I}bano{\v
  g}lu}, {Soydugan}, {Akan}, \& {Demircan}}]{SOY06a}
{Soydugan}, E., {{\.I}bano{\v g}lu}, C., {Soydugan}, F., {Akan}, M.~C., \&
  {Demircan}, O. 2006{\natexlab{a}}, \mnras, 366, 1289

\bibitem[{{Soydugan} {et~al.}(2006{\natexlab{b}}){Soydugan}, {Soydugan},
  {Demircan}, \& {{\.I}bano{\v{g}}lu}}]{SOY06b}
{Soydugan}, E., {Soydugan}, F., {Demircan}, O., \& {{\.I}bano{\v{g}}lu}, C.
  2006{\natexlab{b}}, \mnras, 370, 2013

\bibitem[{{Stassun} {et~al.}(2019){Stassun}, {Oelkers}, {Paegert}, {Torres},
  {Pepper}, {De Lee}, {Collins}, {Latham}, {Muirhead}, {Chittidi},
  {Rojas-Ayala}, {Fleming}, {Rose}, {Tenenbaum}, {Ting}, {Kane}, {Barclay},
  {Bean}, {Brassuer}, {Charbonneau}, {Ge}, {Lissauer}, {Mann}, {McLean},
  {Mullally}, {Narita}, {Plavchan}, {Ricker}, {Sasselov}, {Seager}, {Sharma},
  {Shiao}, {Sozzetti}, {Stello}, {Vanderspek}, {Wallace}, \& {Winn}}]{STA19}
{Stassun}, K.~G., {Oelkers}, R.~J., {Paegert}, M., {et~al.} 2019, \aj, 158, 138

\bibitem[{{Steinmetz} {et~al.}(2020){Steinmetz}, {Guiglion}, {McMillan},
  {Matijevi{\v{c}}}, {Enke}, {Kordopatis}, {Zwitter}, {Valentini}, {Chiappini},
  {Casagrande}, {Wojno}, {Anguiano}, {Bienaym{\'e}}, {Bijaoui}, {Binney},
  {Burton}, {Cass}, {de Laverny}, {Fiegert}, {Freeman}, {Fulbright}, {Gibson},
  {Gilmore}, {Grebel}, {Helmi}, {Kunder}, {Munari}, {Navarro}, {Parker},
  {Ruchti}, {Recio-Blanco}, {Reid}, {Seabroke}, {Siviero}, {Siebert}, {Stupar},
  {Watson}, {Williams}, {Wyse}, {Anders}, {Antoja}, {Birko}, {Bland-Hawthorn},
  {Bossini}, {Garc{\'\i}a}, {Carrillo}, {Chaplin}, {Elsworth}, {Famaey},
  {Gerhard}, {Jofre}, {Just}, {Mathur}, {Miglio}, {Minchev}, {Monari},
  {Mosser}, {Ritter}, {Rodrigues}, {Scholz}, {Sharma}, {Sysoliatina}, \& {RAVE
  Collaboration}}]{STE20}
{Steinmetz}, M., {Guiglion}, G., {McMillan}, P.~J., {et~al.} 2020, \aj, 160, 83

\bibitem[{{Stevens} {et~al.}(2017){Stevens}, {Stassun}, \& {Gaudi}}]{STE17}
{Stevens}, D.~J., {Stassun}, K.~G., \& {Gaudi}, B.~S. 2017, \aj, 154, 259

\bibitem[{{Strohmeier} {et~al.}(1966){Strohmeier}, {Fischer}, \& {Ott}}]{STR66}
{Strohmeier}, W., {Fischer}, H., \& {Ott}, H. 1966, Information Bulletin on
  Variable Stars, 120, 1

\bibitem[{{Strohmeier} {et~al.}(1965){Strohmeier}, {Knigge}, \& {Ott}}]{STR65}
{Strohmeier}, W., {Knigge}, R., \& {Ott}, H. 1965, Information Bulletin on
  Variable Stars, 107, 1

\bibitem[{{Tonry} {et~al.}(2018){Tonry}, {Denneau}, {Flewelling}, {Heinze},
  {Onken}, {Smartt}, {Stalder}, {Weiland}, \& {Wolf}}]{TON18}
{Tonry}, J.~L., {Denneau}, L., {Flewelling}, H., {et~al.} 2018, \apj, 867, 105

\bibitem[{{Tsantaki} {et~al.}(2022){Tsantaki}, {Pancino}, {Marrese},
  {Marinoni}, {Rainer}, {Sanna}, {Turchi}, {Randich}, {Gallart}, {Battaglia},
  \& {Masseron}}]{TSA22}
{Tsantaki}, M., {Pancino}, E., {Marrese}, P., {et~al.} 2022, \aap, 659, A95

\bibitem[{{Uytterhoeven} {et~al.}(2011){Uytterhoeven}, {Moya},
  {Grigahc{\`e}ne}, {Guzik}, {Guti{\'e}rrez-Soto}, {Smalley}, {Handler},
  {Balona}, {Niemczura}, {Fox Machado}, {Benatti}, {Chapellier}, {Tkachenko},
  {Szab{\'o}}, {Su{\'a}rez}, {Ripepi}, {Pascual}, {Mathias},
  {Mart{\'{\i}}n-Ru{\'{\i}}z}, {Lehmann}, {Jackiewicz}, {Hekker},
  {Gruberbauer}, {Garc{\'{\i}}a}, {Dumusque}, {D{\'{\i}}az-Fraile}, {Bradley},
  {Antoci}, {Roth}, {Leroy}, {Murphy}, {De Cat}, {Cuypers}, {Kjeldsen},
  {Christensen-Dalsgaard}, {Breger}, {Pigulski}, {Kiss}, {Still}, {Thompson},
  \& {van Cleve}}]{UYT11}
{Uytterhoeven}, K., {Moya}, A., {Grigahc{\`e}ne}, A., {et~al.} 2011, \aap, 534,
  A125

\bibitem[{{Verberne} {et~al.}(2024){Verberne}, {Koposov}, {Rossi}, {Marchetti},
  {Kuijken}, \& {Penoyre}}]{VER24}
{Verberne}, S., {Koposov}, S.~E., {Rossi}, E.~M., {et~al.} 2024, \aap, 684, A29

\bibitem[{{Volkov} \& {Volkova}(2010)}]{VOL10b}
{Volkov}, I. \& {Volkova}, N. 2010, in Astronomical Society of the Pacific
  Conference Series, Vol. 435, Binaries - Key to Comprehension of the Universe,
  ed. A.~{Pr{\v s}a} \& M.~{Zejda}, 323

\bibitem[{{Volkov}(1990)}]{VOL90}
{Volkov}, I.~M. 1990, Information Bulletin on Variable Stars, 3493

\bibitem[{{von Zeipel}(1924)}]{ZEI24}
{von Zeipel}, H. 1924, \mnras, 84, 665

\bibitem[{{Warner} {et~al.}(2003){Warner}, {Kaye}, \& {Guzik}}]{WAR03}
{Warner}, P.~B., {Kaye}, A.~B., \& {Guzik}, J.~A. 2003, \apj, 593, 1049

\bibitem[{{Wilson} \& {Devinney}(1971)}]{WIL71}
{Wilson}, R.~E. \& {Devinney}, E.~J. 1971, \apj, 166, 605

\bibitem[{{Zasche} {et~al.}(2009){Zasche}, {Liakos}, {Niarchos}, {Wolf},
  {Manimanis}, \& {Gazeas}}]{ZAS09}
{Zasche}, P., {Liakos}, A., {Niarchos}, P., {et~al.} 2009, \na, 14, 121

\bibitem[{{Zhang} {et~al.}(2013){Zhang}, {Luo}, \& {Fu}}]{ZHA13}
{Zhang}, X.-B., {Luo}, C.-Q., \& {Fu}, J.-N. 2013, \apj, 777, 77

\bibitem[{{Zhevakin}(1963)}]{ZHE63}
{Zhevakin}, S.~A. 1963, \araa, 1, 367

\bibitem[{{Zhou}(2001)}]{ZHO01a}
{Zhou}, A.-Y. 2001, Information Bulletin on Variable Stars, 5087

\end{thebibliography}

\onecolumn
\begin{appendix}

\section{TESS light curves}
\label{Sec:LCTESS}

\begin{figure*}
\centering
\includegraphics[width=18cm]{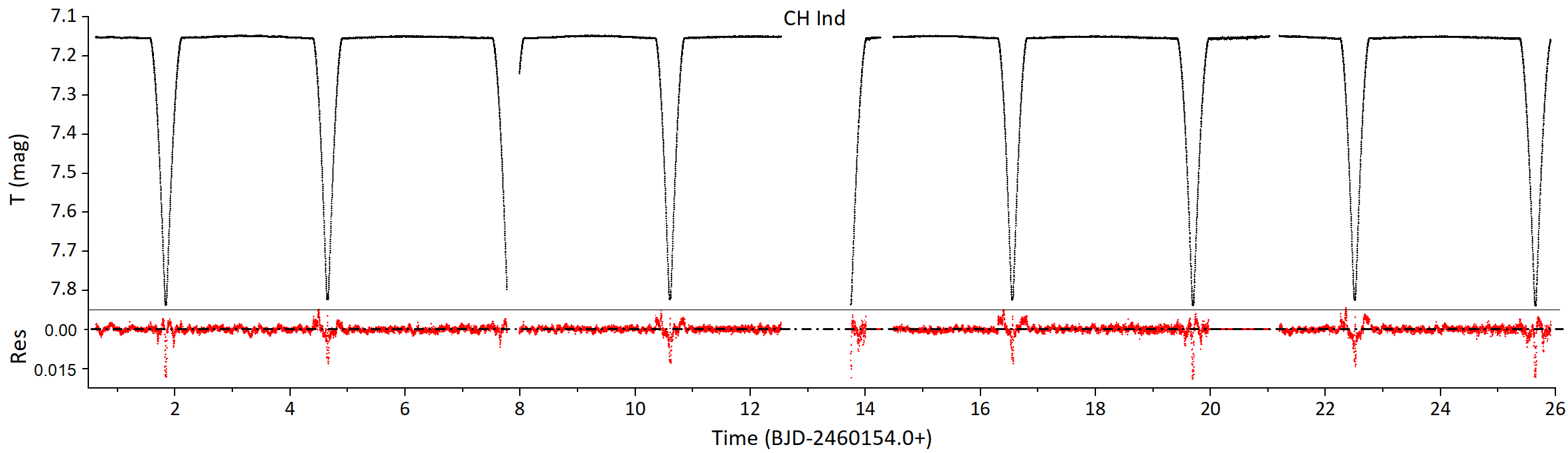}\\
\includegraphics[width=18cm]{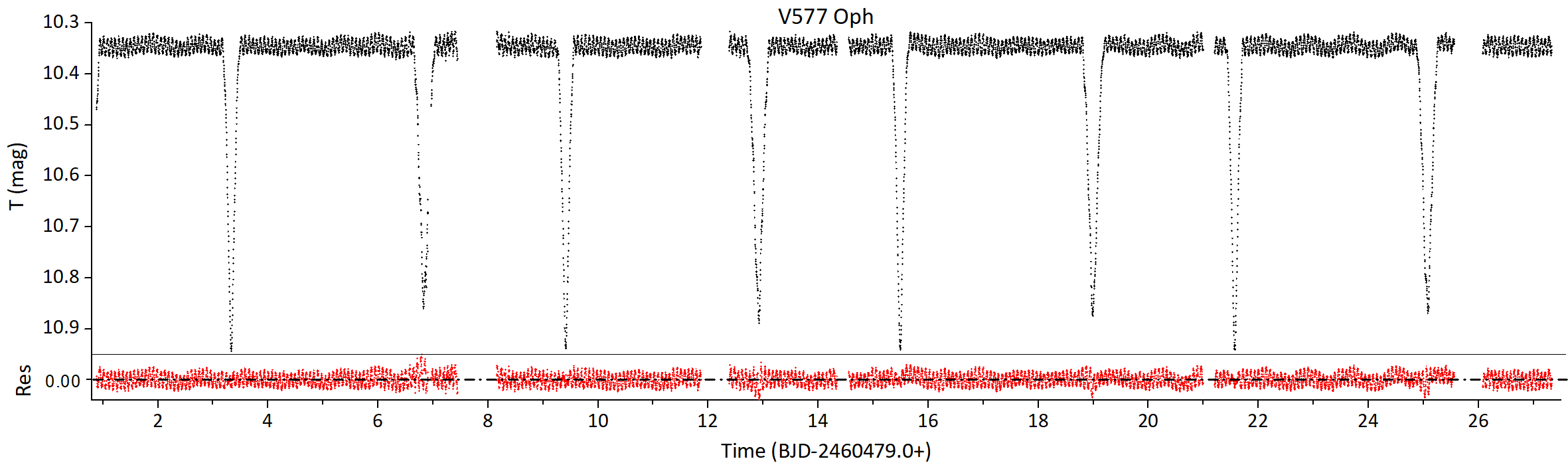}\\
\includegraphics[width=18cm]{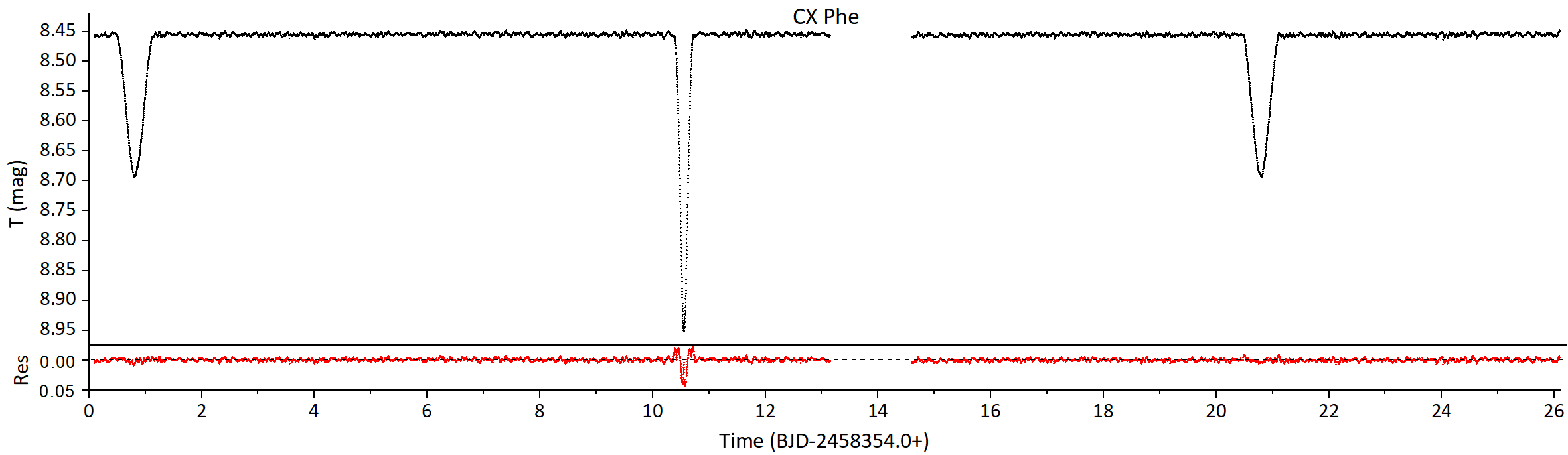}\\
\includegraphics[width=18cm]{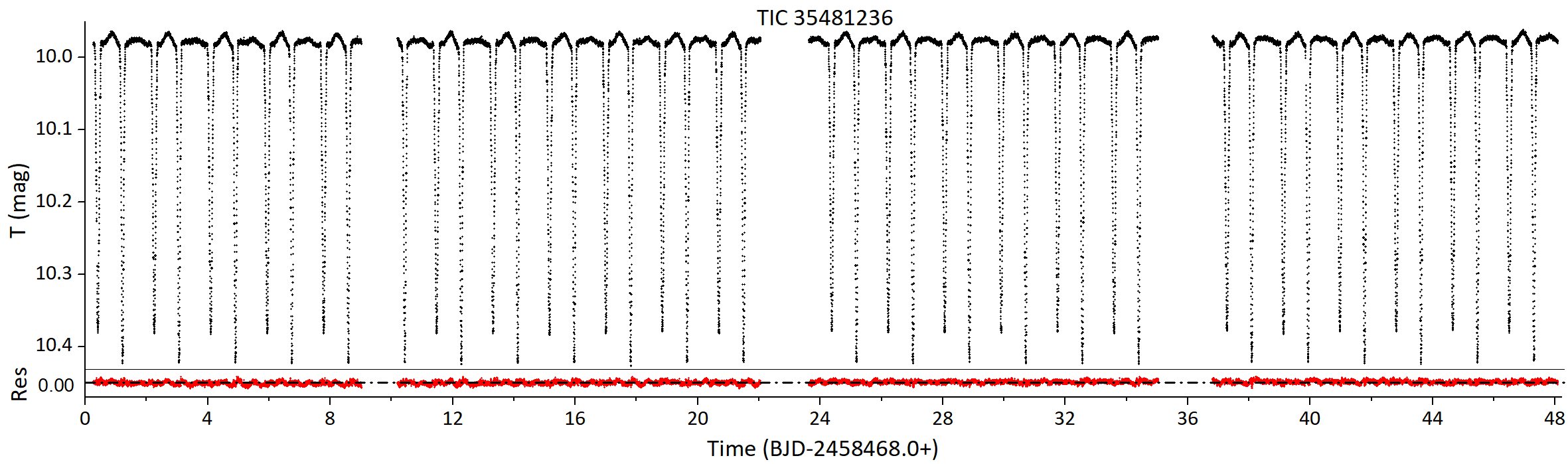}\\
\caption{$TESS$ light curves (upper panels) and residuals of the binary modelling (lower panels) of all systems studied.}
\label{fig:LCs}
\end{figure*}

\section{Times of minima}
\label{Sec:MIN}

Table~\ref{Tab:MIN} contains the timings of primary (I) and secondary (II) minima of all systems, which are based on the TESS data. The minima timings were calculated with the `Batch minima' software \citep{NEL05} using the \citet{KWE56} method.

\begin{table*}
\begin{center}											
\caption{Calculated times of minima of all studied cases.}	
\label{Tab:MIN}	
\begin{tabular}{cc cc cc cc cc }														
\hline\hline																			
BJD -2457000	&	Type	&	BJD -2457000	&	Type	&	BJD -2457000	&	Type	&	BJD -2457000	&	Type	&	BJD -2457000	&	Type	\\
\hline																			
\multicolumn{2}{c}{CH Ind}			&	3504.0787(10)	&	II	&	1472.91328(16)	&	 I	&	1487.66098(6)	&	 I	&	1505.28951(5)	&	II	\\
\cline{1-2}			\cline{3-4}																
3155.84288(1)	&	 I	&	\multicolumn{2}{c}{CX Phe}			&	1473.95100(5)	&	II	&	1488.69864(4)	&	II	&	1506.09526(6)	&	 I	\\
			\cline{3-4}																
3158.65493(1)	&	II	&	1354.8204(3)	&	II	&	1474.75676(3)	&	 I	&	1489.50423(5)	&	 I	&	1507.13300(6)	&	II	\\
3164.60744(1)	&	II	&	1364.5610(1)	&	 I	&	1475.79454(4)	&	II	&	1492.38527(4)	&	II	&	1507.93868(4)	&	 I	\\
3170.56013(1)	&	II	&	1374.7973(4)	&	II	&	1476.60013(4)	&	 I	&	1493.19096(4)	&	 I	&	1508.97636(4)	&	II	\\
3173.70063(2)	&	 I	&	2093.9286(2)	&	II	&	1478.44357(4)	&	 I	&	1494.22877(3)	&	II	&	1509.78213(4)	&	 I	\\
3176.51278(1)	&	II	&	2103.6688(4)	&	 I	&	1479.48139(5)	&	II	&	1495.03477(4)	&	 I	&	1510.81970(4)	&	II	\\
3179.65317(2)	&	 I	&	2113.9003(5)	&	II	&	1480.28676(4)	&	 I	&	1496.07228(4)	&	II	&	1511.62530(5)	&	 I	\\
\cline{1-2}																			
\multicolumn{2}{c}{V577 Oph}			&	3192.6029(3)	&	II	&	1481.32474(6)	&	II	&	1496.87795(4)	&	 I	&	1512.66310(5)	&	II	\\
\cline{1-2}																			
3482.3375(2)	&	 I	&	3202.3433(5)	&	 I	&	1482.13038(4)	&	 I	&	1497.91555(5)	&	II	&	1513.46885(3)	&	 I	\\
			\cline{3-4}																
3485.8481(14)	&	II	&	\multicolumn{2}{c}{TIC 35481236}			&	1483.16842(5)	&	II	&	1498.72148(4)	&	 I	&	1514.50638(5)	&	II	\\
			\cline{3-4}																
3488.4174(3)	&	 I	&	1468.42075(6)	&	II	&	1483.97389(4)	&	 I	&	1499.75917(5)	&	II	&	1515.31206(3)	&	 I	\\
3491.9223(9)	&	II	&	1469.22649(5)	&	 I	&	1485.01158(4)	&	II	&	1500.56502(5)	&	 I	&		&		\\
3494.4959(2)	&	 I	&	1470.26417(4)	&	II	&	1485.81722(5)	&	 I	&	1501.60262(4)	&	II	&		&		\\
3500.5758(2)	&	 I	&	1472.10784(4)	&	II	&	1486.85501(6)	&	II	&	1502.40795(5)	&	 I	&		&		\\
\hline																																					
									
\end{tabular}
\end{center}
\end{table*}


\section{Combination pulsation frequencies}
\label{Sec:PulsMod}

\begin{figure*}[h!]
\centering
\begin{tabular}{cc}		
\includegraphics[width=8.3cm]{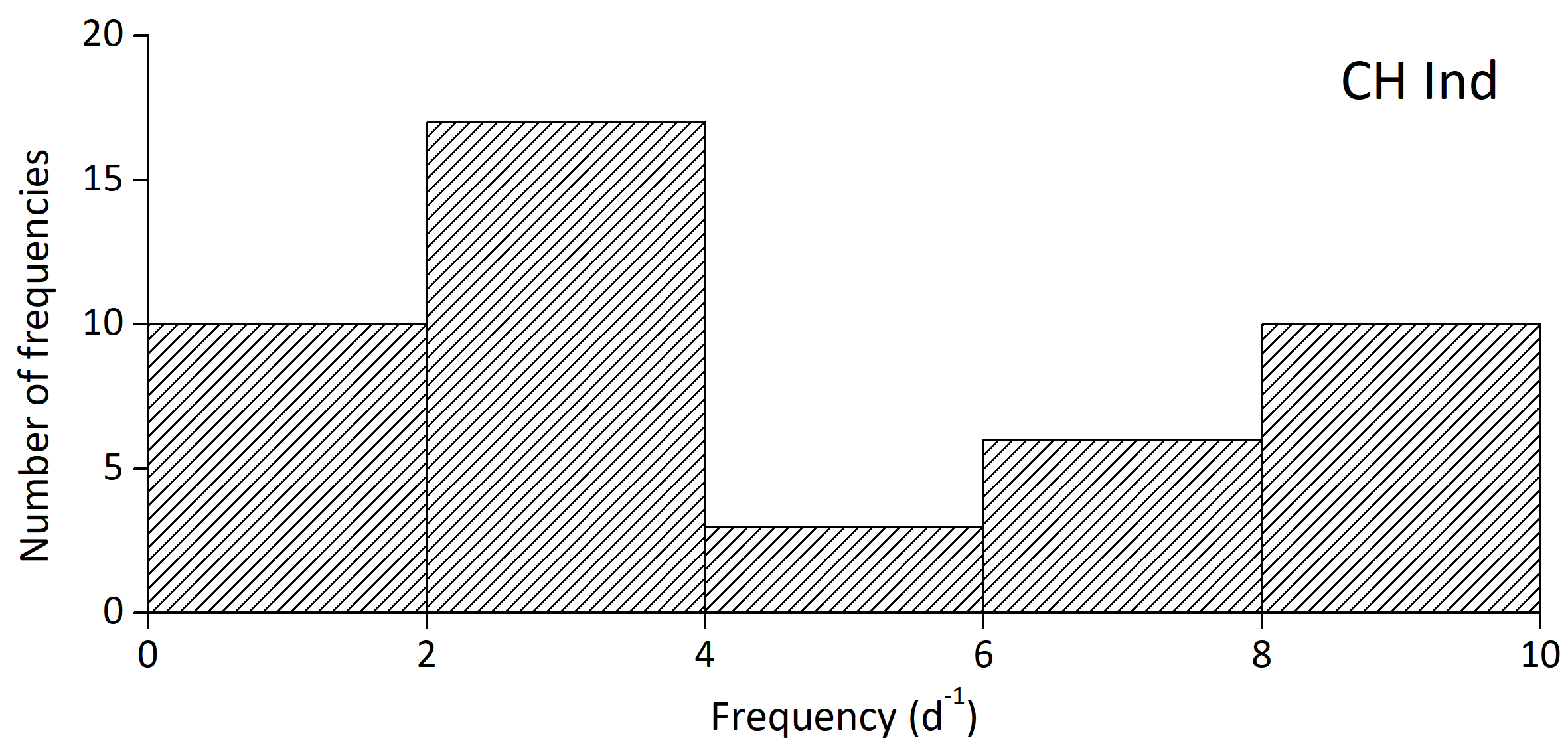}&\includegraphics[width=8.3cm]{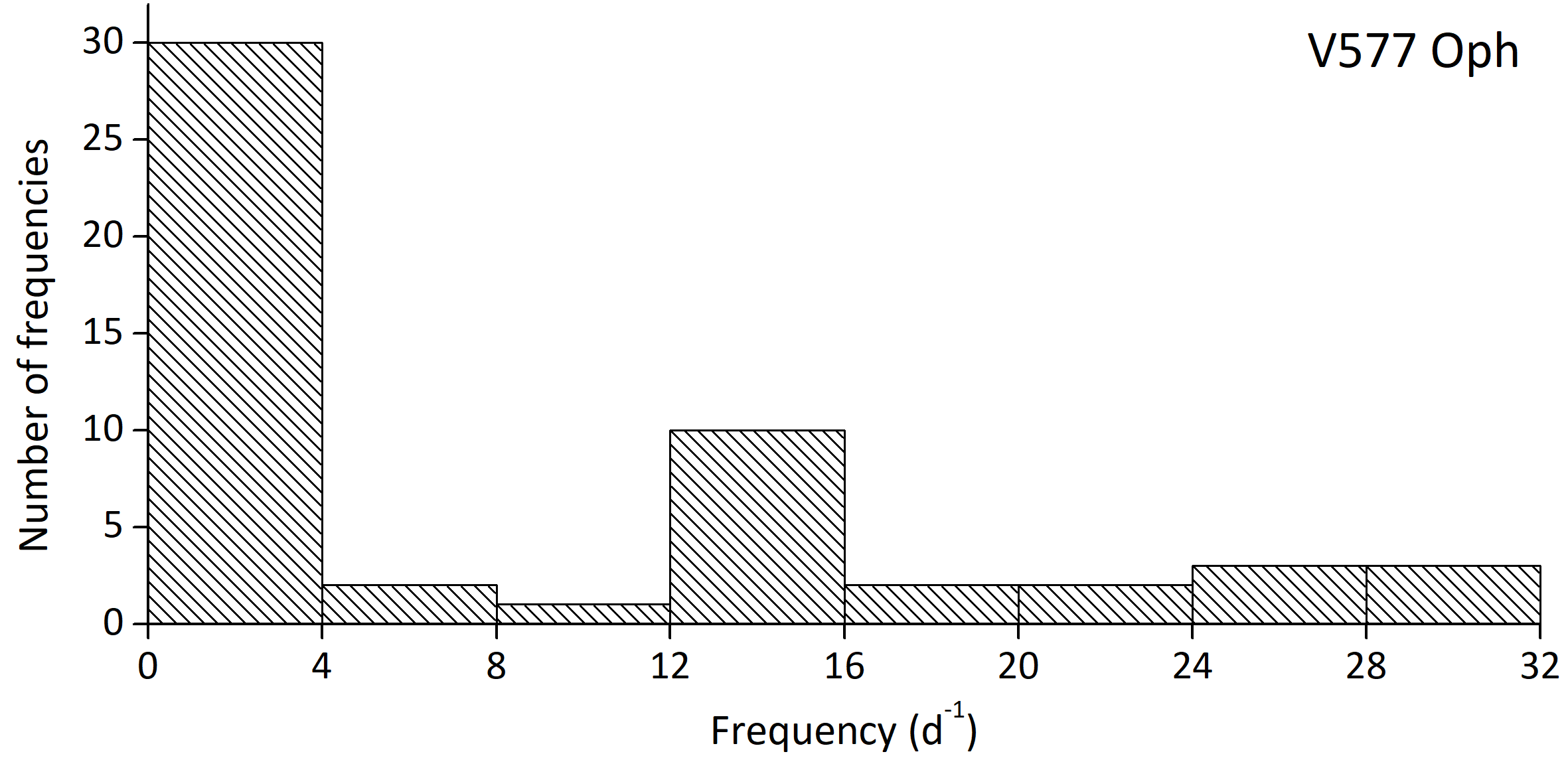}\\
\includegraphics[width=8.3cm]{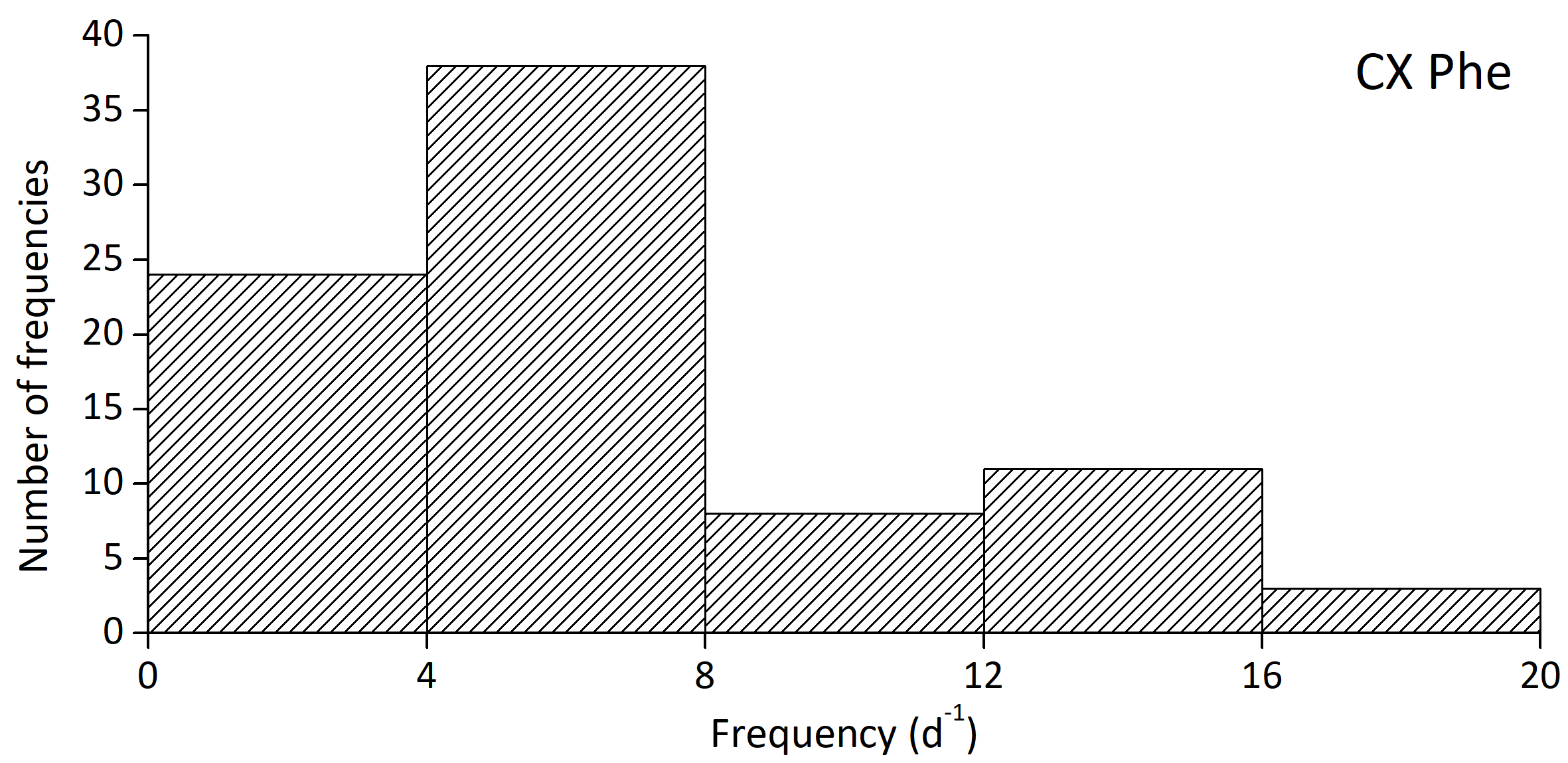}&\includegraphics[width=8.3cm]{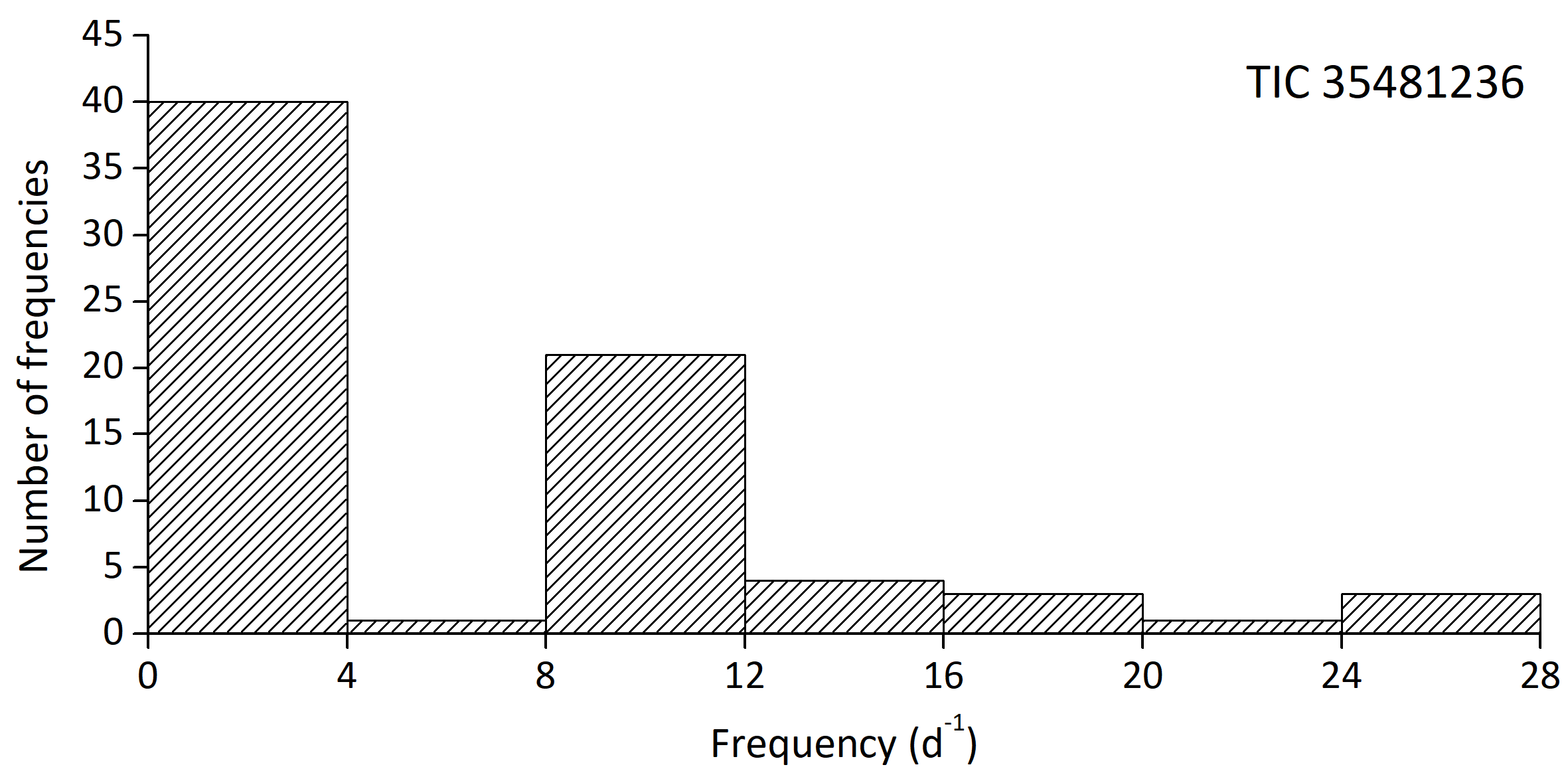}\\
\end{tabular}
\caption{Frequencies distributions of all systems studied.}
\label{fig:FD}
\end{figure*}

Table~\ref{tab:fcombo} contains the increasing number of the frequency ($i$), its value ($f_{\rm i}$), amplitude ($A$), phase ($\Phi$), $S/N$, and the most possible frequency combination. This table is complementary to Table~\ref{Tab:FreqsInd} and their integration is the complete model of the pulsational behaviour of each studied system (Sect.~\ref{Sec:Puls}). Fig.~\ref{fig:FD} plots the distributions of the detected frequencies of all systems studied.

\begin{table*}
\begin{center}											
\caption{Combination pulsation frequencies of all studied systems.}	
\label{tab:fcombo}	
\scalebox{0.945}{
\begin{tabular}{cc cc cc cc cc cc}														
\hline\hline																							
$i$	&	  $f_{\rm i}$	&	$A$	&	  $\Phi$	&	S/N	&	Combination	&	$i$	&	  $f_{\rm i}$	&	$A$	&	  $\Phi$	&	S/N	&	Combination	\\
	&	     (d$^{-1}$)	&	(mmag)	&	(2$\pi$~rad)	&		&		&		&	     (d$^{-1}$)	&	(mmag)	&	(2$\pi$~rad)	&		&		\\
\hline																							
\multicolumn{6}{c}{CH~Ind}											&	30	&	1.632(2)	&	0.046(5)	&	0.35(2)	&	7.3	&	$10f_{\rm orb}$ 	\\
\cline{1-6}																							
3	&	8.2365(9)	&	0.119(5)	&	0.324(6)	&	19.0	&	$3f_2$	&	31	&	9.235(2)	&	0.043(5)	&	0.73(2)	&	6.8	&	$f_1+2f_{\rm orb}$ 	\\
4	&	7.7292(9)	&	0.115(5)	&	0.715(6)	&	18.4	&	$3f_2-3f_{\rm orb}$ 	&	32	&	3.873(3)	&	0.041(5)	&	0.38(2)	&	6.5	&	$f_2+7f_{\rm orb}$ 	\\
5	&	3.0563(9)	&	0.112(5)	&	0.049(7)	&	17.9	&	$f_2+2f_{\rm orb}$	&	33	&	2.892(3)	&	0.039(5)	&	0.27(2)	&	6.3	&	$f_1-2f_2-f_{\rm orb}$ 	\\
6	&	8.566(1)	&	0.091(5)	&	0.787(8)	&	14.5	&	$3f_2-4f_{\rm orb}$	&	34	&	3.564(3)	&	0.039(5)	&	0.41(2)	&	6.3	&	$f_2+5f_{\rm orb}$ 	\\
7	&	0.113(1)	&	0.089(5)	&	0.679(8)	&	14.3	&	$f_1-3f_2-f_{\rm orb}$  	&	35	&	3.593(2)	&	0.047(5)	&	0.69(2)	&	7.4	&	$\sim f_{34}$	\\
8	&	2.813(1)	&	0.088(5)	&	0.101(8)	&	14.1	&	$f_1-2f_2-f_{\rm orb}$	&	36	&	4.543(3)	&	0.038(5)	&	0.22(2)	&	6.1	&	$f_2+11f_{\rm orb}$	\\
9	&	1.040(1)	&	0.089(5)	&	0.636(8)	&	14.2	&	$\sim 6f_{\rm orb}$	&	37	&	5.371(3)	&	0.039(5)	&	0.36(2)	&	6.1	&	$f_2+16f_{\rm orb}$ 	\\
10	&	3.112(1)	&	0.079(5)	&	0.880(9)	&	12.6	&	$\sim f_5$	&	38	&	8.056(3)	&	0.038(5)	&	0.54(2)	&	6.1	&	$f_1-5f_{\rm orb}$	\\
11	&	2.651(1)	&	0.079(5)	&	0.337(9)	&	12.6	&	$4f_2-f_1+f_{\rm orb}$ 	&	39	&	6.727(3)	&	0.038(5)	&	0.77(2)	&	6.1	&	$f_1-13f_{\rm orb}$	\\
12	&	9.405(1)	&	0.078(5)	&	0.178(9)	&	12.4	&	$3f_2-7f_{\rm orb}$ 	&	40	&	6.049(3)	&	0.038(5)	&	0.14(2)	&	6.0	&	$2f_2+4f_{\rm orb}$   	\\
13	&	9.580(2)	&	0.068(5)	&	0.53(1)	&	10.9	&	 $3f_2-8f_{\rm orb}$ 	&	41	&	1.667(3)	&	0.037(5)	&	0.60(2)	&	5.9	&	$10f_{\rm orb}$	\\
14	&	7.056(1)	&	0.070(5)	&	0.90(1)	&	11.1	&	$3f_2-7f_{\rm orb}$ 	&	42	&	3.797(3)	&	0.038(5)	&	0.20(2)	&	6.1	&	$f_2+6f_{\rm orb}$	\\
15	&	0.846(2)	&	0.067(5)	&	0.35(1)	&	10.7	&	$5f_{\rm orb}$	&	43	&	3.469(3)	&	0.037(5)	&	0.67(2)	&	6.0	&	$f_2+4f_{\rm orb}$	\\
16	&	2.329(2)	&	0.067(5)	&	0.57(1)	&	10.7	&	$14f_{\rm orb}$ 	&	44	&	1.182(3)	&	0.034(5)	&	0.75(2)	&	5.5	&	$7f_{\rm orb}$ 	\\
17	&	0.180(2)	&	0.064(5)	&	0.43(1)	&	10.2	&	$\sim f_{\rm orb}$	&	45	&	2.570(3)	&	0.035(5)	&	0.55(2)	&	5.6	&	$f_{2}-f_{\rm orb}$	\\
18	&	0.793(2)	&	0.057(5)	&	0.25(1)	&	9.2	&	$\sim 5f_{\rm orb}$	&	46	&	8.732(3)	&	0.033(5)	&	0.11(2)	&	5.3	&	$f_1-f_{\rm orb}$ 	\\
												\cline{7-12}											
19	&	0.939(2)	&	0.060(5)	&	0.16(1)	&	9.6	&	$3f_2-f_1+7f_{\rm orb}$    	&	\multicolumn{6}{c}{V577~Oph}											\\
												\cline{7-12}											
20	&	5.549(2)	&	0.056(5)	&	0.01(1)	&	8.9	&	$2f_2$ 	&	2	&	1.1522(3)	&	2.82(4)	&	0.461(2)	&	55.1	&	7$f_{\rm orb}$	\\
21	&	9.069(2)	&	0.055(5)	&	0.15(1)	&	8.8	&	$f_1+f_{\rm orb}$ 	&	4	&	14.5788(6)	&	1.35(4)	&	0.954(5)	&	26.3	&	$f_1+f_{\rm orb}$	\\
22	&	6.226(2)	&	0.057(5)	&	0.74(1)	&	9.0	&	$3f_2-12f_{\rm orb}$	&	5	&	11.3186(6)	&	1.29(4)	&	0.064(5)	&	25.3	&	$f_1-2f_3$	\\
23	&	2.132(2)	&	0.054(5)	&	0.81(1)	&	8.6	&	$\sim 13f_{\rm orb}$	&	6	&	1.4054(7)	&	1.24(4)	&	0.424(5)	&	24.3	&	$f_3-f_{\rm orb}$ 	\\
24	&	2.477(2)	&	0.055(5)	&	0.15(1)	&	8.8	&	$f_2-2f_{\rm orb}$ 	&	7	&	1.1812(7)	&	1.15(4)	&	0.736(6)	&	22.6	&	$\sim f_2$ 	\\
25	&	3.234(2)	&	0.052(5)	&	0.39(1)	&	8.3	&	$f_2+3f_{\rm orb}$ 	&	8	&	0.9290(8)	&	1.03(4)	&	0.081(6)	&	20.1	&	$\sim 6f_{\rm orb}$ 	\\
26	&	7.889(2)	&	0.050(5)	&	0.87(1)	&	7.9	&	$f_1-6f_{\rm orb}$ 	&	9	&	25.7111(8)	&	1.00(4)	&	0.532(6)	&	19.6	&	$2f_1-2f_3$ 	\\
27	&	8.392(2)	&	0.049(5)	&	0.65(2)	&	7.8	&	$f_1-3f_{\rm orb}$	&	10	&	1.4654(9)	&	0.95(4)	&	0.840(7)	&	18.6	&	$\sim9f_{\rm orb}$ 	\\
28	&	2.781(2)	&	0.049(5)	&	0.26(2)	&	7.8	&	$\sim f_{2}$	&	11	&	1.2342(9)	&	0.89(4)	&	0.475(7)	&	17.4	&	$f_3-2f_{\rm orb}$ 	\\
29	&	0.679(2)	&	0.046(5)	&	0.91(2)	&	7.3	&	$4f_{\rm orb}$ 	&	12	&	1.727(1)	&	0.80(4)	&	0.420(8)	&	15.6	&	$f_3+f_{\rm orb}$ 	\\
\hline						
\end{tabular}}
\end{center}
\end{table*}

\begin{table*}
\begin{center}											
\caption*{Table~\ref{tab:fcombo} (cont'd)}
\label{Tab:FreqsDep}	
\scalebox{0.95}{
\begin{tabular}{cc cc cc cc cc cc}														
\hline\hline																							
$i$	&	  $f_{\rm i}$	&	$A$	&	  $\Phi$	&	S/N	&	Combination	&	$i$	&	  $f_{\rm i}$	&	$A$	&	  $\Phi$	&	S/N	&	Combination	\\
	&	     (d$^{-1}$)	&	(mmag)	&	(2$\pi$~rad)	&		&		&		&	     (d$^{-1}$)	&	(mmag)	&	(2$\pi$~rad)	&		&		\\
\hline																							
\multicolumn{6}{c}{V577~Oph cont'd}											&	24	&	7.459(1)	&	0.21(1)	&	0.735(7)	&	12.1	&	$f_4+5f_{\rm orb}$ 	\\
\cline{1-6}																							
13	&	1.347(1)	&	0.75(4)	&	0.710(9)	&	14.6	&	$\sim f_6$	&	25	&	8.538(1)	&	0.20(1)	&	0.167(7)	&	11.9	&	$f_3-f_4+6f_{\rm orb}$  	\\
14	&	0.505(1)	&	0.74(4)	&	0.918(9)	&	14.4	&	$3f_{\rm orb}$ 	&	26	&	12.449(1)	&	0.19(1)	&	0.947(7)	&	11.5	&	$f_1+f_4$	\\
15	&	3.763(1)	&	0.70(4)	&	0.151(9)	&	13.6	&	$\sim f_3+13f_{\rm orb}$   	&	27	&	0.266(1)	&	0.19(1)	&	0.153(7)	&	11.4	&	$5f_{\rm orb}$	\\
16	&	0.321(1)	&	0.68(4)	&	0.753(9)	&	13.2	&	$2f_{\rm orb}$ 	&	28	&	0.317(1)	&	0.20(1)	&	0.365(7)	&	11.9	&	$\sim 6f_{\rm orb}$	\\
17	&	14.105(1)	&	0.65(4)	&	0.82(1)	&	12.8	&	$f_1-2f_{\rm orb}$ 	&	29	&	0.457(1)	&	0.20(1)	&	0.460(7)	&	11.7	&	$9f_{\rm orb}$	\\
18	&	13.378(1)	&	0.62(4)	&	0.32(1)	&	12.1	&	$f_1-6f_{\rm orb}$ 	&	30	&	6.090(1)	&	0.19(1)	&	0.158(7)	&	11.4	&	$f_1+18f_{\rm orb}$ 	\\
19	&	0.801(1)	&	0.61(4)	&	0.11(1)	&	12.0	&	$5f_{\rm orb} $	&	31	&	7.771(1)	&	0.19(1)	&	0.011(7)	&	11.2	&	$f_4+11f_{\rm orb}$ 	\\
20	&	3.608(1)	&	0.58(4)	&	0.16(1)	&	11.3	&	$2f_3+3f_{\rm orb}$ 	&	32	&	4.949(1)	&	0.19(1)	&	0.956(7)	&	11.1	&	$f_1-4f_{\rm orb}$	\\
21	&	0.721(2)	&	0.51(4)	&	0.55(1)	&	9.9	&	$f_3-5f_{\rm orb}$ 	&	33	&	8.150(1)	&	0.19(1)	&	0.553(7)	&	11.1	&	$f_3+f_4-f_2$  	\\
22	&	13.148(2)	&	0.47(4)	&	0.32(1)	&	9.2	&	 $f_1-f_3+2f_{\rm orb}$ 	&	34	&	3.348(1)	&	0.18(1)	&	0.869(7)	&	10.7	&	$f_5-f_3+11f_{\rm orb}$  	\\
23	&	1.499(2)	&	0.47(4)	&	0.63(1)	&	9.2	&	$9f_{\rm orb}$  	&	35	&	2.014(1)	&	0.18(1)	&	0.035(7)	&	10.8	&	$\sim f_{15}$	\\
24	&	0.999(2)	&	0.47(4)	&	0.24(1)	&	9.1	&	$6f_{\rm orb}$	&	36	&	5.453(1)	&	0.18(1)	&	0.499(7)	&	10.7	&	$f_1+4f_{\rm orb}$	\\
25	&	1.050(2)	&	0.50(4)	&	0.67(1)	&	9.8	&	$f_3-3f_{\rm orb}$ 	&	37	&	6.058(1)	&	0.18(1)	&	0.028(8)	&	10.4	&	$f_4-23f_{\rm orb}$ 	\\
26	&	27.546(2)	&	0.47(4)	&	0.39(1)	&	9.1	&	 $2f_1-f_3+2f_{\rm orb}$ 	&	38	&	7.256(1)	&	0.18(1)	&	0.039(8)	&	10.4	&	$f_4+f_{\rm orb}$ 	\\
27	&	16.810(2)	&	0.46(4)	&	0.61(1)	&	8.9	&	$f_3+f_1+5f_{\rm orb}$ 	&	39	&	7.705(1)	&	0.18(1)	&	0.820(8)	&	10.4	&	$f_4+10f_{\rm orb}$ 	\\
28	&	28.796(2)	&	0.45(4)	&	0.28(1)	&	8.8	&	$2f_1$	&	40	&	4.718(1)	&	0.17(1)	&	0.232(8)	&	10.3	&	$f_1-9f_{\rm orb}$	\\
29	&	3.652(2)	&	0.45(4)	&	0.95(1)	&	8.7	&	$\sim f_{20}$	&	41	&	5.688(1)	&	0.17(1)	&	0.406(8)	&	10.0	&	$f_1+10f_{\rm orb}$	\\
30	&	1.930(2)	&	0.44(4)	&	0.39(1)	&	8.6	&	$2f_{3}-7f_{\rm orb}$ 	&	42	&	15.697(1)	&	0.17(1)	&	0.304(8)	&	9.9	&	$f_3+5f_{\rm orb}$ 	\\
31	&	3.945(2)	&	0.43(4)	&	0.40(1)	&	8.5	&	$2f_3+5f_{\rm orb}$ 	&	43	&	1.907(1)	&	0.17(1)	&	0.806(8)	&	9.8	&	$\sim f_{15}$	\\
32	&	27.771(2)	&	0.41(4)	&	0.33(2)	&	8.1	&	$2f_{1}-6f_{\rm orb}$ 	&	44	&	6.869(1)	&	0.16(1)	&	0.139(9)	&	9.4	&	$f_4-f_{\rm orb}$ 	\\
33	&	0.120(2)	&	0.40(4)	&	0.82(2)	&	7.7	&	$f_{3}-9f_{\rm orb}$ 	&	45	&	1.460(1)	&	0.16(1)	&	0.938(9)	&	9.3	&	$f_4-f_1-11f_{\rm orb}$ 	\\
34	&	15.867(2)	&	0.38(4)	&	0.09(2)	&	7.4	&	$f_1+9f_{\rm orb}$ 	&	46	&	7.114(1)	&	0.15(1)	&	0.878(9)	&	9.1	&	$f_4-2f_{\rm orb}$ 	\\
35	&	14.059(2)	&	0.38(4)	&	0.26(2)	&	7.5	&	$f_1-2f_{\rm orb}$ 	&	47	&	0.744(1)	&	0.15(1)	&	0.287(9)	&	9.1	&	$15f_{\rm orb}$ 	\\
36	&	2.304(2)	&	0.38(4)	&	0.50(2)	&	7.4	&	$14f_{\rm orb}$ 	&	48	&	8.646(1)	&	0.15(1)	&	0.401(9)	&	8.8	&	$f_4+28f_{\rm orb}$ 	\\
37	&	1.830(2)	&	0.39(4)	&	0.59(2)	&	7.7	&	$f_3+2f_{\rm orb}$ 	&	49	&	6.232(1)	&	0.14(1)	&	0.747(9)	&	8.4	&	$f_1+21f_{\rm orb}$ 	\\
38	&	1.788(2)	&	0.35(4)	&	0.51(2)	&	6.8	&	$2f_{3}-8f_{\rm orb}$ 	&	50	&	1.344(1)	&	0.14(1)	&	0.30(1)	&	8.3	&	$27f_{\rm orb}$ 	\\
39	&	18.333(2)	&	0.35(4)	&	0.05(2)	&	6.8	&	$f_1+2f_3+5f_{\rm orb}$ 	&	51	&	5.636(1)	&	0.14(1)	&	0.16(1)	&	8.2	&	$f_1+9f_{\rm orb}$ 	\\
40	&	6.341(3)	&	0.33(4)	&	0.57(2)	&	6.4	&	$3f_{3}+10f_{\rm orb}$ 	&	52	&	4.805(1)	&	0.14(1)	&	0.95(1)	&	8.4	&	$f_1-8f_{\rm orb}$ 	\\
41	&	14.221(3)	&	0.31(4)	&	0.60(2)	&	6.0	&	$f_1-f_{\rm orb}$ 	&	53	&	5.012(1)	&	0.14(1)	&	0.93(1)	&	8.2	&	$f_1-3f_{\rm orb}$	\\
42	&	12.720(3)	&	0.31(4)	&	0.11(2)	&	6.0	&	$f_1-10f_{\rm orb}$ 	&	54	&	6.581(1)	&	0.13(1)	&	0.98(1)	&	7.5	&	$f_4-13f_{\rm orb}$ 	\\
43	&	3.885(3)	&	0.31(4)	&	0.30(2)	&	6.0	&	$2f_{3}+5f_{\rm orb}$ 	&	55	&	6.474(1)	&	0.13(1)	&	0.20(1)	&	7.6	&	$f_4-15f_{\rm orb}$ 	\\
44	&	4.048(3)	&	0.32(4)	&	0.00(2)	&	6.3	&	$2f_{3}+6f_{\rm orb}$ 	&	56	&	4.562(1)	&	0.13(1)	&	0.41(1)	&	7.5	&	$f_1-13f_{\rm orb}$ 	\\
45	&	29.748(3)	&	0.30(4)	&	0.80(2)	&	5.9	&	$2f_{1}+6f_{\rm orb}$ 	&	57	&	7.617(1)	&	0.12(1)	&	0.36(1)	&	7.2	&	$f_4-8f_{\rm orb}$ 	\\
46	&	2.822(3)	&	0.30(4)	&	0.59(2)	&	5.9	&	$f_3+8f_{\rm orb}$ 	&	58	&	7.948(1)	&	0.12(1)	&	0.96(1)	&	7.2	&	$f_4+15f_{\rm orb}$ 	\\
47	&	3.349(3)	&	0.29(4)	&	0.25(2)	&	5.7	&	$2f_3+2f_{\rm orb}$ 	&	59	&	1.049(1)	&	0.12(1)	&	0.05(1)	&	7.1	&	$21f_{\rm orb}$ 	\\
48	&	0.048(3)	&	0.28(4)	&	0.79(2)	&	5.5	&	$f_3-9f_{\rm orb}$ 	&	60	&	18.023(2)	&	0.12(1)	&	0.17(1)	&	6.8	&	$f_5+14f_{\rm orb}$ 	\\
49	&	13.877(3)	&	0.28(4)	&	0.98(2)	&	5.5	&	$f_1-3f_{\rm orb}$ 	&	61	&	4.359(2)	&	0.11(1)	&	0.88(1)	&	6.6	&	$f_1-17f_{\rm orb}$ 	\\
50	&	23.203(3)	&	0.27(4)	&	0.05(2)	&	5.3	&	$2f_1-3f_3-6f_{\rm orb}$ 	&	62	&	0.626(2)	&	0.11(1)	&	0.05(1)	&	6.6	&	$\sim 12f_{\rm orb}$	\\
51	&	21.450(3)	&	0.27(4)	&	0.05(2)	&	5.3	&	$2f_1-4f_3-7f_{\rm orb}$ 	&	63	&	5.876(2)	&	0.11(1)	&	0.38(1)	&	6.5	&	$f_1+14f_{\rm orb}$ 	\\
52	&	1.089(3)	&	0.27(4)	&	0.92(2)	&	5.2	&	$\sim f_{25}$	&	64	&	12.552(2)	&	0.11(1)	&	0.82(1)	&	6.4	&	$f_5-f_1+6f_{\rm orb}$ 	\\
53	&	28.765(3)	&	0.26(4)	&	0.46(2)	&	5.1	&	$\sim f_{28}$	&	65	&	3.997(2)	&	0.11(1)	&	0.42(1)	&	6.3	&	$f_1-24f_{\rm orb}$ 	\\
\cline{1-6}																							
\multicolumn{6}{c}{CX~Phe}											&	66	&	6.131(2)	&	0.10(1)	&	0.22(1)	&	6.2	&	$f_4-22f_{\rm orb}$ 	\\
\cline{1-6}																							
6	&	14.0271(4)	&	0.50(1)	&	0.005(3)	&	29.3	&	$f_2-9f_{\rm orb}$ 	&	67	&	0.111(2)	&	0.10(1)	&	0.61(1)	&	6.2	&	$2f_{\rm orb}$	\\
7	&	2.0969(4)	&	0.48(1)	&	0.674(3)	&	28.6	&	$f_2-f_1-f_4$	&	68	&	0.419(2)	&	0.11(1)	&	0.27(1)	&	6.3	&	$8f_{\rm orb}$	\\
8	&	8.4514(4)	&	0.48(1)	&	0.001(3)	&	28.2	&	$2f_1-f_5+f_3$ 	&	69	&	1.856(2)	&	0.10(1)	&	0.54(1)	&	6.2	&	$\sim f_{43}$	\\
9	&	15.0618(4)	&	0.47(1)	&	0.940(3)	&	27.9	&	$f_3-8f_{\rm orb}$ 	&	70	&	7.894(2)	&	0.10(1)	&	0.88(1)	&	6.2	&	$f_4+14f_{\rm orb}$ 	\\
10	&	0.0400(4)	&	0.41(1)	&	0.478(3)	&	24.3	&	$\sim f_{\rm orb}$ 	&	71	&	12.847(2)	&	0.10(1)	&	0.48(1)	&	6.0	&	$f_1+f_4+10f_{\rm orb}$  	\\
11	&	0.1608(4)	&	0.42(1)	&	0.808(3)	&	24.7	&	$3f_{\rm orb}$	&	72	&	7.418(2)	&	0.10(1)	&	0.78(1)	&	5.9	&	$f_4+4f_{\rm orb}$ 	\\
12	&	16.2529(5)	&	0.36(1)	&	0.435(4)	&	21.5	&	$f_3+16f_{\rm orb}$ 	&	73	&	7.362(1)	&	0.12(1)	&	0.57(1)	&	7.0	&	$f_4+3f_{\rm orb}$ 	\\
13	&	5.251(1)	&	0.31(1)	&	0.699(4)	&	18.5	&	$f_1+f_{\rm orb}$ 	&	74	&	3.207(2)	&	0.10(1)	&	0.12(1)	&	5.9	&	$2f_1-f_4+f_{\rm orb}$ 	\\
14	&	5.841(1)	&	0.31(1)	&	0.527(4)	&	18.1	&	$f_1+13f_{\rm orb}$ 	&	75	&	4.241(2)	&	0.10(1)	&	0.20(1)	&	5.9	&	$f_1-19f_{\rm orb}$ 	\\
15	&	1.959(1)	&	0.30(1)	&	0.703(5)	&	17.7	&	 $f_4-f_1-f_{\rm orb}$ 	&	76	&	8.824(2)	&	0.10(1)	&	0.13(1)	&	5.8	&	$2f_5-2f_1-f_3$ 	\\
16	&	7.004(1)	&	0.30(1)	&	0.966(5)	&	17.6	&	$f_4-f_{\rm orb}$ 	&	77	&	13.909(2)	&	0.10(1)	&	0.48(1)	&	5.7	&	$f_2-12f_{\rm orb}$	\\
17	&	7.799(1)	&	0.29(1)	&	0.312(5)	&	17.2	&	$f_4+12f_{\rm orb}$ 	&	78	&	1.625(2)	&	0.09(1)	&	0.38(1)	&	5.5	&	$f_4-f_1-f_{\rm orb}$  	\\
18	&	10.038(1)	&	0.27(1)	&	0.021(5)	&	16.1	&	$f_5-f_2+f_4$ 	&	79	&	0.528(2)	&	0.09(1)	&	0.27(1)	&	5.4	&	$11f_{\rm orb}$ 	\\
19	&	6.546(1)	&	0.27(1)	&	0.798(5)	&	15.9	&	$f_4-13f_{\rm orb}$  	&	80	&	12.930(2)	&	0.09(1)	&	0.51(1)	&	5.4	&	$f_4+f_1+10f_{\rm orb}$ 	\\
20	&	0.984(1)	&	0.27(1)	&	0.321(5)	&	15.8	&	$f_3-f_2$	&	81	&	6.785(2)	&	0.09(1)	&	0.80(1)	&	5.4	&	$f_2-f_4-10f_{\rm orb}$	\\
21	&	0.212(1)	&	0.27(1)	&	0.275(5)	&	16.0	&	$4f_{\rm orb}$	&	82	&	8.253(2)	&	0.09(1)	&	0.09(2)	&	5.1	&	$f_3-f_4$	\\
22	&	6.652(1)	&	0.23(1)	&	0.179(6)	&	13.4	&	$f_4-11f_{\rm orb}$ 	&	83	&	5.924(2)	&	0.09(1)	&	0.57(2)	&	5.1	&	$f_1+15f_{\rm orb}$ 	\\
23	&	8.943(1)	&	0.22(1)	&	0.173(6)	&	13.0	&	$f_3-f_4+14f_{\rm orb}$ 	&	84	&	13.371(2)	&	0.08(1)	&	0.02(2)	&	5.0	&	$f_2-23f_{\rm orb}$ 	\\
\hline																																																																								
\end{tabular}}
\end{center}
\end{table*}

\begin{table*}
\begin{center}											
\caption*{Table~\ref{tab:fcombo} (cont'd)}
\label{Tab:FreqsDep}	
\scalebox{0.95}{
\begin{tabular}{cc cc cc cc cc cc}														
\hline\hline																							
$i$	&	  $f_{\rm i}$	&	$A$	&	  $\Phi$	&	S/N	&	Combination	&	$i$	&	  $f_{\rm i}$	&	$A$	&	  $\Phi$	&	S/N	&	Combination	\\
	&	     (d$^{-1}$)	&	(mmag)	&	(2$\pi$~rad)	&		&		&		&	     (d$^{-1}$)	&	(mmag)	&	(2$\pi$~rad)	&		&		\\
\hline																							
\multicolumn{12}{c}{TIC 35481236}																							\\
\hline																							
1	&	2.1702(1)	&	1.17(1)	&	0.835(2)	&	64.7	&	$4f_{\rm orb}$	&	39	&	0.0651(9)	&	0.15(1)	&	0.66(1)	&	8.3	&	$f_1-3f_8+6f_{\rm orb}$ 	\\
2	&	0.0142(1)	&	0.93(1)	&	0.193(2)	&	51.4	&	???	&	40	&	1.045(1)	&	0.14(1)	&	0.72(1)	&	7.9	&	$\sim 2f_{\rm orb}$	\\
3	&	1.6285(2)	&	0.79(1)	&	0.593(2)	&	43.9	&	$3f_{\rm orb}$	&	41	&	10.486(1)	&	0.14(1)	&	0.55(1)	&	7.6	&	$\sim f_8+16f_{\rm orb}$ 	\\
4	&	0.5568(3)	&	0.52(1)	&	0.225(4)	&	28.8	&	$\sim f_{\rm orb}$	&	42	&	8.954(1)	&	0.13(1)	&	0.14(1)	&	7.4	&	$2f_8+10f_{\rm orb}$ 	\\
7	&	1.0869(3)	&	0.45(1)	&	0.093(4)	&	24.7	&	$2f_{\rm orb}$	&	43	&	10.036(1)	&	0.14(1)	&	0.80(1)	&	7.5	&	$2f_8+12f_{\rm orb}$ 	\\
9	&	3.4562(4)	&	0.30(1)	&	0.601(6)	&	16.8	&	$\sim f_8+3f_{\rm orb}$ 	&	44	&	1.259(1)	&	0.13(1)	&	0.55(1)	&	7.3	&	$f_8-f_{\rm orb}$	\\
10	&	19.5424(4)	&	0.30(1)	&	0.669(6)	&	16.7	&	$f_5-10f_{\rm orb}$ 	&	45	&	10.763(1)	&	0.13(1)	&	0.88(1)	&	7.2	&	$6f_8$ 	\\
11	&	1.8605(5)	&	0.28(1)	&	0.492(7)	&	15.4	&	$f_8+5f_2$ 	&	46	&	1.826(1)	&	0.13(1)	&	0.84(1)	&	7.1	&	$f_8+f_2$	\\
12	&	2.0163(5)	&	0.26(1)	&	0.687(7)	&	14.2	&	$7f_{\rm orb}-f_8$ 	&	47	&	0.080(1)	&	0.13(1)	&	0.66(1)	&	7.0	&	$\sim f_{39}$	\\
13	&	2.1528(5)	&	0.26(1)	&	0.968(7)	&	14.1	&	$4f_{\rm orb}$	&	48	&	8.134(1)	&	0.12(1)	&	0.40(2)	&	6.9	&	$15f_{\rm orb}$	\\
14	&	9.5299(5)	&	0.25(1)	&	0.462(7)	&	13.8	&	$2f_8+11f_{\rm orb}$ 	&	49	&	11.864(1)	&	0.12(1)	&	0.77(2)	&	6.8	&	$6f_8+2f_{\rm orb}$ 	\\
15	&	9.3741(5)	&	0.26(1)	&	0.132(7)	&	14.2	&	$f_8+14f_{\rm orb}$ 	&	50	&	0.495(1)	&	0.12(1)	&	0.59(2)	&	6.6	&	$3f_8-9f_{\rm orb}$ 	\\
16	&	0.6096(5)	&	0.25(1)	&	0.390(8)	&	13.7	&	$f_{\rm orb}+2f_2$	&	51	&	18.574(1)	&	0.12(1)	&	0.40(2)	&	6.4	&	$f_6-4f_{\rm orb}$	\\
17	&	3.2548(6)	&	0.23(1)	&	0.856(8)	&	12.8	&	$6f_{\rm orb}$	&	52	&	1.191(1)	&	0.11(1)	&	0.09(2)	&	6.4	&	$\sim f_8-f_{\rm orb}$ 	\\
18	&	0.7184(6)	&	0.23(1)	&	0.158(8)	&	12.6	&	$f_8-2f_{\rm orb}$	&	53	&	13.061(1)	&	0.11(1)	&	0.30(2)	&	6.3	&	$f_5-22f_{\rm orb}$ 	\\
19	&	1.7368(6)	&	0.23(1)	&	0.963(8)	&	12.9	&	$f_8-2f_2$	&	54	&	11.438(1)	&	0.11(1)	&	0.09(2)	&	6.3	&	$21f_{\rm orb}$	\\
20	&	4.3414(6)	&	0.22(1)	&	0.416(9)	&	12.0	&	$8f_{\rm orb}$	&	55	&	10.739(1)	&	0.11(1)	&	0.23(2)	&	6.2	&	$3f_8+10f_{\rm orb}$ 	\\
21	&	0.1274(7)	&	0.20(1)	&	0.544(9)	&	11.3	&	$f_8-3f_{\rm orb}$ 	&	56	&	3.064(1)	&	0.11(1)	&	0.55(2)	&	6.2	&	$2f_8-f_{\rm orb}$ 	\\
22	&	2.8897(7)	&	0.20(1)	&	0.183(9)	&	11.3	&	$f_8+2f_{\rm orb}$ 	&	57	&	1.381(1)	&	0.11(1)	&	0.70(2)	&	6.2	&	$2f_8-4f_{\rm orb}$ 	\\
23	&	3.1571(7)	&	0.19(1)	&	0.82(1)	&	10.5	&	$3f_8-4f_{\rm orb}$ 	&	58	&	10.015(1)	&	0.11(1)	&	0.95(2)	&	6.1	&	$2f_8+12f_{\rm orb}$ 	\\
24	&	10.4223(7)	&	0.19(1)	&	0.47(1)	&	10.4	&	$f_8+16f_{\rm orb}$ 	&	59	&	11.703(1)	&	0.11(1)	&	0.95(2)	&	6.1	&	$2f_8+15f_{\rm orb}$ 	\\
25	&	25.6006(7)	&	0.19(1)	&	0.17(1)	&	10.4	&	$ f_6+9f_{\rm orb}$ 	&	60	&	9.308(1)	&	0.11(1)	&	0.10(2)	&	5.9	&	$17f_{\rm orb}$ 	\\
26	&	1.9332(7)	&	0.18(1)	&	0.86(1)	&	10.0	&	$2f_8-3f_{\rm orb}$ 	&	61	&	12.165(1)	&	0.11(1)	&	0.30(2)	&	5.8	&	$f_8+19f_{\rm orb}$ 	\\
27	&	1.1555(8)	&	0.18(1)	&	0.44(1)	&	9.9	&	$2f_{\rm orb}+4f_2$ 	&	62	&	0.218(1)	&	0.11(1)	&	0.48(2)	&	5.8	&	$f_8-3f_{\rm orb}$ 	\\
28	&	19.6577(8)	&	0.18(1)	&	0.93(1)	&	9.8	&	$f_6-2f_{\rm orb}$	&	63	&	2.334(1)	&	0.10(1)	&	0.98(2)	&	5.7	&	$f_8+f_{\rm orb}$	\\
29	&	26.6106(8)	&	0.17(1)	&	0.06(1)	&	9.6	&	$f_5+3f_{\rm orb}$	&	64	&	3.703(1)	&	0.10(1)	&	0.80(2)	&	5.6	&	$3f_8-3f_{\rm orb}$ 	\\
30	&	2.1068(8)	&	0.17(1)	&	0.15(1)	&	9.4	&	$3f_8-6f_{\rm orb}$ 	&	65	&	11.375(1)	&	0.10(1)	&	0.64(2)	&	5.5	&	$21f_{\rm orb}$ 	\\
31	&	10.3724(8)	&	0.17(1)	&	0.83(1)	&	9.3	&	$f_6-19f_{\rm orb}$ 	&	66	&	1.330(1)	&	0.10(1)	&	0.24(2)	&	5.5	&	$\sim 2f_8-4f_{\rm orb}$ 	\\
32	&	2.6311(8)	&	0.17(1)	&	0.43(1)	&	9.2	&	$3f_8-5f_{\rm orb}$ 	&	67	&	10.927(1)	&	0.10(1)	&	0.86(2)	&	5.4	&	$f_6-18f_{\rm orb}$ 	\\
33	&	11.0737(8)	&	0.17(1)	&	0.60(1)	&	9.2	&	$f_8+17f_{\rm orb}$ 	&	68	&	10.992(1)	&	0.10(1)	&	0.57(2)	&	5.3	&	$f_8+17f_{\rm orb}$ 	\\
34	&	10.5322(8)	&	0.17(1)	&	1.00(1)	&	9.2	&	$f_8+16f_{\rm orb}$ 	&	69	&	12.312(1)	&	0.10(1)	&	0.03(2)	&	5.3	&	$2f_8+16f_{\rm orb}$ 	\\
35	&	3.8012(8)	&	0.16(1)	&	0.70(1)	&	9.1	&	$7f_{\rm orb}$	&	70	&	1.104(1)	&	0.09(1)	&	0.35(2)	&	5.1	&	$\sim 2f_{\rm orb}$	\\
36	&	2.3712(9)	&	0.16(1)	&	0.42(1)	&	8.7	&	$f_8+f_{\rm orb}$	&	71	&	2.692(1)	&	0.10(1)	&	0.43(2)	&	5.7	&	$\sim 5f_{\rm orb}$	\\
37	&	0.5449(9)	&	0.15(1)	&	0.28(1)	&	8.4	&	$\sim f_{\rm orb}$	&	72	&	0.305(1)	&	0.09(1)	&	0.65(2)	&	5.2	&	$2f_8-6f_{\rm orb}$ 	\\
38	&	12.3756(9)	&	0.15(1)	&	0.76(1)	&	8.3	&	$3f_8+13f_{\rm orb}$ 	&	73	&	9.574(1)	&	0.09(1)	&	0.90(2)	&	5.0	&	$2f_8+11f_{\rm orb}$ 	\\
\hline																														
\end{tabular}}
\end{center}
\end{table*}

\end{appendix}
\end{document}